\documentclass[a4paper,twocolumn,traditabstract]{aa}
\usepackage{graphicx}
\usepackage[breaklinks, colorlinks, citecolor=blue]{hyperref}
\usepackage{natbib}
\usepackage{stfloats}
\usepackage{subfigure}
\usepackage{amsmath}
\usepackage{epsfig}
\usepackage{psfrag}
\usepackage{multirow}
\usepackage{color}
\usepackage{txfonts}

\usepackage{url}

\newcommand{\healpix}{{\tt HEALPix}}

\def\pdeg{\ifmmode $\setbox0=\hbox{$^{\circ}$}\rlap{\hskip.11\wd0 .}$^{\circ}
 \else \setbox0=\hbox{$^{\circ}$}\rlap{\hskip.11\wd0 .}$^{\circ}$\fi}

\def\setsymbol#1#2{\expandafter\def\csname #1\endcsname{#2}}
\def\getsymbol#1{\csname #1\endcsname}

\def\Planck{\textit{Planck}}

\def\all2013resultspapers{\nocite{planck2013-p01, planck2013-p02, planck2013-p02a, planck2013-p02d, planck2013-p02b, planck2013-p03, planck2013-p03c, planck2013-p03f, planck2013-p03d, planck2013-p03e, planck2013-p01a, planck2013-p06, planck2013-p03a, planck2013-pip88, planck2013-p08, planck2013-p11, planck2013-p12, planck2013-p13, planck2013-p14, planck2013-p15, planck2013-p05b, planck2013-p17, planck2013-p09, planck2013-p09a, planck2013-p20, planck2013-p19, planck2013-pipaberration, planck2013-p05, planck2013-p05a, planck2013-pip56, planck2013-p06b}}

\newbox\tablebox    \newdimen\tablewidth
\def\leaderfil{\leaders\hbox to 5pt{\hss.\hss}\hfil}
\def\tablenote#1 #2\par{\begingroup \parindent=0.8em
    \abovedisplayshortskip=0pt\belowdisplayshortskip=0pt
    \noindent
    $$\hss\vbox{\hsize\tablewidth \hangindent=\parindent \hangafter=1 \noindent
    \hbox to \parindent{$^#1$\hss}\strut#2\strut\par}\hss$$
    \endgroup}

\def\L2{\ifmmode L_2\else $L_2$\fi}

\def\DeltaT{\ifmmode \Delta T\else $\Delta T$\fi}
\def\deltat{\ifmmode \Delta t\else $\Delta t$\fi}
\def\fknee{\ifmmode f_{\rm knee}\else $f_{\rm knee}$\fi}
\def\Fmax{\ifmmode F_{\rm max}\else $F_{\rm max}$\fi}
\def\solar{\ifmmode{\rm M}_{\mathord\odot}\else${\rm M}_{\mathord\odot}$\fi}
\def\Msolar{\ifmmode{\rm M}_{\mathord\odot}\else${\rm M}_{\mathord\odot}$\fi}
\def\Lsolar{\ifmmode{\rm L}_{\mathord\odot}\else${\rm L}_{\mathord\odot}$\fi}

\def\inv{\ifmmode^{-1}\else$^{-1}$\fi}
\def\mo{\ifmmode^{-1}\else$^{-1}$\fi}
\def\sup#1{\ifmmode ^{\rm #1}\else $^{\rm #1}$\fi}
\def\expo#1{\ifmmode \times 10^{#1}\else $\times 10^{#1}$\fi}
\def\,{\thinspace}
\def\lsim{\mathrel{\raise .4ex\hbox{\rlap{$<$}\lower 1.2ex\hbox{$\sim$}}}}
\def\gsim{\mathrel{\raise .4ex\hbox{\rlap{$>$}\lower 1.2ex\hbox{$\sim$}}}}

\def\simprop{\mathrel{\raise .4ex\hbox{\rlap{$\propto$}\lower 1.2ex\hbox{$\sim$}}}}
\def\deg{\ifmmode^\circ\else$^\circ$\fi}
\def\pdeg{\ifmmode $\setbox0=\hbox{$^{\circ}$}\rlap{\hskip.11\wd0 .}$^{\circ}
          \else \setbox0=\hbox{$^{\circ}$}\rlap{\hskip.11\wd0 .}$^{\circ}$\fi}
\def\arcs{\ifmmode {^{\scriptstyle\prime\prime}}
          \else $^{\scriptstyle\prime\prime}$\fi}
\def\arcm{\ifmmode {^{\scriptstyle\prime}}
          \else $^{\scriptstyle\prime}$\fi}
\newdimen\sa  \newdimen\sb
\def\parcs{\sa=.07em \sb=.03em
     \ifmmode \hbox{\rlap{.}}^{\scriptstyle\prime\kern -\sb\prime}\hbox{\kern -\sa}
     \else \rlap{.}$^{\scriptstyle\prime\kern -\sb\prime}$\kern -\sa\fi}
\def\parcm{\sa=.08em \sb=.03em
     \ifmmode \hbox{\rlap{.}\kern\sa}^{\scriptstyle\prime}\hbox{\kern-\sb}
     \else \rlap{.}\kern\sa$^{\scriptstyle\prime}$\kern-\sb\fi}
\def\ra[#1 #2 #3.#4]{#1\sup{h}#2\sup{m}#3\sup{s}\llap.#4}
\def\dec[#1 #2 #3.#4]{#1\deg#2\arcm#3\arcs\llap.#4}
\def\deco[#1 #2 #3]{#1\deg#2\arcm#3\arcs}
\def\rra[#1 #2]{#1\sup{h}#2\sup{m}}

\def\dots{\relax\ifmmode \ldots\else $\ldots$\fi}
\def\WHzsr{\ifmmode $W\,Hz\mo\,sr\mo$\else W\,Hz\mo\,sr\mo\fi}
\def\mHz{\ifmmode $\,mHz$\else \,mHz\fi}
\def\GHz{\ifmmode $\,GHz$\else \,GHz\fi}
\def\mKs{\ifmmode $\,mK\,s$^{1/2}\else \,mK\,s$^{1/2}$\fi}
\def\muKs{\ifmmode \,\mu$K\,s$^{1/2}\else \,$\mu$K\,s$^{1/2}$\fi}
\def\muKRJs{\ifmmode \,\mu$K$_{\rm RJ}$\,s$^{1/2}\else \,$\mu$K$_{\rm RJ}$\,s$^{1/2}$\fi}
\def\muKHz{\ifmmode \,\mu$K\,Hz$^{-1/2}\else \,$\mu$K\,Hz$^{-1/2}$\fi}
\def\MJysr{\ifmmode \,$MJy\,sr\mo$\else \,MJy\,sr\mo\fi}
\def\MJysrmK{\ifmmode \,$MJy\,sr\mo$\,mK$_{\rm CMB}\mo\else \,MJy\,sr\mo\,mK$_{\rm CMB}\mo$\fi}
\def\microns{\ifmmode \,\mu$m$\else \,$\mu$m\fi}

\def\muK{\ifmmode \,\mu$K$\else \,$\mu$\hbox{K}\fi}
\def\microK{\ifmmode \,\mu$K$\else \,$\mu$\hbox{K}\fi}
\def\muW{\ifmmode \,\mu$W$\else \,$\mu$\hbox{W}\fi}
\def\kms{\ifmmode $\,km\,s$^{-1}\else \,km\,s$^{-1}$\fi}
\def\kmsMpc{\ifmmode $\,\kms\,Mpc\mo$\else \,\kms\,Mpc\mo\fi}
\setsymbol{LFI:center:frequency:70GHz:units}{70.3\,GHz}
\setsymbol{LFI:center:frequency:44GHz:units}{44.1\,GHz}
\setsymbol{LFI:center:frequency:30GHz:units}{28.5\,GHz}

\setsymbol{LFI:center:frequency:70GHz}{70.3}
\setsymbol{LFI:center:frequency:44GHz}{44.1}
\setsymbol{LFI:center:frequency:30GHz}{28.5}

\setsymbol{LFI:center:frequency:LFI18:Rad:M:units}{71.7\GHz}
\setsymbol{LFI:center:frequency:LFI19:Rad:M:units}{67.5\GHz}
\setsymbol{LFI:center:frequency:LFI20:Rad:M:units}{69.2\GHz}
\setsymbol{LFI:center:frequency:LFI21:Rad:M:units}{70.4\GHz}
\setsymbol{LFI:center:frequency:LFI22:Rad:M:units}{71.5\GHz}
\setsymbol{LFI:center:frequency:LFI23:Rad:M:units}{70.8\GHz}
\setsymbol{LFI:center:frequency:LFI24:Rad:M:units}{44.4\GHz}
\setsymbol{LFI:center:frequency:LFI25:Rad:M:units}{44.0\GHz}
\setsymbol{LFI:center:frequency:LFI26:Rad:M:units}{43.9\GHz}
\setsymbol{LFI:center:frequency:LFI27:Rad:M:units}{28.3\GHz}
\setsymbol{LFI:center:frequency:LFI28:Rad:M:units}{28.8\GHz}
\setsymbol{LFI:center:frequency:LFI18:Rad:S:units}{70.1\GHz}
\setsymbol{LFI:center:frequency:LFI19:Rad:S:units}{69.6\GHz}
\setsymbol{LFI:center:frequency:LFI20:Rad:S:units}{69.5\GHz}
\setsymbol{LFI:center:frequency:LFI21:Rad:S:units}{69.5\GHz}
\setsymbol{LFI:center:frequency:LFI22:Rad:S:units}{72.8\GHz}
\setsymbol{LFI:center:frequency:LFI23:Rad:S:units}{71.3\GHz}
\setsymbol{LFI:center:frequency:LFI24:Rad:S:units}{44.1\GHz}
\setsymbol{LFI:center:frequency:LFI25:Rad:S:units}{44.1\GHz}
\setsymbol{LFI:center:frequency:LFI26:Rad:S:units}{44.1\GHz}
\setsymbol{LFI:center:frequency:LFI27:Rad:S:units}{28.5\GHz}
\setsymbol{LFI:center:frequency:LFI28:Rad:S:units}{28.2\GHz}

\setsymbol{LFI:center:frequency:LFI18:Rad:M}{71.7}
\setsymbol{LFI:center:frequency:LFI19:Rad:M}{67.5}
\setsymbol{LFI:center:frequency:LFI20:Rad:M}{69.2}
\setsymbol{LFI:center:frequency:LFI21:Rad:M}{70.4}
\setsymbol{LFI:center:frequency:LFI22:Rad:M}{71.5}
\setsymbol{LFI:center:frequency:LFI23:Rad:M}{70.8}
\setsymbol{LFI:center:frequency:LFI24:Rad:M}{44.4}
\setsymbol{LFI:center:frequency:LFI25:Rad:M}{44.0}
\setsymbol{LFI:center:frequency:LFI26:Rad:M}{43.9}
\setsymbol{LFI:center:frequency:LFI27:Rad:M}{28.3}
\setsymbol{LFI:center:frequency:LFI28:Rad:M}{28.8}
\setsymbol{LFI:center:frequency:LFI18:Rad:S}{70.1}
\setsymbol{LFI:center:frequency:LFI19:Rad:S}{69.6}
\setsymbol{LFI:center:frequency:LFI20:Rad:S}{69.5}
\setsymbol{LFI:center:frequency:LFI21:Rad:S}{69.5}
\setsymbol{LFI:center:frequency:LFI22:Rad:S}{72.8}
\setsymbol{LFI:center:frequency:LFI23:Rad:S}{71.3}
\setsymbol{LFI:center:frequency:LFI24:Rad:S}{44.1}
\setsymbol{LFI:center:frequency:LFI25:Rad:S}{44.1}
\setsymbol{LFI:center:frequency:LFI26:Rad:S}{44.1}
\setsymbol{LFI:center:frequency:LFI27:Rad:S}{28.5}
\setsymbol{LFI:center:frequency:LFI28:Rad:S}{28.2}

\setsymbol{LFI:white:noise:sensitivity:70GHz:units}{134.7\muKs}
\setsymbol{LFI:white:noise:sensitivity:44GHz:units}{164.7\muKs}
\setsymbol{LFI:white:noise:sensitivity:30GHz:units}{143.4\muKs}

\setsymbol{LFI:white:noise:sensitivity:70GHz}{134.7}
\setsymbol{LFI:white:noise:sensitivity:44GHz}{164.7}
\setsymbol{LFI:white:noise:sensitivity:30GHz}{143.4}

\setsymbol{LFI:white:noise:sensitivity:LFI18:Rad:M:units}{512.0\muKs}
\setsymbol{LFI:white:noise:sensitivity:LFI19:Rad:M:units}{581.4\muKs}
\setsymbol{LFI:white:noise:sensitivity:LFI20:Rad:M:units}{590.8\muKs}
\setsymbol{LFI:white:noise:sensitivity:LFI21:Rad:M:units}{455.2\muKs}
\setsymbol{LFI:white:noise:sensitivity:LFI22:Rad:M:units}{492.0\muKs}
\setsymbol{LFI:white:noise:sensitivity:LFI23:Rad:M:units}{507.7\muKs}
\setsymbol{LFI:white:noise:sensitivity:LFI24:Rad:M:units}{462.2\muKs}
\setsymbol{LFI:white:noise:sensitivity:LFI25:Rad:M:units}{413.6\muKs}
\setsymbol{LFI:white:noise:sensitivity:LFI26:Rad:M:units}{478.6\muKs}
\setsymbol{LFI:white:noise:sensitivity:LFI27:Rad:M:units}{277.7\muKs}
\setsymbol{LFI:white:noise:sensitivity:LFI28:Rad:M:units}{312.3\muKs}
\setsymbol{LFI:white:noise:sensitivity:LFI18:Rad:S:units}{465.7\muKs}
\setsymbol{LFI:white:noise:sensitivity:LFI19:Rad:S:units}{555.6\muKs}
\setsymbol{LFI:white:noise:sensitivity:LFI20:Rad:S:units}{623.2\muKs}
\setsymbol{LFI:white:noise:sensitivity:LFI21:Rad:S:units}{564.1\muKs}
\setsymbol{LFI:white:noise:sensitivity:LFI22:Rad:S:units}{534.4\muKs}
\setsymbol{LFI:white:noise:sensitivity:LFI23:Rad:S:units}{542.4\muKs}
\setsymbol{LFI:white:noise:sensitivity:LFI24:Rad:S:units}{399.2\muKs}
\setsymbol{LFI:white:noise:sensitivity:LFI25:Rad:S:units}{392.6\muKs}
\setsymbol{LFI:white:noise:sensitivity:LFI26:Rad:S:units}{418.6\muKs}
\setsymbol{LFI:white:noise:sensitivity:LFI27:Rad:S:units}{302.9\muKs}
\setsymbol{LFI:white:noise:sensitivity:LFI28:Rad:S:units}{285.3\muKs}

\setsymbol{LFI:white:noise:sensitivity:LFI18:Rad:M}{512.0}
\setsymbol{LFI:white:noise:sensitivity:LFI19:Rad:M}{581.4}
\setsymbol{LFI:white:noise:sensitivity:LFI20:Rad:M}{590.8}
\setsymbol{LFI:white:noise:sensitivity:LFI21:Rad:M}{455.2}
\setsymbol{LFI:white:noise:sensitivity:LFI22:Rad:M}{492.0}
\setsymbol{LFI:white:noise:sensitivity:LFI23:Rad:M}{507.7}
\setsymbol{LFI:white:noise:sensitivity:LFI24:Rad:M}{462.2}
\setsymbol{LFI:white:noise:sensitivity:LFI25:Rad:M}{413.6}
\setsymbol{LFI:white:noise:sensitivity:LFI26:Rad:M}{478.6}
\setsymbol{LFI:white:noise:sensitivity:LFI27:Rad:M}{277.7}
\setsymbol{LFI:white:noise:sensitivity:LFI28:Rad:M}{312.3}
\setsymbol{LFI:white:noise:sensitivity:LFI18:Rad:S}{465.7}
\setsymbol{LFI:white:noise:sensitivity:LFI19:Rad:S}{555.6}
\setsymbol{LFI:white:noise:sensitivity:LFI20:Rad:S}{623.2}
\setsymbol{LFI:white:noise:sensitivity:LFI21:Rad:S}{564.1}
\setsymbol{LFI:white:noise:sensitivity:LFI22:Rad:S}{534.4}
\setsymbol{LFI:white:noise:sensitivity:LFI23:Rad:S}{542.4}
\setsymbol{LFI:white:noise:sensitivity:LFI24:Rad:S}{399.2}
\setsymbol{LFI:white:noise:sensitivity:LFI25:Rad:S}{392.6}
\setsymbol{LFI:white:noise:sensitivity:LFI26:Rad:S}{418.6}
\setsymbol{LFI:white:noise:sensitivity:LFI27:Rad:S}{302.9}
\setsymbol{LFI:white:noise:sensitivity:LFI28:Rad:S}{285.3}

\setsymbol{LFI:knee:frequency:70GHz:units}{29.5\mHz}
\setsymbol{LFI:knee:frequency:44GHz:units}{56.2\mHz}
\setsymbol{LFI:knee:frequency:30GHz:units}{113.7\mHz}

\setsymbol{LFI:knee:frequency:70GHz}{29.5}
\setsymbol{LFI:knee:frequency:44GHz}{56.2}
\setsymbol{LFI:knee:frequency:30GHz}{113.7}

\setsymbol{LFI:knee:frequency:LFI18:Rad:M:units}{16.3\mHz}
\setsymbol{LFI:knee:frequency:LFI19:Rad:M:units}{15.1\mHz}
\setsymbol{LFI:knee:frequency:LFI20:Rad:M:units}{18.7\mHz}
\setsymbol{LFI:knee:frequency:LFI21:Rad:M:units}{37.2\mHz}
\setsymbol{LFI:knee:frequency:LFI22:Rad:M:units}{12.7\mHz}
\setsymbol{LFI:knee:frequency:LFI23:Rad:M:units}{34.6\mHz}
\setsymbol{LFI:knee:frequency:LFI24:Rad:M:units}{46.2\mHz}
\setsymbol{LFI:knee:frequency:LFI25:Rad:M:units}{24.9\mHz}
\setsymbol{LFI:knee:frequency:LFI26:Rad:M:units}{67.6\mHz}
\setsymbol{LFI:knee:frequency:LFI27:Rad:M:units}{187.4\mHz}
\setsymbol{LFI:knee:frequency:LFI28:Rad:M:units}{122.2\mHz}
\setsymbol{LFI:knee:frequency:LFI18:Rad:S:units}{17.7\mHz}
\setsymbol{LFI:knee:frequency:LFI19:Rad:S:units}{22.0\mHz}
\setsymbol{LFI:knee:frequency:LFI20:Rad:S:units}{8.7\mHz}
\setsymbol{LFI:knee:frequency:LFI21:Rad:S:units}{25.9\mHz}
\setsymbol{LFI:knee:frequency:LFI22:Rad:S:units}{15.8\mHz}
\setsymbol{LFI:knee:frequency:LFI23:Rad:S:units}{129.8\mHz}
\setsymbol{LFI:knee:frequency:LFI24:Rad:S:units}{100.9\mHz}
\setsymbol{LFI:knee:frequency:LFI25:Rad:S:units}{38.9\mHz}
\setsymbol{LFI:knee:frequency:LFI26:Rad:S:units}{58.9\mHz}
\setsymbol{LFI:knee:frequency:LFI27:Rad:S:units}{104.4\mHz}
\setsymbol{LFI:knee:frequency:LFI28:Rad:S:units}{40.7\mHz}

\setsymbol{LFI:knee:frequency:LFI18:Rad:M}{16.3}
\setsymbol{LFI:knee:frequency:LFI19:Rad:M}{15.1}
\setsymbol{LFI:knee:frequency:LFI20:Rad:M}{18.7}
\setsymbol{LFI:knee:frequency:LFI21:Rad:M}{37.2}
\setsymbol{LFI:knee:frequency:LFI22:Rad:M}{12.7}
\setsymbol{LFI:knee:frequency:LFI23:Rad:M}{34.6}
\setsymbol{LFI:knee:frequency:LFI24:Rad:M}{46.2}
\setsymbol{LFI:knee:frequency:LFI25:Rad:M}{24.9}
\setsymbol{LFI:knee:frequency:LFI26:Rad:M}{67.6}
\setsymbol{LFI:knee:frequency:LFI27:Rad:M}{187.4}
\setsymbol{LFI:knee:frequency:LFI28:Rad:M}{122.2}
\setsymbol{LFI:knee:frequency:LFI18:Rad:S}{17.7}
\setsymbol{LFI:knee:frequency:LFI19:Rad:S}{22.0}
\setsymbol{LFI:knee:frequency:LFI20:Rad:S}{8.7}
\setsymbol{LFI:knee:frequency:LFI21:Rad:S}{25.9}
\setsymbol{LFI:knee:frequency:LFI22:Rad:S}{15.8}
\setsymbol{LFI:knee:frequency:LFI23:Rad:S}{129.8}
\setsymbol{LFI:knee:frequency:LFI24:Rad:S}{100.9}
\setsymbol{LFI:knee:frequency:LFI25:Rad:S}{38.9}
\setsymbol{LFI:knee:frequency:LFI26:Rad:S}{58.9}
\setsymbol{LFI:knee:frequency:LFI27:Rad:S}{104.4}
\setsymbol{LFI:knee:frequency:LFI28:Rad:S}{40.7}

\setsymbol{LFI:slope:70GHz:units}{$-1.03$\mHz}
\setsymbol{LFI:slope:44GHz:units}{$-0.89$\mHz}
\setsymbol{LFI:slope:30GHz:units}{$-0.87$\mHz}

\setsymbol{LFI:slope:70GHz}{$-1.03$}
\setsymbol{LFI:slope:44GHz}{$-0.89$}
\setsymbol{LFI:slope:30GHz}{$-0.87$}

\setsymbol{LFI:slope:LFI18:Rad:M:units}{$-1.04$\mHz}
\setsymbol{LFI:slope:LFI19:Rad:M:units}{$-1.09$\mHz}
\setsymbol{LFI:slope:LFI20:Rad:M:units}{$-0.69$\mHz}
\setsymbol{LFI:slope:LFI21:Rad:M:units}{$-1.56$\mHz}
\setsymbol{LFI:slope:LFI22:Rad:M:units}{$-1.01$\mHz}
\setsymbol{LFI:slope:LFI23:Rad:M:units}{$-0.96$\mHz}
\setsymbol{LFI:slope:LFI24:Rad:M:units}{$-0.83$\mHz}
\setsymbol{LFI:slope:LFI25:Rad:M:units}{$-0.91$\mHz}
\setsymbol{LFI:slope:LFI26:Rad:M:units}{$-0.95$\mHz}
\setsymbol{LFI:slope:LFI27:Rad:M:units}{$-0.87$\mHz}
\setsymbol{LFI:slope:LFI28:Rad:M:units}{$-0.88$\mHz}
\setsymbol{LFI:slope:LFI18:Rad:S:units}{$-1.15$\mHz}
\setsymbol{LFI:slope:LFI19:Rad:S:units}{$-1.00$\mHz}
\setsymbol{LFI:slope:LFI20:Rad:S:units}{$-0.95$\mHz}
\setsymbol{LFI:slope:LFI21:Rad:S:units}{$-0.92$\mHz}
\setsymbol{LFI:slope:LFI22:Rad:S:units}{$-1.01$\mHz}
\setsymbol{LFI:slope:LFI23:Rad:S:units}{$-0.95$\mHz}
\setsymbol{LFI:slope:LFI24:Rad:S:units}{$-0.73$\mHz}
\setsymbol{LFI:slope:LFI25:Rad:S:units}{$-1.16$\mHz}
\setsymbol{LFI:slope:LFI26:Rad:S:units}{$-0.79$\mHz}
\setsymbol{LFI:slope:LFI27:Rad:S:units}{$-0.82$\mHz}
\setsymbol{LFI:slope:LFI28:Rad:S:units}{$-0.91$\mHz}

\setsymbol{LFI:slope:LFI18:Rad:M}{$-1.04$}
\setsymbol{LFI:slope:LFI19:Rad:M}{$-1.09$}
\setsymbol{LFI:slope:LFI20:Rad:M}{$-0.69$}
\setsymbol{LFI:slope:LFI21:Rad:M}{$-1.56$}
\setsymbol{LFI:slope:LFI22:Rad:M}{$-1.01$}
\setsymbol{LFI:slope:LFI23:Rad:M}{$-0.96$}
\setsymbol{LFI:slope:LFI24:Rad:M}{$-0.83$}
\setsymbol{LFI:slope:LFI25:Rad:M}{$-0.91$}
\setsymbol{LFI:slope:LFI26:Rad:M}{$-0.95$}
\setsymbol{LFI:slope:LFI27:Rad:M}{$-0.87$}
\setsymbol{LFI:slope:LFI28:Rad:M}{$-0.88$}
\setsymbol{LFI:slope:LFI18:Rad:S}{$-1.15$}
\setsymbol{LFI:slope:LFI19:Rad:S}{$-1.00$}
\setsymbol{LFI:slope:LFI20:Rad:S}{$-0.95$}
\setsymbol{LFI:slope:LFI21:Rad:S}{$-0.92$}
\setsymbol{LFI:slope:LFI22:Rad:S}{$-1.01$}
\setsymbol{LFI:slope:LFI23:Rad:S}{$-0.95$}
\setsymbol{LFI:slope:LFI24:Rad:S}{$-0.73$}
\setsymbol{LFI:slope:LFI25:Rad:S}{$-1.16$}
\setsymbol{LFI:slope:LFI26:Rad:S}{$-0.79$}
\setsymbol{LFI:slope:LFI27:Rad:S}{$-0.82$}
\setsymbol{LFI:slope:LFI28:Rad:S}{$-0.91$}

\setsymbol{LFI:FWHM:70GHz:units}{13\parcm01}
\setsymbol{LFI:FWHM:44GHz:units}{27\parcm92}
\setsymbol{LFI:FWHM:30GHz:units}{32\parcm65}

\setsymbol{LFI:FWHM:70GHz}{13.01}
\setsymbol{LFI:FWHM:44GHz}{27.92}
\setsymbol{LFI:FWHM:30GHz}{32.65}

\setsymbol{LFI:FWHM:LFI18:units}{13\parcm39}
\setsymbol{LFI:FWHM:LFI19:units}{13\parcm01}
\setsymbol{LFI:FWHM:LFI20:units}{12\parcm75}
\setsymbol{LFI:FWHM:LFI21:units}{12\parcm74}
\setsymbol{LFI:FWHM:LFI22:units}{12\parcm87}
\setsymbol{LFI:FWHM:LFI23:units}{13\parcm27}
\setsymbol{LFI:FWHM:LFI24:units}{22\parcm98}
\setsymbol{LFI:FWHM:LFI25:units}{30\parcm46}
\setsymbol{LFI:FWHM:LFI26:units}{30\parcm31}
\setsymbol{LFI:FWHM:LFI27:units}{32\parcm65}
\setsymbol{LFI:FWHM:LFI28:units}{32\parcm66}

\setsymbol{LFI:FWHM:LFI18}{13.39}
\setsymbol{LFI:FWHM:LFI19}{13.01}
\setsymbol{LFI:FWHM:LFI20}{12.75}
\setsymbol{LFI:FWHM:LFI21}{12.74}
\setsymbol{LFI:FWHM:LFI22}{12.87}
\setsymbol{LFI:FWHM:LFI23}{13.27}
\setsymbol{LFI:FWHM:LFI24}{22.98}
\setsymbol{LFI:FWHM:LFI25}{30.46}
\setsymbol{LFI:FWHM:LFI26}{30.31}
\setsymbol{LFI:FWHM:LFI27}{32.65}
\setsymbol{LFI:FWHM:LFI28}{32.66}

\setsymbol{LFI:FWHM:uncertainty:LFI18:units}{0.170\arcm}
\setsymbol{LFI:FWHM:uncertainty:LFI19:units}{0.174\arcm}
\setsymbol{LFI:FWHM:uncertainty:LFI20:units}{0.170\arcm}
\setsymbol{LFI:FWHM:uncertainty:LFI21:units}{0.156\arcm}
\setsymbol{LFI:FWHM:uncertainty:LFI22:units}{0.164\arcm}
\setsymbol{LFI:FWHM:uncertainty:LFI23:units}{0.171\arcm}
\setsymbol{LFI:FWHM:uncertainty:LFI24:units}{0.652\arcm}
\setsymbol{LFI:FWHM:uncertainty:LFI25:units}{1.075\arcm}
\setsymbol{LFI:FWHM:uncertainty:LFI26:units}{1.131\arcm}
\setsymbol{LFI:FWHM:uncertainty:LFI27:units}{1.266\arcm}
\setsymbol{LFI:FWHM:uncertainty:LFI28:units}{1.287\arcm}

\setsymbol{LFI:FWHM:uncertainty:LFI18}{0.170}
\setsymbol{LFI:FWHM:uncertainty:LFI19}{0.174}
\setsymbol{LFI:FWHM:uncertainty:LFI20}{0.170}
\setsymbol{LFI:FWHM:uncertainty:LFI21}{0.156}
\setsymbol{LFI:FWHM:uncertainty:LFI22}{0.164}
\setsymbol{LFI:FWHM:uncertainty:LFI23}{0.171}
\setsymbol{LFI:FWHM:uncertainty:LFI24}{0.652}
\setsymbol{LFI:FWHM:uncertainty:LFI25}{1.075}
\setsymbol{LFI:FWHM:uncertainty:LFI26}{1.131}
\setsymbol{LFI:FWHM:uncertainty:LFI27}{1.266}
\setsymbol{LFI:FWHM:uncertainty:LFI28}{1.287}

\setsymbol{HFI:center:frequency:100GHz:units}{100\,GHz}
\setsymbol{HFI:center:frequency:143GHz:units}{143\,GHz}
\setsymbol{HFI:center:frequency:217GHz:units}{217\,GHz}
\setsymbol{HFI:center:frequency:353GHz:units}{353\,GHz}
\setsymbol{HFI:center:frequency:545GHz:units}{545\,GHz}
\setsymbol{HFI:center:frequency:857GHz:units}{857\,GHz}

\setsymbol{HFI:center:frequency:100GHz}{100}
\setsymbol{HFI:center:frequency:143GHz}{143}
\setsymbol{HFI:center:frequency:217GHz}{217}
\setsymbol{HFI:center:frequency:353GHz}{353}
\setsymbol{HFI:center:frequency:545GHz}{545}
\setsymbol{HFI:center:frequency:857GHz}{857}

\setsymbol{HFI:Ndetectors:100GHz}{8}
\setsymbol{HFI:Ndetectors:143GHz}{11}
\setsymbol{HFI:Ndetectors:217GHz}{12}
\setsymbol{HFI:Ndetectors:353GHz}{12}
\setsymbol{HFI:Ndetectors:545GHz}{3}
\setsymbol{HFI:Ndetectors:857GHz}{4}

\setsymbol{HFI:FWHM:Maps:100GHz:units}{9\parcm88}
\setsymbol{HFI:FWHM:Maps:143GHz:units}{7\parcm18}
\setsymbol{HFI:FWHM:Maps:217GHz:units}{4\parcm87}
\setsymbol{HFI:FWHM:Maps:353GHz:units}{4\parcm65}
\setsymbol{HFI:FWHM:Maps:545GHz:units}{4\parcm72}
\setsymbol{HFI:FWHM:Maps:857GHz:units}{4\parcm39}
\setsymbol{HFI:FWHM:Maps:100GHz}{9.88}
\setsymbol{HFI:FWHM:Maps:143GHz}{7.18}
\setsymbol{HFI:FWHM:Maps:217GHz}{4.87}
\setsymbol{HFI:FWHM:Maps:353GHz}{4.65}
\setsymbol{HFI:FWHM:Maps:545GHz}{4.72}
\setsymbol{HFI:FWHM:Maps:857GHz}{4.39}

\setsymbol{HFI:beam:ellipticity:Maps:100GHz}{1.15}
\setsymbol{HFI:beam:ellipticity:Maps:143GHz}{1.01}
\setsymbol{HFI:beam:ellipticity:Maps:217GHz}{1.06}
\setsymbol{HFI:beam:ellipticity:Maps:353GHz}{1.05}
\setsymbol{HFI:beam:ellipticity:Maps:545GHz}{1.14}
\setsymbol{HFI:beam:ellipticity:Maps:857GHz}{1.19}

\setsymbol{HFI:FWHM:Mars:100GHz:units}{9\parcm37}
\setsymbol{HFI:FWHM:Mars:143GHz:units}{7\parcm04}
\setsymbol{HFI:FWHM:Mars:217GHz:units}{4\parcm68}
\setsymbol{HFI:FWHM:Mars:353GHz:units}{4\parcm43}
\setsymbol{HFI:FWHM:Mars:545GHz:units}{3\parcm80}
\setsymbol{HFI:FWHM:Mars:857GHz:units}{3\parcm67}

\setsymbol{HFI:FWHM:Mars:100GHz}{9.37}
\setsymbol{HFI:FWHM:Mars:143GHz}{7.04}
\setsymbol{HFI:FWHM:Mars:217GHz}{4.68}
\setsymbol{HFI:FWHM:Mars:353GHz}{4.43}
\setsymbol{HFI:FWHM:Mars:545GHz}{3.80}
\setsymbol{HFI:FWHM:Mars:857GHz}{3.67}

\setsymbol{HFI:beam:ellipticity:Mars:100GHz}{1.18}
\setsymbol{HFI:beam:ellipticity:Mars:143GHz}{1.03}
\setsymbol{HFI:beam:ellipticity:Mars:217GHz}{1.14}
\setsymbol{HFI:beam:ellipticity:Mars:353GHz}{1.09}
\setsymbol{HFI:beam:ellipticity:Mars:545GHz}{1.25}
\setsymbol{HFI:beam:ellipticity:Mars:857GHz}{1.03}

\setsymbol{HFI:CMB:relative:calibration:100GHz}{$\lsim 1\%$}
\setsymbol{HFI:CMB:relative:calibration:143GHz}{$\lsim 1\%$}
\setsymbol{HFI:CMB:relative:calibration:217GHz}{$\lsim 1\%$}
\setsymbol{HFI:CMB:relative:calibration:353GHz}{$\lsim 1\%$}
\setsymbol{HFI:CMB:relative:calibration:545GHz}{}
\setsymbol{HFI:CMB:relative:calibration:857GHz}{}

\setsymbol{HFI:CMB:absolute:calibration:100GHz}{$\lsim 2\%$}
\setsymbol{HFI:CMB:absolute:calibration:143GHz}{$\lsim 2\%$}
\setsymbol{HFI:CMB:absolute:calibration:217GHz}{$\lsim 2\%$}
\setsymbol{HFI:CMB:absolute:calibration:353GHz}{$\lsim 2\%$}
\setsymbol{HFI:CMB:absolute:calibration:545GHz}{}
\setsymbol{HFI:CMB:absolute:calibration:857GHz}{}

\setsymbol{HFI:FIRAS:gain:calibration:accuracy:statistical:100GHz}{}
\setsymbol{HFI:FIRAS:gain:calibration:accuracy:statistical:143GHz}{}
\setsymbol{HFI:FIRAS:gain:calibration:accuracy:statistical:217GHz}{}
\setsymbol{HFI:FIRAS:gain:calibration:accuracy:statistical:353GHz}{2.5\%}
\setsymbol{HFI:FIRAS:gain:calibration:accuracy:statistical:545GHz}{1\%}
\setsymbol{HFI:FIRAS:gain:calibration:accuracy:statistical:857GHz}{0.5\%}

\setsymbol{HFI:FIRAS:gain:calibration:accuracy:systematic:100GHz}{}
\setsymbol{HFI:FIRAS:gain:calibration:accuracy:systematic:143GHz}{}
\setsymbol{HFI:FIRAS:gain:calibration:accuracy:systematic:217GHz}{}
\setsymbol{HFI:FIRAS:gain:calibration:accuracy:systematic:353GHz}{}
\setsymbol{HFI:FIRAS:gain:calibration:accuracy:systematic:545GHz}{7\%}
\setsymbol{HFI:FIRAS:gain:calibration:accuracy:systematic:857GHz}{7\%}

\setsymbol{HFI:FIRAS:zero:point:accuracy:100GHz:units}{0.8\MJysr}
\setsymbol{HFI:FIRAS:zero:point:accuracy:143GHz:units}{}
\setsymbol{HFI:FIRAS:zero:point:accuracy:217GHz:units}{}
\setsymbol{HFI:FIRAS:zero:point:accuracy:353GHz:units}{1.4\MJysr}
\setsymbol{HFI:FIRAS:zero:point:accuracy:545GHz:units}{2.2\MJysr}
\setsymbol{HFI:FIRAS:zero:point:accuracy:857GHz:units}{1.7\MJysr}

\setsymbol{HFI:FIRAS:zero:point:accuracy:100GHz}{0.8}
\setsymbol{HFI:FIRAS:zero:point:accuracy:143GHz}{}
\setsymbol{HFI:FIRAS:zero:point:accuracy:217GHz}{}
\setsymbol{HFI:FIRAS:zero:point:accuracy:353GHz}{1.4}
\setsymbol{HFI:FIRAS:zero:point:accuracy:545GHz}{2.2}
\setsymbol{HFI:FIRAS:zero:point:accuracy:857GHz}{1.7}

\setsymbol{HFI:unit:conversion:100GHz:units}{0.2415\MJysrmK}
\setsymbol{HFI:unit:conversion:143GHz:units}{0.3694\MJysrmK}
\setsymbol{HFI:unit:conversion:217GHz:units}{0.4811\MJysrmK}
\setsymbol{HFI:unit:conversion:353GHz:units}{0.2883\MJysrmK}
\setsymbol{HFI:unit:conversion:545GHz:units}{0.05826\MJysrmK}
\setsymbol{HFI:unit:conversion:857GHz:units}{0.002238\MJysrmK}

\setsymbol{HFI:unit:conversion:100GHz}{0.2415}
\setsymbol{HFI:unit:conversion:143GHz}{0.3694}
\setsymbol{HFI:unit:conversion:217GHz}{0.4811}
\setsymbol{HFI:unit:conversion:353GHz}{0.2883}
\setsymbol{HFI:unit:conversion:545GHz}{0.05826}
\setsymbol{HFI:unit:conversion:857GHz}{0.002238}

\setsymbol{HFI:colour:correction:alpha=-2:V1.01:100GHz}{0.9893}
\setsymbol{HFI:colour:correction:alpha=-2:V1.01:143GHz}{0.9759}
\setsymbol{HFI:colour:correction:alpha=-2:V1.01:217GHz}{1.0007}
\setsymbol{HFI:colour:correction:alpha=-2:V1.01:353GHz}{1.0028}
\setsymbol{HFI:colour:correction:alpha=-2:V1.01:545GHz}{1.0019}
\setsymbol{HFI:colour:correction:alpha=-2:V1.01:857GHz}{0.9889}

\setsymbol{HFI:colour:correction:alpha=0:V1.01:100GHz}{1.0008}
\setsymbol{HFI:colour:correction:alpha=0:V1.01:143GHz}{1.0148}
\setsymbol{HFI:colour:correction:alpha=0:V1.01:217GHz}{0.9909}
\setsymbol{HFI:colour:correction:alpha=0:V1.01:353GHz}{0.9888}
\setsymbol{HFI:colour:correction:alpha=0:V1.01:545GHz}{0.9878}
\setsymbol{HFI:colour:correction:alpha=0:V1.01:857GHz}{1.0014}

\providecommand{\sorthelp}[1]{}

\begin{document}

\title{Polarization measurements analysis}
\subtitle{I. Impact of the full covariance matrix on \\
 polarization fraction and angle measurements}
\author{L. Montier\inst{1,2}, S. Plaszczynski\inst{3}, F. Levrier\inst{4},
 M. Tristram\inst{3}, D. Alina\inst{1,2}, I. Ristorcelli\inst{1,2},
 J.-P. Bernard\inst{1,2}}
\institute{ 
\inst{1} Universit\'{e} de Toulouse, UPS-OMP, IRAP, F-31028 Toulouse cedex 4,
 France\\
\inst{2} CNRS, IRAP, 9 Av. colonel Roche, BP 44346, F-31028 Toulouse cedex 4,
 France\\
\inst{3} Laboratoire de l'Acc\'{e}l\'{e}rateur Lin\'{e}aire, Universit\'{e}
 Paris-Sud 11, CNRS/IN2P3, Orsay, France\\
\inst{4} LERMA/LRA - ENS Paris et Observatoire de Paris, 24 rue Lhormond,
 75231 Paris Cedex 05, France
}

\authorrunning{Montier et al.}
\titlerunning{Polarization measurements analysis I}

\abstract{
With the forthcoming release of high precision polarization measurements, such
as from the {\it Planck\/} satellite, the metrology of polarization
needs to improve.  In particular, it is important to take into account
full knowledge of the noise properties when estimating polarization fraction
and polarization angle, which suffer from well-known biases.
While strong simplifying assumptions have usually been made in polarization
analysis, we present a method for including the full covariance matrix of the
Stokes parameters in estimates for the distributions of the polarization
fraction and angle.
We thereby quantify the impact of the noise properties on the biases in the
observational quantities.
We derive analytical expressions for the probability density functions of these
quantities, taking into account the full complexity of the covariance matrix,
including the Stokes $I$ intensity components. 
We perform Monte Carlo simulations to explore the impact of the noise
properties on the statistical variance and bias of
the polarization fraction and angle.
We show that for low variations (${<}\,10$\%) of the effective ellipticity
between the $Q$ and $U$ components around the symmetrical case 
the covariance matrix may be simplified as is usually done,
with negligible impact on the bias. 
For signal-to-noise ratios on intensity lower than 10
the uncertainty on the total intensity is shown to drastically increase
the uncertainty of the polarization fraction but not the relative bias
of the polarization fraction,
while a 10\% correlation between the intensity and the polarized components 
does not significantly affect the bias of the polarization fraction. 
We compare estimates of the uncertainties affecting polarization measurements,
addressing limitations of estimates of the S/N,
and we show how to build conservative confidence intervals for polarization
fraction and angle simultaneously.
This study, which is the first of a set of papers dedicated to the analysis
of polarization measurements, focuses on the basic polarization fraction and
angle measurements.  It covers the noise regime where the complexity of the
covariance matrix may be largely neglected in order to perform further
analysis.  A companion paper focuses on the best estimators of the
polarization fraction and angle, and their associated uncertainties.

}

\maketitle

\section{Introduction}
\label{sec:introduction}

Linear polarization measurements are usually decomposed into their Stokes
components ($I$, $Q$, and $U$), from which one can derive
polarization fraction ($p$) and angle ($\psi$).
However, these are known to be potentially
biased quantities, as first discussed by \citet{Serkowski1958}.
At its most fundamental level this arises because $p$ is constrained to be
positive, while $\psi$ is a non-linear function of the ratio of $Q$ and $U$,
and hence even if $Q$ and $U$ are Gaussian distributed, $p$ and $\psi$ will
not be so simple.

While it is advisable to work as much as possible with the Stokes parameters
to avoid such issues, it is sometimes more convenient to 
use the coordinates $p$ and $\psi$ when connecting polarization data to
physical models and interpretations.   For instance, we may be interested in 
the maximum fraction of polarization $p$ observed in our Galaxy or the correlation 
between the polarization fraction and the structure of the magnetic field, 
which is not easy to carry out over large regions of the sky when using the Stokes parameters.
Thus, many authors,
e.g., \citet{Wardle1974}, \citet{Simmons1985} and more recently
\citet{Vaillancourt2006} and \citet{Quinn2012}, have suggested ways of
dealing with polarization fraction estimates to try to correct for the
biases.  \citet{Vinokur1965}
was the first to focus on the polarization angle, with later papers by
\citet{Clarke1993} and \citet{Naghizadeh1993}.
In all such studies there have been strong assumptions made about the
noise properties of the polarization measurements.
The noise on the $Q$ and $U$ components are usually considered to be fully
symmetric and with no correlation between them, and furthermore the intensity
is always assumed to be perfectly known.  These assumptions,
which we will call the ``canonical simplifications,'' can be useful in
practice, in that they allow for rapid progress, but on the other hand
they are often simply not correct assumptions to make.

Our work is motivated by the need to understand polarization emission data
at microwave to sub-millimetre wavelengths, although the analysis is general
enough to be applied to any kind of polarization data.  Nevertheless, the
details of experimental setup design cannot be ignored, since they affect how
correlated are the data.
Because the computation of the Stokes parameters and their associated
uncertainties strongly depends on the instrumental design, technical efforts
have been made to limit the impact of the instrumental systematics.
For example, single-dish instruments such as STOKES \citep{Platt1991}, 
Hertz \citep{Schleuning1997}, SPARO \citep{Renbarger2004}
or SCU-Pol \citep{Greaves2003} had to face strong systematics due to
noise correlation between orthogonal components
and atmospheric turbulence, 
while the SHARP optics \citep{Li2008} allowed the SHARC-II facility
\citep{Dowell1998}
at the Caltech Submillimeter Observatory to be converted into a dual-dish
experiment to avoid such noise correlation issues. 
Nevertheless polarization measurements obtained until now
were limited by systematics and statistical uncertainties.
Even in some of the most recent studies no correction for the bias of the
polarization fraction was applied \citep[e.g.,][]{Dotson2010}, or
only high signal-to-noise ratio (hereafter S/N)
data were used for analysis ($p/\sigma>3$) 
in order to avoid the issue \citep[e.g.,][]{Vaillancourt2012}.
One naturally wonders wether this common choice of S/N
greater than 3 is relevant for all experiments, and 
how the noise correlation between orthogonal Stokes components or noise
asymmetry between the Stokes parameters could impact this choice.

A major motivation for studying polarized emission in microwaves is the
extraction of the weak polarization of the cosmic microwave background.
It has been demonstrated by the balloon-borne Archeops \citep{Benoit2004}
experiment and via polarization observations by the {\it WMAP\/} satellite
\citep{page2007} that the polarized cosmological signal is dominated by
Galactic foregrounds at large scales and intermediate latitude (with
a polarization fraction of 3--10\%).
Thus the characterization of polarized Galactic dust emission in the
submillimetre range has become one of the challenges for the coming decade.
This is in order to study the role of magnetic fields for
the dynamics of the interstellar medium and star formation, as well as for
characterizing the foregrounds for the cosmological polarization signal.
The limitations of instrumental specifications and data analysis are
therefore being continually challenged.
Fully mapping the polarization fraction and angle at large scales is going
to be a major outcome of these studies for Galactic science in the near future.
This makes it increasingly important to address the issues of biasing of
polarization measures.

With new experiments like the {\it Planck\/} \footnote{\Planck~(\url{http://www.esa.int/Planck}) is a
project of the European Space Agency (ESA) with instruments
  provided by two scientific consortia funded by ESA member states (in
  particular the lead countries France and Italy), with contributions
  from NASA (USA) and telescope reflectors provided by a collaboration
  between ESA and a scientific consortium led and funded by Denmark.} satellite \citep{Tauber2010}, 
the balloon-borne experiments BLAST-Pol \citep{Fissel2010} 
and PILOT \citep{Bernard2007}, or ground based facilities with polarization
capability, such as ALMA \citep{Perez2013}, 
SMA \citep{Girart2006}, NOEMA \citep[at Plateau de Bure,][]{Boissier2009},
and XPOL \citep[at the IRAM 30-m telescope,][]{Thum2008}, we are entering
a new era for Galactic polarization studies, when much better control of
the systematics is being achieved.  Comprehensive characterization of the
instrumental noise means that it becomes crucial to fully account for
knowledge of the noise properties between orthogonal components
when analysing these polarization measurements. 
Because the {\it Planck\/} data exhibit large-scale variations over the whole 
sky, in terms of S/N and covariance matrix,
the impact of the full complexity of the noise will have to be corrected
in order to obtain a uniform survey of the polarization fraction and angle
-- something that is essential to perform large-scale modelling of our
Galaxy.

This paper is the first part of an ensemble of papers dedicated to the
analysis of polarization measurements, 
presenting the methods for handling complex polarized data with a high level
of inhomogeneity in terms of S/N or covariance matrix configurations.
We aim here to present the formalism for discussing polarization fraction
and angle, while taking into account the full covariance matrix.  We will
quantify how much the naive measurements of polarization 
fraction and angle are impacted by the noise covariance, and the extent
to which the non-diagonal terms of the covariance matrix may be neglected. 
Two other studies, focused on the best estimators of the true polarization
parameters, will be presented in the second and third parts of this set.
Throughout, we will make use of two basic assumptions: (i) that the circular
polarization (i.e., Stokes $V$) can be neglected; and (ii) that
the noise on the other Stokes parameters can be assumed to be Gaussian.

The paper is organized as follows.  We first derive in Sect.~\ref{sec:pdf}
the full expressions for the probability density functions of polarization
fraction and angle measurements, using the full covariance matrix.
In Sect.~\ref{sec:impact_fullcov} we explore the impact of the complexity of
the covariance matrix on polarization measurement estimates, and we provide
conservative domains of the covariance matrix where the canonical
simplification remains valid.  We finally address the question of the S/N
estimate in Sect.~\ref{sec:sigma_p}, where we compare four estimators for the
polarization measurement uncertainty.

\begin{figure*}[tp!]
\begin{tabular}{cc}
 \includegraphics*[width=9cm,viewport=100 100 500 400]{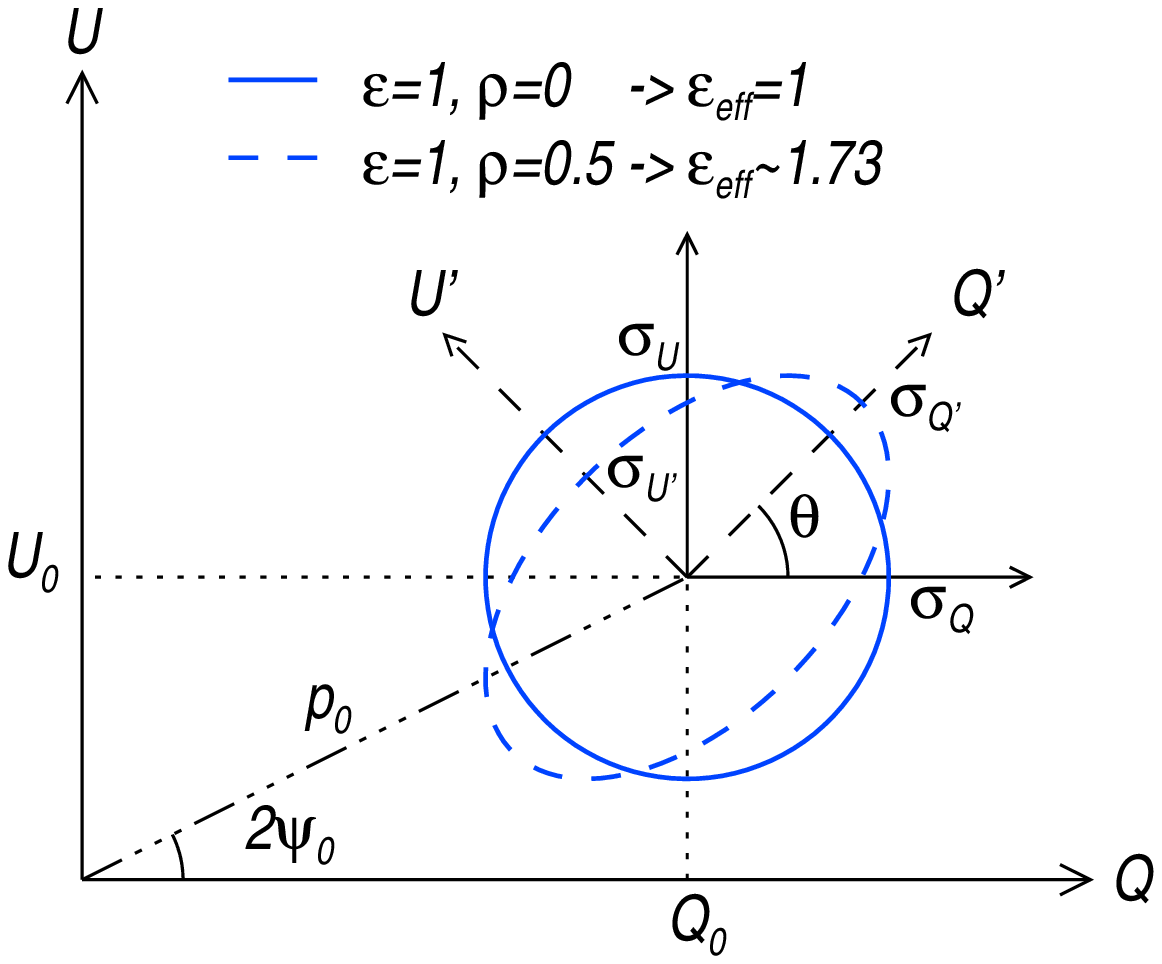} &
 \includegraphics*[width=9cm,viewport=100 100 500 400]{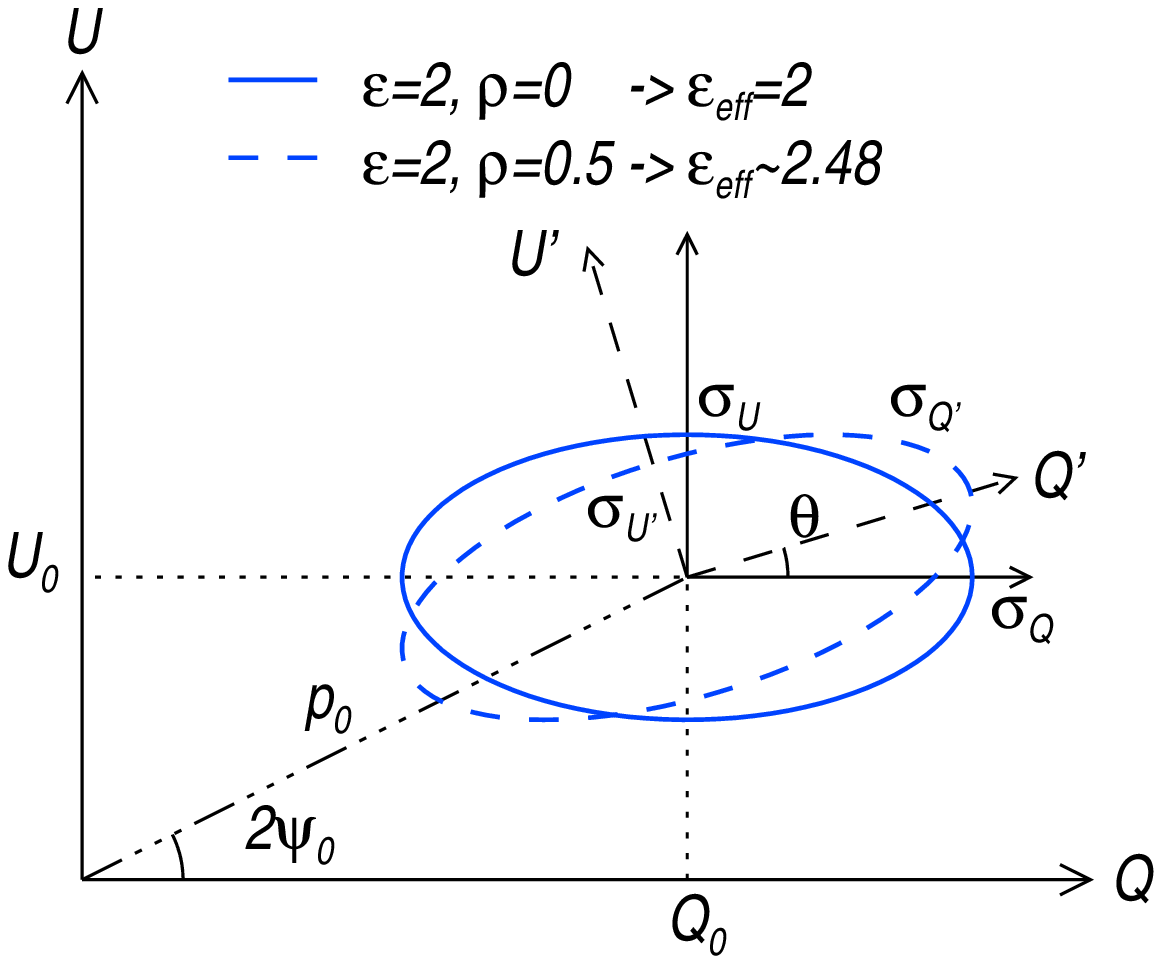} \\
 \end{tabular}
 \caption{Illustrations of the noise distribution in the ($Q$, $U$) plane.
The blue solid and dashed lines represent the 1$\,\sigma$
probability contours around the true polarization values ($Q_0$, $U_0$),
also parameterized by ($p_0$, $\psi_0$). 
{\it Left\/}: the canonical case ($\varepsilon$\,{=}\,1, $\rho$\,{=}\,0) is
shown as a solid line.  The dashed line shows the introduction of a
correlation $\rho$\,{=}\,0.5, leading to an effective ellipticity
($\varepsilon_{\rm eff}\,{>}\,1$) rotated by an angle $\theta$. 
{\it Right\/}: same transformation, starting from the elliptical case
($\varepsilon$\,{=}\,2, $\rho$\,{=}\,0).
 }
 \label{fig:geometry_diagram}
\end{figure*}

\section{($p,\psi$) probability density functions}
\label{sec:pdf}

\subsection{Notation}
\label{sec:notations}

The goal of this paper is to characterize the distribution of naive
polarization measurements, given the true polarization parameters and their
associated noise estimates.  Let us denote the true values by
($I_0$, $Q_0$, $U_0$), representing the
the true total intensity and Stokes linear polarization parameters, and
with $P_0\,{=}\,\sqrt{Q_0^2 + U_0^2}$.
The quantities ($I$, $Q$, $U$) are the same for the measured values.
The polarization fraction and polarization angle are defined by
\begin{equation}
p_0 \equiv \frac{\sqrt{Q_0^2 + U_0^2}}{I_0}, \quad
 \psi_0 \equiv \frac{1}{2} \mathrm{atan} \left( \frac{U_0}{Q_0} \right) 
\end{equation}
for the true values and 
\begin{equation}
p \equiv \frac{\sqrt{Q^2 + U^2}}{I}, \quad
 \psi \equiv \frac{1}{2} \mathrm{atan} \left( \frac{U}{Q} \right) 
\label{eq:definition_ppsi}
\end{equation}
for the measurements.  The true Stokes parameters can be expressed by 
$Q_0\,{\equiv}\,p_0\,I_0\,\mathrm{cos} (2\psi_0)$
and
$U_0\,{\equiv}\,p_0 \, I_0 \, \mathrm{sin} (2\psi_0)$,
while for the measurements
$Q\,{\equiv}\,p\,I\,\mathrm{cos} (2\psi) $
and 
$U\,{\equiv}\,p \, I \, \mathrm{sin} (2\psi)$.
Although the true intensity $I_0$ is strictly positive, the measured
intensity $I$ may be negative due to noise, 
thus $I_0$ can take values between 0 and $+\infty$, while $I$ ranges
between $-\infty$ and $+\infty$.
The measured Stokes parameters $Q$ and $U$ are real, finite quantities,
spanning from $-\infty$ to $+\infty$ and with the addition of noise do not
necessarily satisfy the relation $Q^2+U^2 \le I^2$ obeyed by the underlying
quantities, i.e., $Q_0^2+U_0^2 \le I_0^2$. 
The true polarization fraction $p_0$ can take values in the range 0 to 1,
while the measured polarization fraction $p$ ranges between
$-\infty$ and $+\infty$.  Finally we define $\psi_0$ and $\psi$ such that they
are both defined in the range $[-\pi/2,+\pi/2]$.

Previous studies of polarization measurements usually made strong assumptions
concerning the noise properties, in particular:
(i) correlations between the total and polarized intensities were neglected;
(ii) correlated noise between $Q$ and $U$ was also neglected;
and (iii) equal noise was assumed on $Q$ and $U$ measurements.
We propose instead in this paper to use the full covariance matrix
defined by
\begin{equation}
\tens{\Sigma}\equiv\left(\begin{array}{ccc}
\sigma_{\rm II} & \sigma_{\rm IQ} & \sigma_{\rm IU} \\
\sigma_{\rm IQ} & \sigma_{\rm QQ} & \sigma_{\rm QU} \\
\sigma_{\rm IU} & \sigma_{\rm QU} & \sigma_{\rm UU} \\
\end{array}\right)
 = \left(\begin{array}{ccc}
{\sigma_{\rm I}}^2 & \rho_{\rm Q} \sigma_{\rm I}\sigma_{\rm Q} &
 \rho_{\rm U} \sigma_{\rm I}\sigma_{\rm U} \\
\rho_{\rm Q} \sigma_{\rm I}\sigma_{\rm Q} & {\sigma_{\rm Q}}^2 &
 \rho \sigma_{\rm Q}\sigma_{\rm U} \\
\rho_{\rm U} \sigma_{\rm I}\sigma_{\rm U} &\rho \sigma_{\rm Q}\sigma_{\rm U} &
 {\sigma_{\rm U}}^2 \\
\end{array}\right) \, ,
\end{equation}
where $\sigma_{XY}$
is the covariance of the two random variables $X$ and $Y$,
and the following quantities are usually introduced in the literature to
simplify the notation:
\begin{equation}
\varepsilon \equiv \frac{\sigma_{\rm Q}}{\sigma_{\rm U}}\, ; \quad
\rho \equiv \frac{\sigma_{\rm QU}}{\sigma_{\rm Q}\,\sigma_{\rm U}} \,; \quad
\rho_{\rm Q} \equiv \frac{\sigma_{\rm IQ}}{\sigma_{\rm I}\,\sigma_{\rm Q}} \, ;
 \quad
\rho_{\rm U} \equiv \frac{\sigma_{\rm IU}}{\sigma_{\rm I}\,\sigma_{\rm U}} \, .
\end{equation}
Here $\varepsilon$ is the ellipticity between the $Q$ and $U$ noise components,
and $\rho$ (which lies between $-1$ and $+1$) is the 
correlation between the $Q$ and $U$ noise components.  Similarly
$\rho_{\rm Q}$ and $\rho_{\rm U}$ are the correlations between the noise in
intensity $I$ and the $Q$ and $U$ components, respectively. 

The parameterization just described could be misleading, however,
since the ellipticity $\varepsilon$ does not represent the effective
ellipticity in the ($Q$, $U$) plane
if the correlation is not null.  This is illustrated in
Fig.~\ref{fig:geometry_diagram} for two initial values of the ellipticity
$\varepsilon$.  A new reference frame ($Q^\prime$, $U^\prime$) where
the Stokes parameters are now uncorrelated 
can always be obtained through rotation by an angle
\begin{equation}
\label{eq:theta}
\theta = \frac{1}{2} \mathrm{atan}
 \left( \frac{2 \sigma_{\rm QU}}{\sigma_{\rm Q}^2 - \sigma_{\rm U}^2} \right).
\end{equation}
We can calculate the covariance matrix in the rotated frame by taking the
usual $\tens{R}\,\tens{\Sigma}\,\tens{R}^{\rm T}$.
In this new reference frame, the errors on $Q^\prime$ and $U^\prime$
are uncorrelated and defined as
\begin{equation}
\begin{array}{l}
\sigma_{Q^\prime}^2 = \sigma_{\rm Q}^2 \cos^2\theta 
 + \sigma_{\rm U}^2 \sin^2 \theta + \sigma_{\rm QU} \sin 2\theta \, , \\
\sigma_{U^\prime}^2 = \sigma_{\rm Q}^2 \sin^2\theta
 + \sigma_{\rm U}^2 \cos^2 \theta - \sigma_{\rm QU} \sin 2\theta \, , 
\end{array}
\end{equation}
so that the effective ellipticity $\varepsilon_{\rm eff}$ is now given by
\begin{equation}
\varepsilon_{\rm eff}^2 = \frac{\sigma_{\rm Q}^2
 + \sigma_{\rm U}^2 + {\sigma^\prime}^2}
 {\sigma_{\rm Q}^2 + \sigma_{\rm U}^2 - {\sigma^\prime}^2} \, , 
\end{equation}
where
\begin{equation}
{\sigma^\prime}^2 = \sqrt{\left( \sigma_{\rm Q}^2 - \sigma_{\rm U}^2 \right)^2
 + 4\sigma_{\rm QU}^2} \, .
\end{equation}
When expressed as a function of the ($\varepsilon$, $\rho$) parameters we
obtain
\begin{equation}
\label{eq:epsi_eff}
\varepsilon_{\rm eff}^2 = \frac{1 + \varepsilon^2
 + \sqrt{(\varepsilon^2-1)^2 + 4\rho^2\varepsilon^2}}
 {1 + \varepsilon^2 - \sqrt{(\varepsilon^2-1)^2 + 4\rho^2\varepsilon^2}}
\end{equation}
and 
\begin{equation}
\label{eq:phi}
 \theta = \frac{1}{2} \mathrm{atan}
 \left( \frac{2 \rho \varepsilon}{\varepsilon^2-1} \right) \, .
\end{equation}

This parameterization of the covariance matrix $\tens{\Sigma}$ in terms of
$\varepsilon_{\rm eff}$ and $\theta$ will be preferred in our work
for two reasons.  Firstly,
the shape of the noise distribution in the ($Q$, $U$) space is now contained
in a single parameter, the effective ellipticity $\varepsilon_{\rm eff}$
(${\ge}\,1$), instead of two parameters, $\varepsilon$ and $\rho$.
Secondly, the noise distribution is now independent of the reference frame.
This is also related to the fact that the properties of the noise distribution
do not depend on 3 ($I_0$, $p_0$, $\psi_0$)
plus 6 (from $\tens{\Sigma}$) parameters,
but only on 8, since it actually only depends 
on the difference of the angles $2\psi_0 - \theta$, which greatly
simplifies the analysis.
For what follows we also define $\det(\tens{\Sigma})\,{=}\,\sigma^6$,
the determinant of the covariance matrix.

\subsection{3D probability density functions}
\label{sec:pdf_likelihood}

The probability density function gives the probability to obtain a set of
values ($I$, $Q$, $U$) given the true Stokes parameters ($I_0$, $Q_0$, $U_0$)
and the covariance matrix $\tens{\Sigma}$.  As a short-hand, we refer to this
as the ``3D PDF.''
When Gaussian noise is assumed for each Stokes component, 
this distribution, in the space ($I$, $Q$, $U$) is given by
\begin{equation}
F(X\, | \,X_0, \tens{\Sigma})
 = \sqrt{\frac{\det(\tens{\Sigma}^{-1})}{(2\pi)^3}}
 \ \mathrm{exp} \left \lgroup
 - \frac{(X-X_0)^{\rm T}\, \tens{\Sigma}^{-1} \, (X-X_0)}{2} \right \rgroup,
\label{eq:pdf_IQU}
\end{equation}
where $X$ and $X_0$ are the vectors of the Stokes parameters $[I,Q,U]$
and $[I_0,Q_0,U_0]$, $\tens{\Sigma}^{-1}$ is the inverse of the
covariance matrix (also called the ``precision matrix''), and
$\det(\tens{\Sigma}^{-1})\,{=}\,\sigma^{-6}$ is the determinant of
$\tens{\Sigma}^{-1}$.
This definition ensures that the probability density function is normalized
to 1.
Note that iso-probability surfaces in the ($I$, $Q$, $U$) space are ellipsoids. 

Using normalized polar coordinates, the probability density function
$f(I,p,\psi\, | \,I_0,p_0,\psi_0, \tens{\Sigma})$ can be computed explicitly.
However, the expression (see Eq.~\ref{eq:f_ipphi}) is a little cumbersome,
and so we have put it in Appendix~\ref{sec:PDFexpressions}.
Notice the presence of a factor $2|p|I^2$ in front of the exponential, coming
from the Jacobian of the transformation.

\subsection{2D marginal ($p,\psi$) distribution}
\label{sec:marginal_ppsi_pdf}

We compute the 2D probability density function $f_{\rm 2D}(p,\psi)$ by
marginalizing the probability density
function $f(I,p,\psi)$ (see Eq.~\ref{eq:f_ipphi}) over intensity $I$ on the
range $-\infty$ to $+\infty$. 
The computation is quite straightforward (see Appendix~\ref{appendix:feta}), 
leading to an expression that depends on the sign of $p$, given in
Eq.~\ref{eq:f_2d_ppos} and Eq.~\ref{eq:f_2d_pneg}.  In these expressions
``erf'' is the Gauss error and we have also defined the functions
\begin{equation}
\begin{array}{rcrcl}
\alpha & = & \left(
\begin{array}{c}
1\\
p\cos2\psi\\
p\sin2\psi
\end{array}
\right)^{\rm T}
& \tens{\Sigma}^{-1} &
\left(
\begin{array}{c}
1\\
p\cos2\psi\\
p\sin2\psi
\end{array}
\right), \\ 
\beta & = & \left(
\begin{array}{c}
1\\
p\cos2\psi\\
p\sin2\psi
\end{array}
\right)^{\rm T}
& \tens{\Sigma}^{-1} &
\left(
\begin{array}{c}
1\\
p_0\cos2\psi_0\\
p_0\sin2\psi_0
\end{array}
\right), \\ 
\gamma & = & \left(
\begin{array}{c}
1\\
p_0\cos2\psi_0\\
p_0\sin2\psi_0
\end{array}
\right)^{\rm T}
& \tens{\Sigma}^{-1} &
\left(
\begin{array}{c}
1\\
p_0\cos2\psi_0\\
p_0\sin2\psi_0
\end{array}
\right). \\ 
\end{array}
\end{equation}

In many cases, two further assumptions can be made: 
(i) the correlations between $I$ and ($Q$, $U$) is negligible,
i.e., $\rho_{\rm Q}\,{=}\, \rho_{\rm U}\,{=}\,0$; and 
(ii) the signal-to-noise ration of the intensity $I_0/\sigma_{\rm I}$ is so
large that $I$ can be considered to be perfectly known, yielding $I\,{=}\,I_0$. 
Making such assumptions allows us to reduce the covariance matrix
$\tens{\Sigma}$ to a $2\times2$ matrix,
$\tens{\Sigma}_{\rm p}$, which we define as
\begin{equation}
 \tens{\Sigma}_{\rm p} = \frac{1}{I_0^2} \left \lgroup \begin{array}{cc}
\sigma_{\rm QQ} & \sigma_{\rm QU} \\
\sigma_{\rm QU} & \sigma_{\rm UU} \\
\end{array}\right \rgroup
\quad = \quad
 \frac{ \sigma_{\rm p,G}^2 } {\sqrt{1-\rho^2 }} \left \lgroup
 \begin{array}{cc}
 \varepsilon & \rho \\
 \rho & 1 / \varepsilon \\
 \end{array}
 \right \rgroup,
\end{equation}
where $\sigma_{\rm p,G}$ is defined by
$\det(\tens{\Sigma}_{\rm p})\,{=}\,\sigma_{\rm p,G}^4$, leading to
\begin{equation}
\label{eq:sigpg}
\sigma_{\rm p,G}^2 = \frac{\sigma_{\rm Q}^2} {I_0^2} \,
 \frac{ \sqrt{1-\rho^2}}{\varepsilon} \quad
 \Bigg( = \frac{\sigma_{\rm Q^\prime}^2} {I_0^2}
 \frac{1}{\varepsilon_{\rm eff}} \Bigg)\, .
\end{equation}
This parameter $\sigma_{\rm p,G}$ is linked to the normalization of the 2D
distribution; it represents the 
radius of the equivalent spherical Gaussian distribution that has the same
integrated area as the elliptical Gaussian distribution.
The probability density function $f_{\rm 2D}$ can then be simplified,
as given in Eq.~\ref{eq:f_2d_polar}.
The matching between the two expressions for $f_{\rm 2D}$,
Eqs.~\ref{eq:f_2d_ppos}--\ref{eq:f_2d_pneg}, and Eq.~\ref{eq:f_2d_polar},
when $I_0/\sigma_{\rm I} \rightarrow \infty$, is simply ensured by the
consistency of the determinants of $\tens{\Sigma}$ and $\tens{\Sigma}_{\rm p}$,
when $\rho_{\rm Q}\,{=}\,\rho_{\rm U}\,{=}\,$0:
\begin{equation}
\sigma^6 = \sigma_{\rm I}^2 \sigma_{\rm Q}^2 \sigma_{\rm U}^2
 = \sigma_{\rm I}^2 I_0^4 \sigma_{\rm p,G}^4.
\end{equation}
We also recall that in the canonical case ($\varepsilon_{\rm eff}\,{=}\,1$),
the probability density function can be simplified to
\begin{equation}
f_{\rm 2D} = \frac{p}{\pi \sigma_{\rm p}^2} \, \mathrm{exp}
 \left \{-\frac{1}{2\sigma_{\rm p}^2}
 \left[p^2 + p_0^2 - 2pp_0\cos2(\psi-\psi_0) \right] \right\} \, ,
\end{equation}
where $\sigma_{\rm p,G}$ also simplifies to
$\sigma_{\rm p}\,{=}\,\sigma_{\rm Q}$/$I_0\,{=}\,\sigma_{\rm U}$/$I_0$.
We provide illustrations of the 2D PDFs in Appendix~\ref{sec:illustration_pdf}.

\subsection{1D Marginal $p$ and $\psi$ distributions}
\label{sec:marginal_p_psi_pdf}

The marginal probability density functions of $p$
and $\psi$ can be obtained by integrating the 2D PDF given by
Eq.~\ref{eq:f_2d_polar} 
over $\psi$ (between $-\pi/2$ and $+\pi/2$) and $p$ (between 0 and $+\infty$),
respectively, when assuming the S/N on the intensity to be infinite.
These two probability density functions theoretically depend on $p_0$,
$\psi_0$, and $\tens{\Sigma}_{\rm p}$.
While the expressions obtained in the general case \citep{Aalo2007} are
provided in Appendix~\ref{sec:f_1d_full}, 
the expression for the marginal $p$ distribution reduces to the Rice law
\citep{Rice1945} when $\varepsilon\,{=}\,1$ and $\rho\,{=}\,0$:
\begin{equation}
 R(p\, | \, p_0, \sigma_{\rm p}) = \frac{p}{\sigma_{\rm p}^2}
 \mathrm{exp} \left( -\frac{(p^2 + p_0^2)}{2\sigma_{\rm p}^2}
 \right) \mathcal{I}_0\left( \frac{p p_0}{\sigma_{\rm p}^2}\right) ,
 \label{eq:rice}
\end{equation}
where $\mathcal{I}_0(x)$ is the zeroth-order modified Bessel function of the
first kind \citep{AbramowitzStegun1964}.
This expression no long has a dependence on $\psi_0$. 
With the same assumptions, the marginal $\psi$ distribution
\citep[extensively studied in][]{Naghizadeh1993} is given by
\begin{equation}
 G(\psi\, |\, p_0,\psi_0,\sigma_{\rm p}) = \frac{1}{\sqrt{\pi}}
 \left\{ \frac{1}{\sqrt{\pi}} + \eta_0e^{\eta_0^2}
 \left[ 1+ \mathrm{erf}(\eta_0) \right] \right\}
 e^{-p_0^2I_0^2/2\sigma_{\rm p}^2},
 \end{equation}
where $\eta_0\,{=}\,(p_0 I_0/\sqrt{2}\sigma_{\rm p})\cos2(\psi-\psi_0)$.
This distribution depends on $p_0$, and is symmetric about $\psi_0$.

\section{Impact of the covariance matrix on the bias}
\label{sec:impact_fullcov}

We now quantify how the effective ellipticity of the covariance matrix
impacts the bias of the polarization measurements,
compared to the canonical case.  We would like to determine under what
conditions the covariance matrix may be simplified to its 
canonical expression, in order to minimize computations.
The impact of the correlation and the ellipticity of the covariance matrix
are first explored in the two dimensional ($p$, $\psi$) plane with infinite
intensity S/N.  The cases are then investigated of finite S/N on
intensity and of correlation between total and polarized intensity.

\subsection{Methodology}
\label{sec:bias_methodology}

Given a collection of measurements of the same underlying polarization
parameters ($p_0$, $\psi_0$), we build the statistical bias on $p$ and $\psi$
by averaging the discrepancies $\Delta p\,{=}\,\overline{p}-p_0$ and
$\Delta \psi\,{=}\,\overline{\psi}-\psi_0$ (always defining the quantity
$\psi-\psi_0$ between $-\pi/2$ and $+\pi/2$).
With knowledge of the probability density function
$f_{\rm 2D}(p,\psi\, | \, p_0, \psi_0, \tens{\Sigma}_{\rm p})$, 
we can directly obtain the statistical bias by computing the mean estimates
\begin{equation}
\Delta p \, (p_0, \psi_0, \tens{\Sigma}_{\rm p})=\overline{p} - p_0 
\end{equation}
and 
\begin{equation}
\Delta \psi\,(p_0,\psi_0,\tens{\Sigma}_{\rm p})=\overline{\psi} - \psi_0 \, .
\end{equation}
Here $\overline{p}$ and $ \overline{\psi}$ are the mean estimates from
the probability density function, defined as the first moments of
$f_{\rm 2D}$:
\begin{equation}
\label{eq:mean_p}
\quad \quad \overline{p}=\int_{0}^{+\infty} \int_{\psi_0-\pi/2}^{\psi_0+\pi/2}
 p f_{\rm 2D}(p,\psi \, | \, p_0, \psi_0, \tens{\Sigma}_{\rm p} ) \, dp d\psi;
\end{equation}
and 
\begin{equation}
\quad \quad \overline{\psi}=\int_{0}^{+\infty}
 \int_{\psi_0-\pi/2}^{\psi_0+\pi/2}
 \psi f_{\rm 2D}(p,\psi \, | \, p_0, \psi_0, \tens{\Sigma}_{\rm p} )
 \, dp d\psi \, .
\end{equation}

In order to quantify the importance of this bias, we can compare it to the
dispersion of the polarization fraction and angle measurements,
$\sigma_{p,0}$ and $\sigma_{\psi,0}$.  These are defined as
the second moments of the probability density function
$f_{\rm 2D}$:
\begin{equation}
\label{eq:sigp_0}
 \sigma_{{\rm p},0}^2 = \int_{0}^{+\infty} \int_{-\pi/2}^{\pi/2}
 \left( p-\overline{p}\right)^2
 f_{\rm 2D}(p,\psi \, | \, p_0, \psi_0, \tens{\Sigma}_{\rm p} )
 \, dp d\psi \, ;
\end{equation}
and \begin{equation}
 \sigma_{\psi,0}^2 = \int_{0}^{+\infty} \int_{-\pi/2}^{\pi/2}
 \left(\psi-\overline{\psi}\right)^2
 f_{\rm 2D}(p,\psi \, | \, p_0, \psi_0, \tens{\Sigma}_{\rm p} )
 \, dp d\psi \, .
\end{equation}
Here subscript $0$ signifies that this dispersion has been computed
using full knowledge of the true polarization parameters and the
associated probability density function.

\begin{figure}[t!]
\begin{tabular}{c}
 \includegraphics[width=9cm]{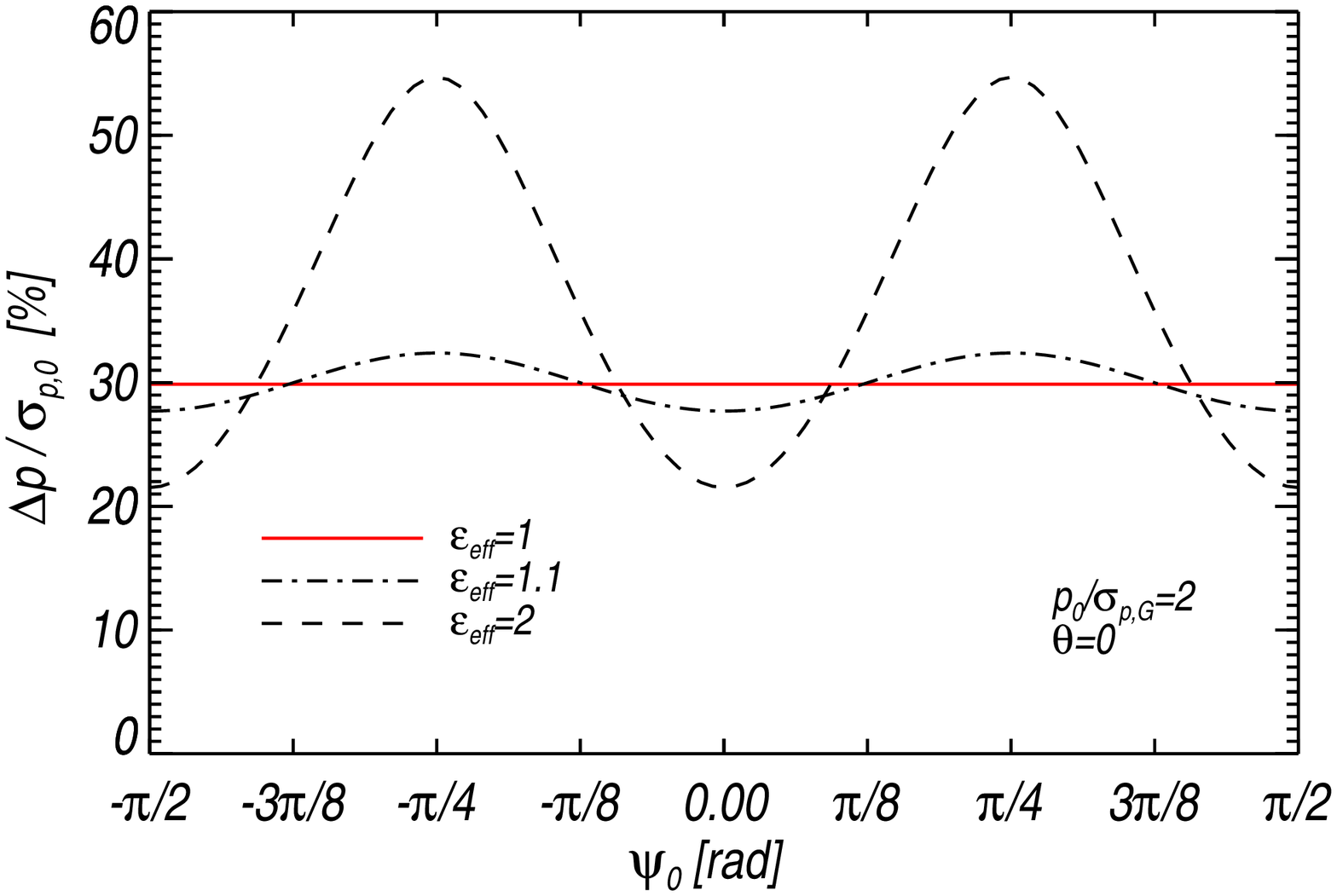} \\
 \includegraphics[width=9cm]{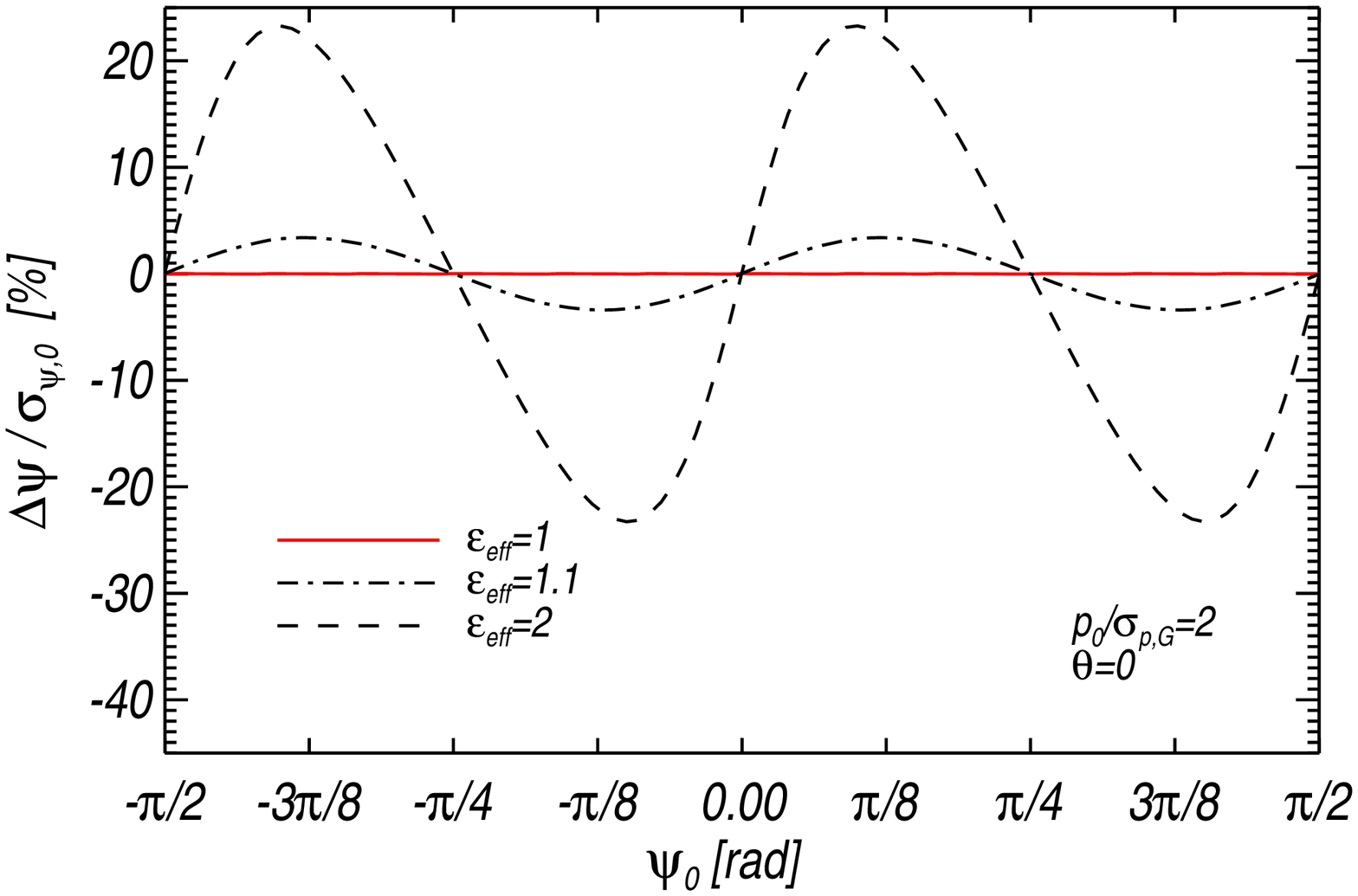} 
 \end{tabular}
 \caption{Impact of the initial true polarization angle $\psi_0$ and
 varying effective ellipticity $\varepsilon_{\rm eff}$
 on the relative polarization fraction bias
 $\Delta p/\sigma_{\rm p,0}$ (top) and the
 relative polarization angle bias $\Delta \psi / \sigma_{\psi,0}$ (bottom). 
 We assume no correlation here, so that $\theta\,{=}\,0$ and we set the
 signal-to-noise ratio to $p_0 / \sigma_{\rm p,G}\,{=}\,2$.
 The canonical case ($\varepsilon_{\rm eff}\,{=}\,1$) is shown by the red line.}
 \label{fig:impact_psi0_biasppsi}
\end{figure}

We choose $\sigma_{\rm p,G}$ introduced in Sect.~\ref{sec:marginal_ppsi_pdf} as our characteristic estimate of the polarization fraction noise
in its relationship to the covariance matrix $\tens{\Sigma}_{\rm p}$.
This will be used to define the signal-to-noise ratio of the polarization
fraction $p_0/\sigma_{\rm p,G}$, 
which is kept constant when exploring the ellipticity and correlation of the 
$Q$--$U$ components.  In Sect.~\ref{sec:sigma_p} we will discuss how robust
this estimate is against the true dispersion $\sigma_{{\rm p},0}$.

We define three specific setups of the covariance matrix to investigate:
(i) the {\it canonical\/} case, $\varepsilon_{\rm eff}\,{=}\,1$, equivalent to
$\varepsilon\,{=}\,1$, $\rho\,{=}\,0$;
the {\it low\/} regime, $1\,{\le}\,\varepsilon_{\rm eff}\,{<}\,1.1$;
and the {\it extreme\/} regime, $1\,{\le}\,\varepsilon_{\rm eff}\,{<}\,2$.
These will be used in the rest of this paper to quantify departures of the
covariance matrix from the canonical case, and to
characterize the impact of the covariance matrix on polarization measurements
in each regime.  It is worth recalling
that to each value of the effective ellipticity $\varepsilon_{\rm eff}$
there corresponds a set of equivalent parameters
$\varepsilon$, $\rho$, and $\theta$.

\begin{figure}
\begin{tabular}{c}
 \includegraphics[width=9cm]{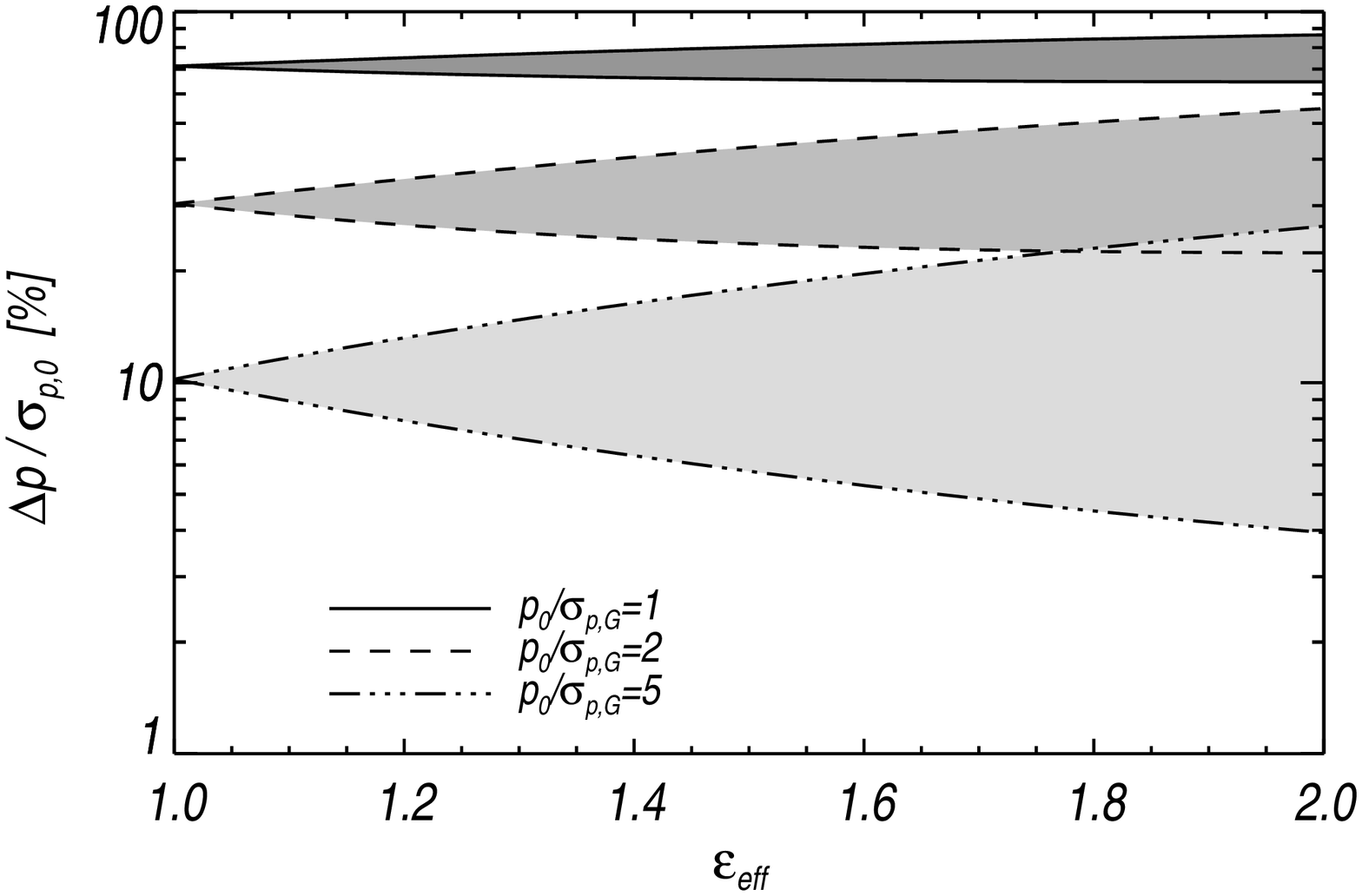} \\
 \includegraphics[width=9cm]{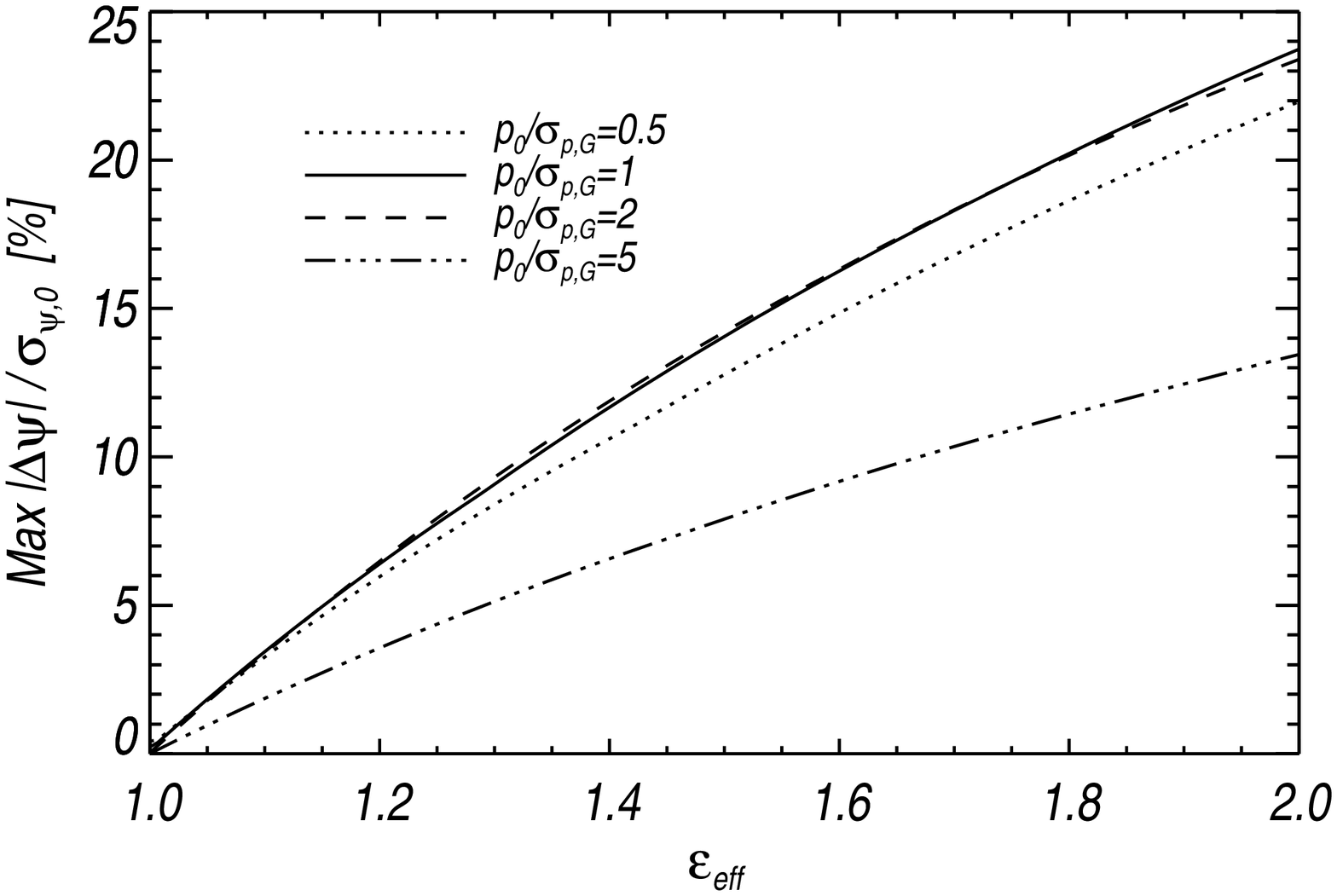}
 \end{tabular}
 \caption{ Impact of the effective ellipticity $\varepsilon_{\rm eff}$ on the
 levels of bias.
 {\it Top\/}: $\Delta p/\sigma_{{\rm p},0}$ as a function of the effective
 ellipticity $\varepsilon_{\rm eff}$, displayed for three levels 
 of the signal-to-noise ratio, $p_0 / \sigma_{\rm p,G}\,{=}\,1$, 2, and 5. 
 The grey shaded regions indicate the whole extent of variability due to
 $\psi_0$ and $\theta$ spanning the range $-\pi/2$ to $\pi/2$. 
 {\it Bottom\/}: maximum $|\Delta \psi| / \sigma_{\psi,0}$ value for
 $\psi_0$ and $\theta$ spanning the range $-\pi/2$ to $\pi/2$, plotted 
 as a function of the effective ellipticity $\varepsilon_{\rm eff}$, displayed
 for four levels of the signal-to-noise ratio, 
 $p_0 / \sigma_{\rm p,G}\,{=}\,0.5$, 1, 2, and 5.
 }
 \label{fig:impact_epsilon_biasppsi}
\end{figure}

\subsection{$Q$--$U$ ellipticity}
\label{sec:bias_ppsi_2d_ellipticity}

We assume here that the intensity is perfectly known and that there is no
correlation between the total intensity $I$ and the polarized intensity,
so that $I\,{=}\,I_0$ and $\rho_{\rm Q}\,{=}\,\rho_{\rm U}\,{=}\,0$. 
In this case we can now refer to Eq.~\ref{eq:f_2d_polar} for the 2D
probability density function.

Contrary to the canonical case, when the effective ellipticity differs from
$\varepsilon_{\rm eff}\,{=}\,1$, 
the statistical biases on the polarization fraction and angle become dependent
on the true polarization angle $\psi_0$, as illustrated in
Fig.~\ref{fig:impact_psi0_biasppsi} for the special case of
$\theta\,{=}\,0$ (no correlation).
For extreme values of the ellipticity (e.g., $\varepsilon_{\rm eff}\,{=}\,2$),
the relative bias on $p$ oscillates between 0.9 and 1.5 times the canonical
bias ($\varepsilon_{\rm eff}\,{=}\,1$, red line).
These oscillations with $\psi_0$ quickly vanish when the ellipticity gets
closer to 1, as shown for $\varepsilon_{\rm eff}\,{=}\,1.1$ in the figure.
The presence of correlations (i.e., $\rho \ne 0$) increases the effective
ellipticity of the noise distribution
associated with a global rotation, as detailed in sect.~\ref{sec:notations}.
Thus correlations induce the same oscillation patterns observed in
Fig.~\ref{fig:impact_psi0_biasppsi} for a null correlation, 
but amplified at the corresponding effective ellipticity
$\varepsilon_{\rm eff}$ and shifted by an angle $\theta/2$, according to
Eqs.~\ref{eq:epsi_eff} and \ref{eq:phi}, respectively.

The top panel of Fig.~\ref{fig:impact_epsilon_biasppsi} shows the dependence
of the polarization fraction bias on the effective ellipticity for
three levels of S/N: $p_0/\sigma_{\rm p,G}\,{=}\,1$, 2, and 5, and including
the whole range of true polarization angle $\psi_0$.
The grey shaded regions indicate the variability interval of
$\Delta p / \sigma_{{\rm p},0}$ for each ellipticity,
for changes in $\psi_0$ over the range $-\pi/2$ to $\pi/2$.
We observe that the higher the S/N, the stronger the relative impact of
the ellipticity compared to the canonical case.
In the {\it low\/} regime the relative bias to the dispersion 
increases from 9\% to 12\% (compared to 10\% in the canonical case)
at a S/N of 5, while it spans from 69\% and 73\% (around the 71\% of the
canonical case) at a S/N of 1.
Hence, in the {\it low\/} regime, the impact of the ellipticity
on the bias
of the polarization fraction represents only about 4\% of the dispersion,
whatever the S/N, which can therefore be neglected.
However, in the {\it extreme\/} regime, the impact of the ellipticity can go
up to 33\% at intermediate S/N (${\sim}\,2$), which can no longer be neglected.

Now concerning the impact on polarization angle -- while no bias occurs in the
canonical case, some oscillations in the bias $\Delta \psi$ 
with $\psi_0$ appear as soon as $\varepsilon_{\rm eff}\,{>}\,1$.  The amplitude
can reach up to 24\% of the dispersion in the {\it extreme\/} regime 
and up to 4\% in the {\it low\/} regime, 
as illustrated in the bottom panel of Fig.~\ref{fig:impact_psi0_biasppsi}.
Again, these oscillations are shifted and amplified in the presence of
correlations between the Stokes parameters, compared to the case with no
correlation.  As a global indicator, in the bottom panel of
Fig.~\ref{fig:impact_epsilon_biasppsi} we provide the maximum bias
${\rm Max} |\Delta \psi|$ normalized by the dispersion $\sigma_{\psi,0}$
over the whole range of $\psi_0$ as a function of the ellipticity.
This quantiry barely exceeds 24\% (i.e., ${\sim}\,9^\circ$) in the worst case,
i.e., for $\varepsilon_{\rm eff}\,{=}\,2$ and low S/N, and 
it falls to below 4\% (i.e., ${\sim}\,1.5^\circ$) in the {\it low\/} regime.
Thus the bias on $\psi$ always remains well below the level of the true 
uncertainty on the polarization angle at the same S/N
(see Sect.~\ref{sec:sigma_p}), 
so that the bias of the polarization angle induced by an ellipticity
$\varepsilon_{\rm eff}\,{>}\,1$ can be neglected to first order for the
{\it low\/} regime of the ellipticity, i.e., when there is less than a
10\% departure from the canonical case.

\subsection{$I$ uncertainty}
\label{sec:bias_ppsi_2d_iuncertainty}

The uncertainty in the total intensity $I$ has two sources: the measurement
uncertainty expressed in the covariance matrix; and an astrophysical component
of the uncertainty due to the imperfect characterization of 
the unpolarized contribution to the total intensity.  This second source
can be seen, for instance, with the cosmic infrared background in
{\it Planck\/} data -- its unpolarized emission can be 
viewed as a systematic uncertainty on the total intensity (dominated by
the Galactic dust thermal emission), when one is interested in the
polarization fraction of the Galactic dust.
To retrieve the actual polarization fraction, it is necessary to compute it
through
\begin{equation}
p=\frac{\sqrt{Q^2+U^2}}{(I-\Delta I)},
\end{equation}
where $\Delta I$ is the unpolarized emission, which is imperfectly known.
The uncertainty $\sigma_{\Delta I}$ on this quantity can be viewed as an
additional uncertainty $\sigma_{\rm I}$ on the total intensity
and therefore the S/N has to be written
$I_0/\sigma_{\rm I}\,{=}\,(I-\Delta I)/\sigma_{\Delta I}$.

In order to consider the effects on polarization quantities,
we first recall that, because of its definition, the measurement of
polarization angle $\psi$ is not impacted by the 
uncertainty on intensity (when no correlation exists between $I$ and
$Q$ and $U$), contrary to the polarization fraction $p$, which is defined
as the ratio of the polarized intensity to the total intensity.  Thus the
uncertainty of the total intensity does not induce any bias on $\psi$.

To quantify the influence of a finite signal-to-noise ratio
$I_0/\sigma_{\rm I}$ on the bias of $p$, we compute the mean polarization
fraction over the PDF:
\begin{equation}
\overline{p}=\iiint \frac{\sqrt{Q^2+U^2}}{I}
 F\left(I,Q,U\,|\,I_0,Q_0,U_0,\tens{\Sigma}\right)dI\,dQ\,dU,
\end{equation}
with $F$ given by Eq.~\ref{eq:pdf_IQU}.  We write it this way, because using
$f_{\rm 2D}$ given by Eqs.~\ref{eq:f_2d_ppos} and \ref{eq:f_2d_pneg} would
lead to both positive and negative logarithmic divergences for
$p\to\pm\infty$ (related to samples for which $I\to0$).  These divergences
can be shown to be artificial by using the Gaussian PDF of $(I,Q,U)$ instead
of $f_{\rm 2D}$.

\begin{figure}
 \psfrag{-----title-----}{$ \widetilde{\Delta p} / p_0 \, [\%]$}
\centerline{\includegraphics[width=9cm]{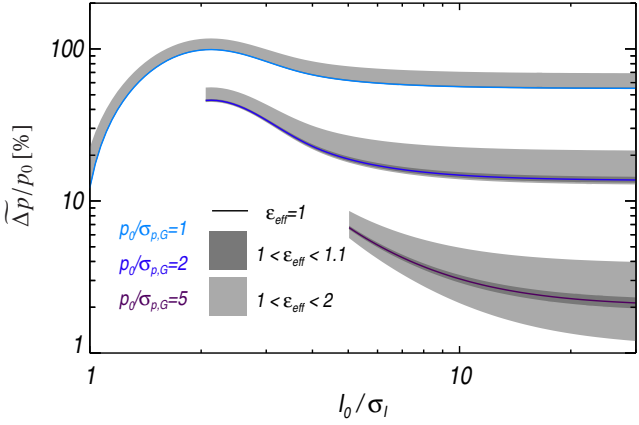}}
\caption{Absolute polarization fraction bias normalized to the true value
$p_0$.
The level of the dispersion scaled to the true value, $\sigma_{{\rm p},0}/p_0$,
is shown as a dashed line for the canonical case as a function of the S/N
$I_0/\sigma_{\rm I}$, plotted for three values of the polarization S/N,
$p_0/\sigma_{\rm p,G}$, and values of the effective ellipticity
$\varepsilon_{\rm eff}$ covering the {\it canonical},
{\it low}, and {\it extreme\/}
regimes of the covariance matrix.  The intensity correlation coefficients are
set to $\rho_{\rm Q}\,{=}\,\rho_{\rm U}\,{=}\,0$.  Notice that we only consider
the domain where $(I_0/\sigma_{\rm I})\,{>}\,(p_0/\sigma_{\rm p,G})$.} 
\label{fig:dp-var-eta-var-epsilon-var-SNRp_absolute}
\end{figure}

The presence of noise in total intensity measurements increases the absolute
bias $\Delta p=\overline{p}-p_0$, as shown in
Fig.~\ref{fig:dp-var-eta-var-epsilon-var-SNRp_absolute},
where $\Delta p$ (scaled by the true value $p_0$) is plotted as a function of
the signal-to-noise ratio $I_0/\sigma_{\rm I}$.  This is shown for three levels
of the polarization S/N ratio $p_0/\sigma_{\rm p,G}\,{=}\,1$, 2, and 5, and the
three regimes of the covariance matrix, indicated as a solid line
({\it canonical\/}),
dark shading ({\it low\/} regime) and light shading ({\it extreme\/} regime),
assuming that $\rho_{\rm Q}\,{=}\,\rho_{\rm U}\,{=}\,0$.

The absolute bias may be enhanced by a factor of 5--10 when the
signal-to-noise ratio on $I$ goes from infinite (i.e., perfectly known $I$)
to about 2.  It then drops again for lower signal-to-noise ratios, which is
the result of the increasing number of negative $p$ samples.  Notice that we
only consider the domain where
$(I_0/\sigma_{\rm I})\,{>}\,(p_0/\sigma_{\rm p,G})$. 

Comparison of the bias to the dispersion $\sigma_{{\rm p},0}$, as was done in
the previous subsection, is not straightforward when the total intensity is
uncertain.  This is because the integral defining $\sigma_{{\rm p},0}$
(see Eq.~\ref{eq:sigp_0}) has positive linear divergences for $p\to\pm\infty$.
Unlike the case of $\overline{p}$, this divergence cannot be alleviated by
working in $(I,Q,U)$ space.

\begin{figure}
 \psfrag{-----title-----}{$\Delta p / \widetilde{\sigma}_{{\rm p},0} \, [\%]$}
\centerline{\includegraphics[width=9cm]{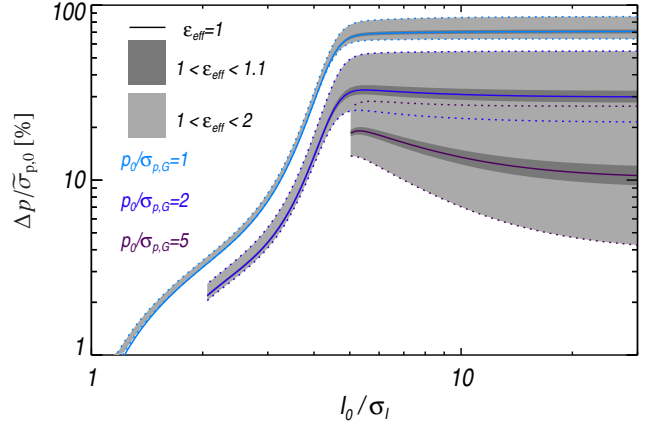}}
\caption{Same as Fig.~\ref{fig:dp-var-eta-var-epsilon-var-SNRp_absolute},
but showing the bias on the polarization fraction relative to the dispersion
proxy $\widetilde{\sigma}_{{\rm p},0}$.  See text for a description of this
quantity.} 
\label{fig:dp-var-eta-var-epsilon-var-SNRp}
\end{figure}

To overcome this
we therefore used a proxy $\widetilde{\sigma}_{{\rm p},0}$, which is the
dispersion of $p$ computed on a subset of $(I,Q,U)$ space that excludes total
intensity values below $\omega I_0$, with $\omega=10^{-7}$.  This threshold is
somewhat arbitrary, as $\widetilde{\sigma}_{{\rm p},0}$ increases linearly
with $1/\omega$.  The value $10^{-7}$ is merely meant to serve as an
illustration.  Figure~\ref{fig:dp-var-eta-var-epsilon-var-SNRp} shows
$\Delta p/\widetilde{\sigma}_{{\rm p},0}$ as a function of $I_0/\sigma_{\rm I}$
for the same values of the polarization signal-to-noise ratio
$p_0/\sigma_{\rm p,G}$ and the same regimes of the covariance matrix as in
Fig.~\ref{fig:dp-var-eta-var-epsilon-var-SNRp_absolute}.  At high S/N for $I$,
we asymptotically recover the values obtained in the top panel of
Fig.~\ref{fig:impact_epsilon_biasppsi}.  As long as $I_0/\sigma_{\rm I}>5$,
the relative bias on $p$ is barely affected by the uncertainty on the
intensity, especially for low polarization S/N, $p_0/\sigma_{\rm p,G}$.
A small trend is still seen in the range $5<I_0/\sigma_{\rm I}<10$ for
$p_0/\sigma_{\rm p,G}\,{=}\,5$ -- the relative bias may be enhanced by a factor
of around 2 in that case, when the S/N on intensity and polarization are of
the same order (${\sim}\,5$).  However, this situation is unlikely to be
observed in astrophysical data, since the uncertainty on total intensity is
usually much smaller than that on polarized intensity. 

Contrary to these high S/N ($I_0/\sigma_{\rm I}\,{>}\,5$) features, which are
quite robust with respect to the choice of threshold $\omega I_0$, the drop in
relative bias at lower intensity S/N, i.e., $I_0/\sigma_{\rm I}\,{<}\,5$, is
essentially due to the divergence of the dispersion of $p$.  Hence this part of
Fig.~\ref{fig:dp-var-eta-var-epsilon-var-SNRp_absolute}
should be taken as nothing more than an illustration of the
divergence at low S/N for $I$.  It should be stressed, however, that this
increase of the dispersion of $p$ has to be carefully taken into account when
dealing with low S/N intensity data, which can be the case well away from
the Galactic plane.

\subsection{Correlation between $I$ and $Q$--$U$}
\label{sec:bias_ppsi_2d_ileakage}

With non-zero noise on total intensity, it becomes 
possible to explore the effects of the coefficients 
$\rho_{\rm Q}$ and $\rho_{\rm U}$, corresponding to correlation
between the intensity $I$ and the $(Q,U)$ plane. 

We first note that the introduction of correlation parameters $\rho_{\rm Q}$
and $\rho_{\rm U}$ that are different from zero directly modifies the
ellipticity $\varepsilon$ and correlation $\rho$ between Stokes $Q$ and $U$. 
Simple considerations on the Cholesky decomposition of the covariance matrix
$\tens{\Sigma}$ (given in Appendix~\ref{sec:appendix-cholesky}) show that
for a given ellipticity $\varepsilon$ and correlation parameter $\rho$,
obtained when $\rho_{\rm Q}\,{=}\,\rho_{\rm U}\,{=}\,0$, the ellipticity
$\varepsilon^\prime$ and correlation $\rho^\prime$ become
\begin{equation}
\label{eq:rhouq_epsirho}
\varepsilon^\prime=\varepsilon\sqrt{\frac{1-\rho_{\rm Q}^2}{1-\rho_{\rm U}^2}}
 \quad \mathrm{and} \quad 
\rho^\prime=\rho_{\rm Q}\rho_{\rm U}+\rho
 \sqrt{\left(1-\rho_{\rm Q}^2\right)\left(1-\rho_{\rm U}^2\right)}
\end{equation}
when $\rho_{\rm Q}$ and $\rho_{\rm U}$ are no longer null.
Consequently, non null $\rho_{\rm Q}$ and $\rho_{\rm U}$ yield similar
impacts as found for a non-canonical effective ellipticity
($\varepsilon_{\rm eff}$$\ne$1), discussed
in Sect.~\ref{sec:bias_ppsi_2d_ellipticity}.
Moreover, in order to investigate the sole impact of non-null 
 $\rho_{\rm Q}$ and $\rho_{\rm U}$ with a finite S/N on the intensity,
we have compared the case $(\varepsilon,\rho,\rho_{\rm Q},\rho_{\rm U})$
to the reference case $(\varepsilon^\prime, \rho^\prime,0,0)$.
We find that the relative change of the polarization fraction bias $\Delta p$
is at most 10--15\% over the whole range of $I_0/\sigma_{\rm I}$ explored
in this work (i.e., $I_0/\sigma_{\rm I}\geqslant 1$). 
 
Concerning the polarization angle bias, the difference between the bias
computed for $(\varepsilon,\rho,\rho_{\rm Q},\rho_{\rm U})$ and that for
the reference case $(\varepsilon^\prime,\rho^\prime,0,0)$ 
is at most $\Delta\psi-\Delta\psi_\mathrm{ref}\sim 4^\circ$,
and essentially goes to zero above $I_0/\sigma_{\rm I}\,{\sim}\,2$--3. 
The dependence of the change in bias with $(\rho_{\rm Q},\rho_{\rm U})$ 
is similar to that for $\Delta p/\Delta p_\mathrm{ref}$,
except that it depends solely on $\rho_{\rm U}$ for $\psi_0\,{=}\,0$
and solely on $\rho_{\rm Q}$ for $\psi_0\,{=}\,\pi/4$.

\section{Polarization uncertainty estimates}
\label{sec:sigma_p}

If we are given the polarization measurements and the noise covariance matrix
of the Stokes parameters, we would like to derive estimates of the
uncertainties associated with the polarization fraction and angle.  These are
required to: (i) define the signal-to-noise ratio of these polarization
measurements; and (ii) quantify how important the bias is compared to the
accuracy of the measurements.  In the most general case the uncertainties
in the polarization fraction and angle do not follow a Gaussian distribution,
so that confidence intervals should be properly used to obtain
an estimate of the associated errors, as is described in
Sec.~\ref{sec:confidence_intervals}.
However, it can sometimes be assumed as a first approximation that the
distributions are Gaussian, in order to derive quick estimates of the
$p$ and $\psi$ uncertainties, defined as the variance of the
2D distribution of the polarization measurements.  We explore below the extent
to which this approximation can be utilized, when using
the most common estimators of these two quantities.

\begin{figure*}
\vspace{1cm}
\begin{tabular}{cc}
 \psfrag{---------------------ytitle---------------------}{$\mathcal{P} \left( p_0 \in \left[ p - \sigma_{\rm p,0} , p + \sigma_{\rm p,0} \right] \right)\, [\%]$}
 \includegraphics[width=9cm]{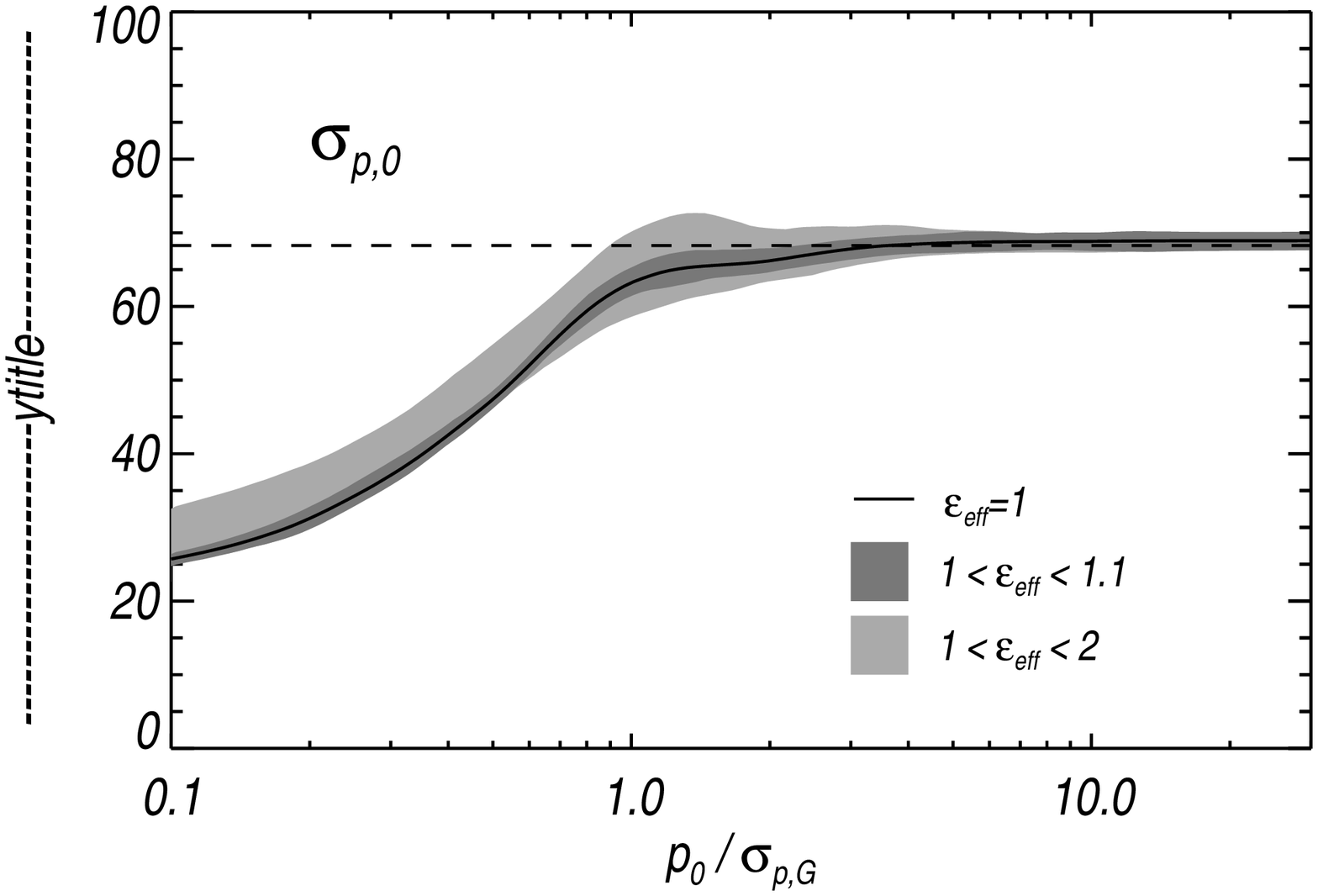} &
 \psfrag{---------------------ytitle---------------------}{$\mathcal{P} \left( p_0 \in \left[ p - \sigma_{\rm p,G} , p + \sigma_{\rm p,G} \right] \right)\, [\%]$}
 \includegraphics[width=9cm]{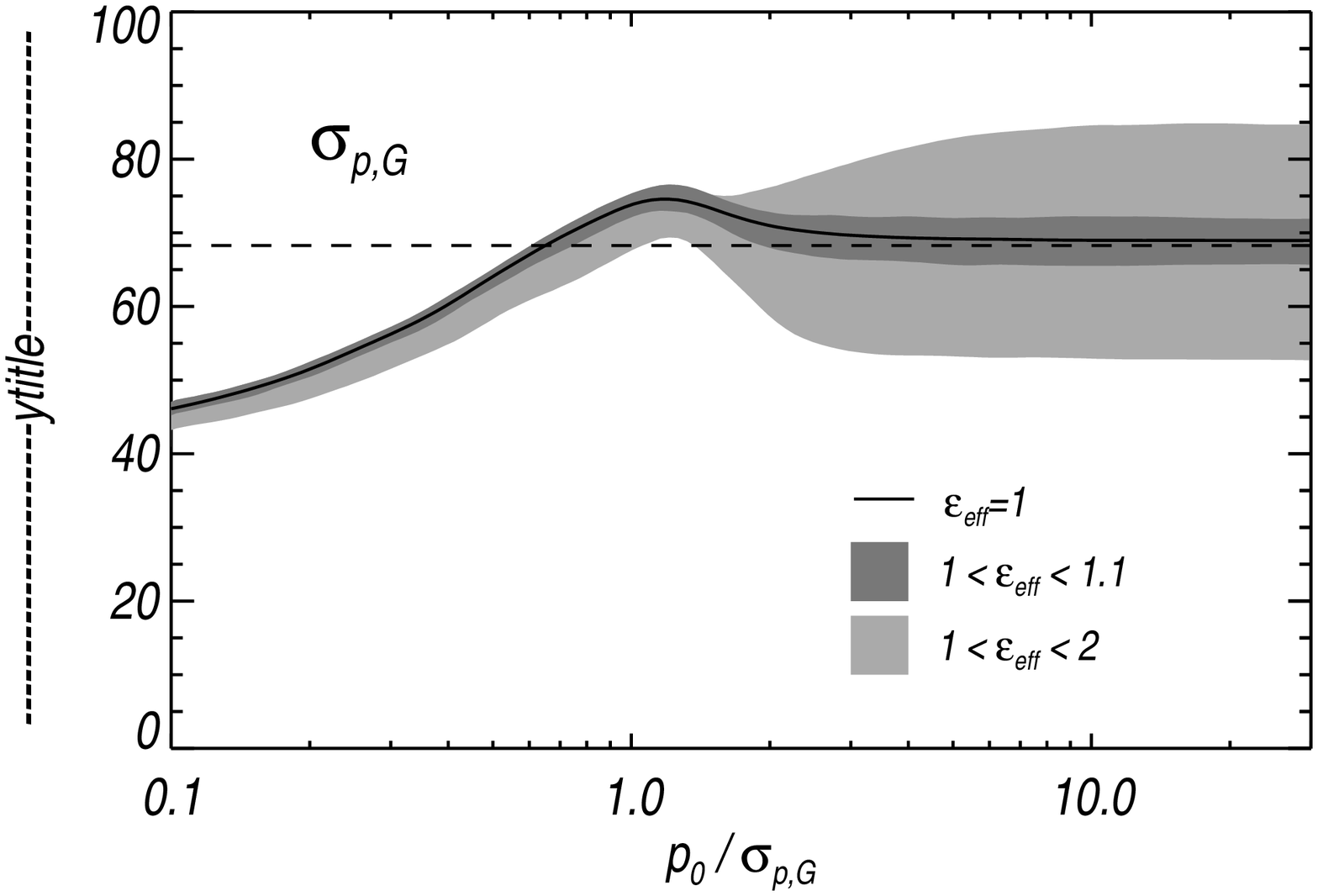} \\
 \psfrag{---------------------ytitle---------------------}{$\mathcal{P} \left( p_0 \in \left[ p - \sigma_{\rm p,C} , p + \sigma_{\rm p,C} \right] \right)\, [\%]$}
 \includegraphics[width=9cm]{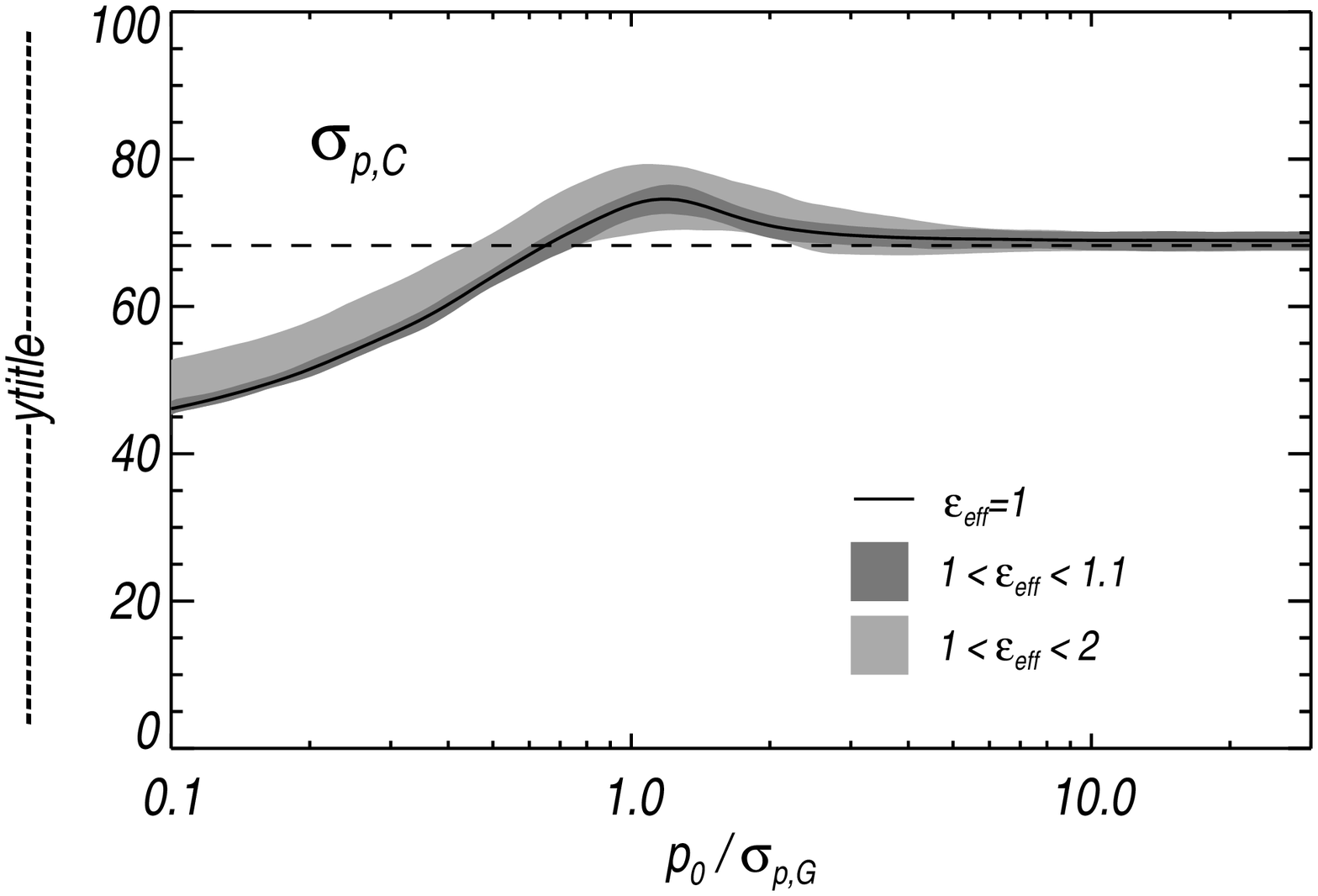} &
 \psfrag{---------------------ytitle---------------------}{$\mathcal{P} \left( p_0 \in \left[ p - \sigma_{\rm p,A} , p + \sigma_{\rm p,A} \right] \right)\, [\%]$}
 \includegraphics[width=9cm]{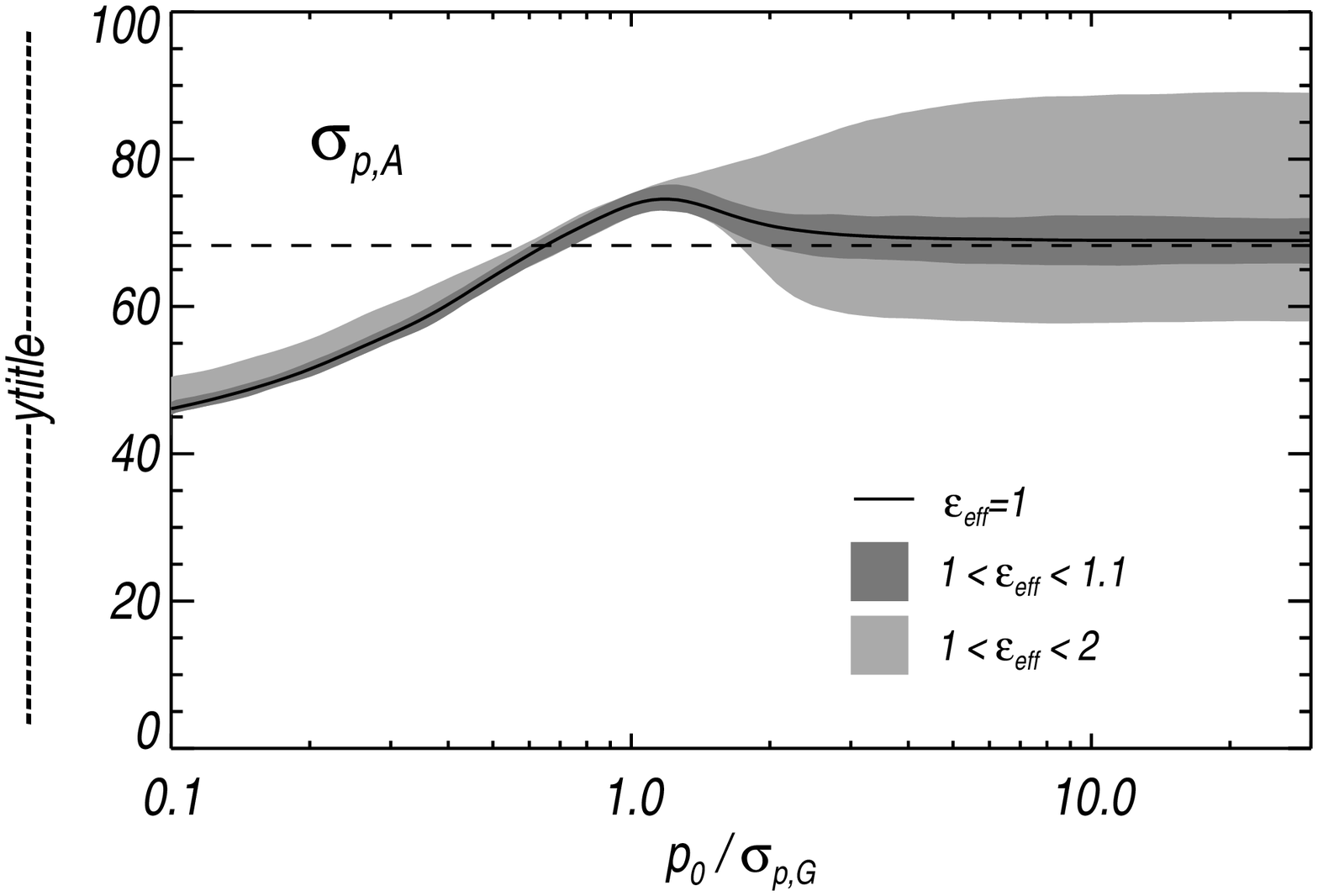} 
 \end{tabular}
 \caption{Probability $\mathcal{P}$ to find the true polarization fraction
 $p_0$ inside the interval $[p-\sigma_{\rm p}^{\rm low},
 p+\sigma_{\rm p}^{\rm up}]$, where
 $\sigma_{\rm p}^{\rm low}$ and $\sigma_{\rm p}^{\rm up}$
 are the 1$\,\sigma$ lower and upper limits, respectively.  We plot this
 for each estimators: true $\sigma_{\rm p,0}$ (top left);
 classical $\sigma_{\rm p,C}$ (bottom left);
 geometric $\sigma_{\rm p,G}$ (top right);
 and arithmetic $\sigma_{\rm p,A}$ (bottom right).
 These are plotted as a function of the S/N $p_0/\sigma_{\rm p,G}$.
 Monte Carlo simulations have been carried out in the {\it canonical\/}
 (solid line),
 {\it low\/} (dark grey), and {\it extreme\/} (light grey) regimes of the
 covariance matrix.  The expected 68.27\% level is shown as a dashed line. }
 \label{fig:uncertainties_sigp}
\end{figure*}

\begin{figure*}
\vspace{1cm}
\begin{tabular}{cc}
 \psfrag{---------------------ytitle---------------------}{$\mathcal{P} \left( \psi_0 \in \left[ \psi - \sigma_{\psi,0} , \psi + \sigma_{\psi,0} \right] \right)\, [\%]$}
 \includegraphics[width=9cm]{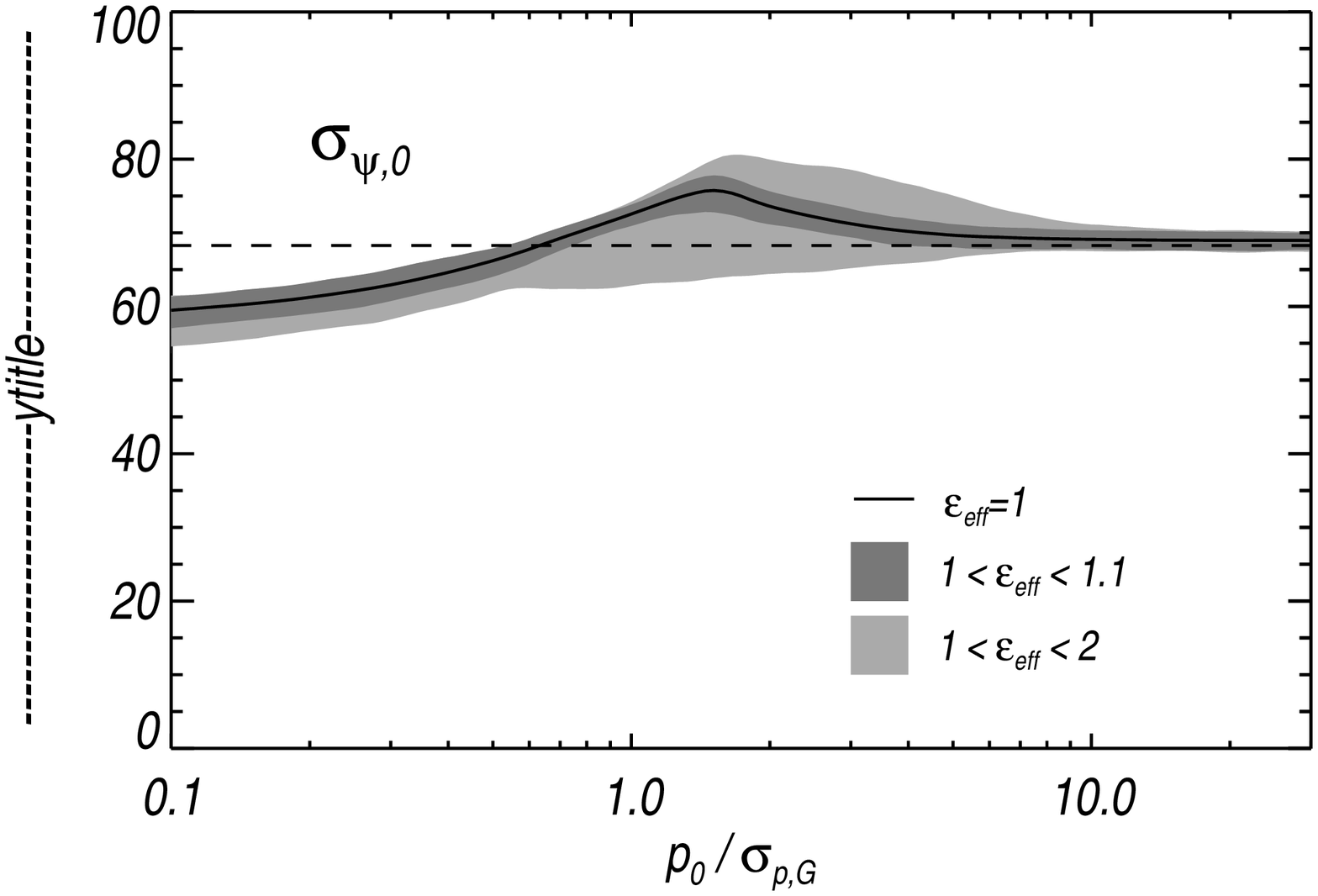} &
 \psfrag{---------------------ytitle---------------------}{$\mathcal{P} \left( \psi_0 \in \left[ \psi - \sigma_{\psi,{\rm C}} , \psi + \sigma_{\psi,{\rm C}} \right] \right)\, [\%]$}
 \includegraphics[width=9cm]{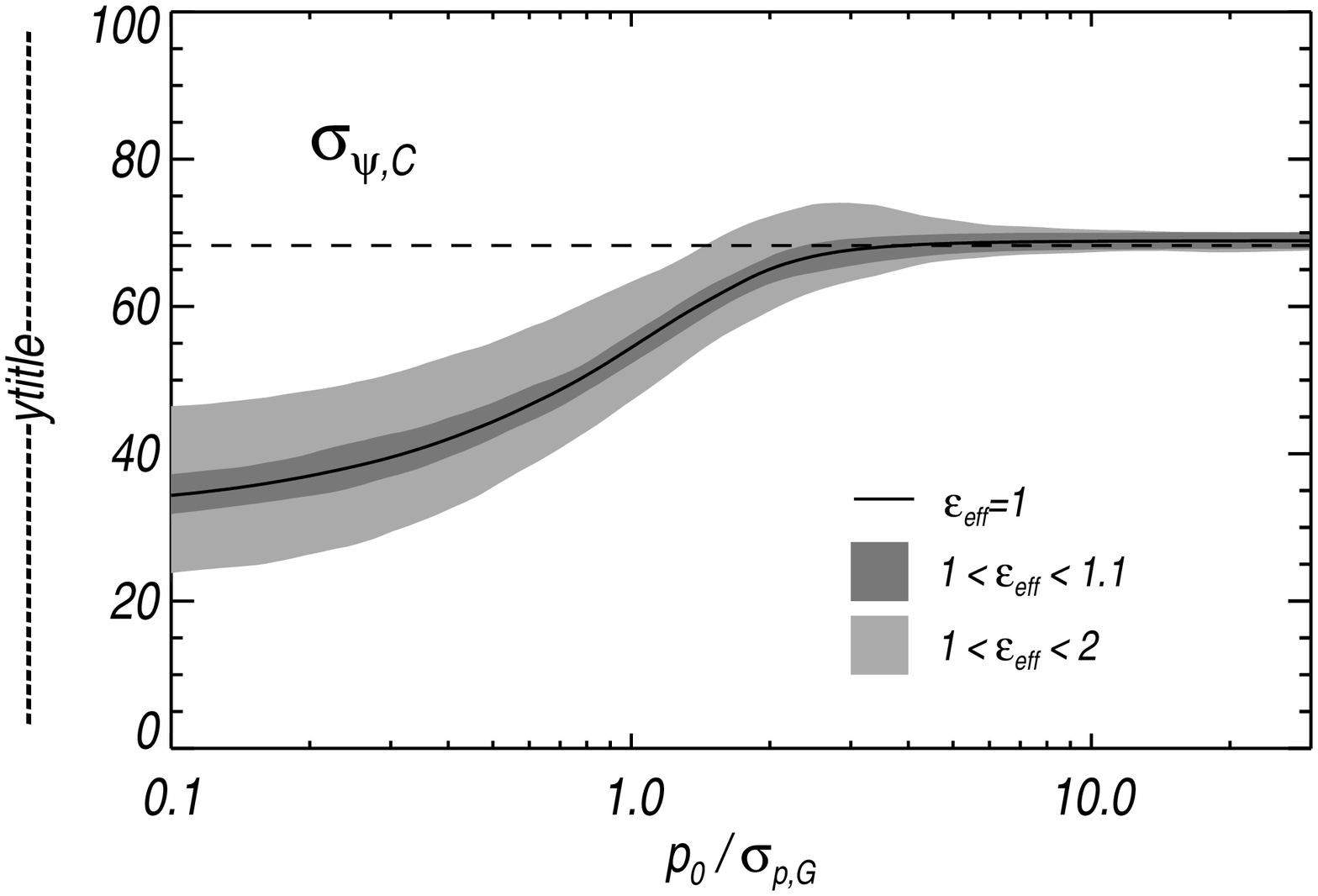} 
 \end{tabular}
 \caption{Same as Fig.~\ref{fig:uncertainties_sigp}, but for the polarization
 angle uncertainty estimators.  {\it Left\/}: $\sigma_{\psi,0}$.
 {\it Right\/}: classical $\sigma_{\psi,{\rm C}}$.}
 \label{fig:uncertainties_sigpsi}
\end{figure*}

\subsection{Standard deviation estimates}

To compare the robustness of the uncertainty estimates, we build 10 000
Monte Carlo simulated measurements in each of the three regimes
of the covariance matrix ({\it canonical}, {\it low}, and {\it extreme\/}),
by varying the S/N of $p$ and the polarization angle $\psi_0$
inside the range $-\pi/2$ and $\pi/2$.  We use the simulations to
compute the posterior fraction of measurements for which 
the true value $p_0$ or $\psi_0$ falls inside the $\pm \sigma$
range around the measurement. 
This provides the probability $\mathcal{P}$ shown in
Figs.~\ref{fig:uncertainties_sigp} and \ref{fig:uncertainties_sigpsi}, 
for $p$ and $\psi$, respectively.
 
\begin{figure}
 \includegraphics[width=9cm]{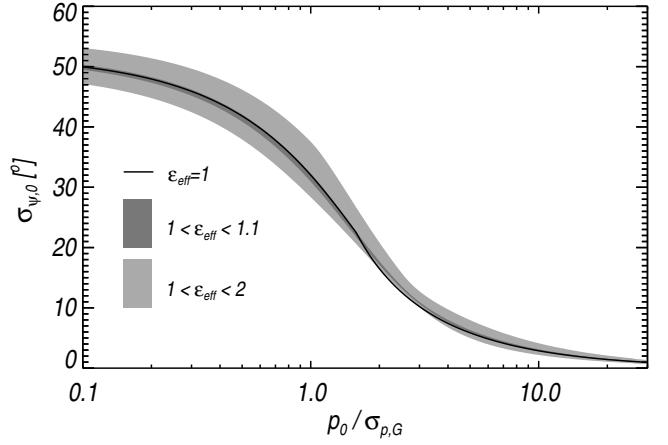} 
 \caption{True polarization angle uncertainty, $\sigma_{\psi,0}$,
 as a function of the S/N, $p_0/\sigma_{\rm p,G}$.
 The three regimes ({\it canonical}, {\it low}, and {\it extreme\/}) of 
 the covariance matrix are explored (solid line, light, and dark grey shaded
 regions, respectively).}
 \label{fig:uncertainties_sigphi_true}
\end{figure}

We first focus on the true uncertainty estimates, as defined
in Sect.~\ref{sec:bias_methodology}.  We observe that the
$\sigma_{{\rm p},0}$ true estimates
(top left of Fig.~\ref{fig:uncertainties_sigp}) 
fall below the Gaussian value ${\rm erf}(\sqrt{2}/2)$ (i.e., 68\%)
once the S/N goes below 3.
The $\sigma_{\psi,0}$ true estimates (left of
Fig.~\ref{fig:uncertainties_sigpsi}) provide conservative probabilities
($\mathcal{P}>68\%$) for S/N$\,{>}\,0.5$.
This is also shown in Fig.~\ref{fig:uncertainties_sigphi_true} as a function
of the S/N, for the {\it canonical}, {\it low}, and {\it extreme\/}
regimes of the covariance matrix.  Notice that it is not strongly dependent
on the ellipticity of the covariance matrix.  It shows a maximum of
$\pi/\sqrt{12}\,{\simeq}\,52^\circ$ at low S/N, and 
converges slowly to 0 at high S/N (still being $\sim10^\circ$ at a
S/N$\,{=}\,3$).
Thus we might imagine using such estimates as reasonably good approximations
to the uncertainties at high S/N (${>}\,3$) for $p$, 
and over almost the entire range of S/N for $\psi$. 
However, these true $p$ and $\psi$ uncertainties, 
$\sigma_{{\rm p},0}$ and $\sigma_{\psi,0}$, respectively, depend on $p_0$
and $\psi_0$, which remain theoretically unknown.
Thus we can only provide specific estimates of those variance quantities,
as detailed below.

\subsection{Geometric and arithmetic estimators}

Two estimates of the polarization fraction uncertainty can be obtained
independently of the measurements themselves, which makes them easy to
compute: (i) the geometric ($\sigma_{\rm p,G}$) estimate; and (ii)
the arithmetic ($\sigma_{\rm p,A}$) estimate.

The geometric estimator was already introduced earlier, when we derived
the expression for the two dimensional ($p,\psi$) 
probability density function $f_{\rm 2D}$.  It is defined via the determinant
of the 2D covariance matrix $\tens{\Sigma}_{\rm p}$ as
$\det(\tens{\Sigma}_{\rm p})\,{=}\,\sigma_{\rm p,G}^4$, with
its expression given in Eq.~\ref{eq:sigpg}. 
We recall that the determinant of the covariance matrix
$\tens{\Sigma}_{\rm p}$ is linked to the area inside a probability contour, 
and independent of the reference frame of the Stokes parameters.
In the canonical case, this estimate gives back the usual expressions,
$\sigma_{\rm p,G}\,{=}\,\sigma_{\rm Q}/I_0\,{=}\,\sigma_{\rm U}/I_0$,
used to quantify the noise on the polarization fraction.
It can be considered as the geometric mean of $\sigma_{\rm Q}$ and
$\sigma_{\rm U}$ when there is no correlation between them, i.e.,
$\sigma_{\rm p,G}^2\,{=}\,\sigma_{\rm Q} \sigma_{\rm U} / I_0^2$. 

The arithmetic estimator is defined as a simple quadratic mean of the variance
in $Q$ and $U$:
\begin{equation}
 \sigma_{{\rm p},A}^2 = \frac{1}{2}
 \frac{\sigma_{\rm Q}^2 + \sigma_{\rm U}^2}{I_0^2}
 = \frac{\sigma_{\rm Q}^2}{I_0^2} \frac{(\varepsilon^2 + 1)}{2 \varepsilon^2}.
\end{equation}
This estimate also gives back
$\sigma_{{\rm p},A}\,{=}\,\sigma_{\rm Q}/I_0\,{=}\,\sigma_{\rm U}/I_0$
in the canonical case.  Furthermore, it is also independent of the reference
frame or whether correlations are present.

The two estimators have very similar behaviour, as can be seen in the
top and bottom right panels of Fig.~\ref{fig:uncertainties_sigp}.
They agree perfectly with a 68\% confidence level for S/N
$p_0/\sigma_{{\rm p},0}\,{>}\,4$ and for standard simplification of
the covariance matrix.  Both estimators provide conservative probability
($\mathcal{P},{>}\,68\%$) in the S/N range 0.5-4.
The impact of the effective ellipticity of the covariance matrix
(grey shaded area) is stronger for larger values of the S/N ($>$2), 
and can yield variations of 30\% in the probability $\mathcal{P}$ for
the {\it extreme\/} regime.  These estimators should be used cautiously for
high ellipticity, but provide quick and conservative estimates in the other
cases.

\subsection{Classical Estimate}
\label{sec:classical_estimate}

The classical determination of the uncertainties proposed by
\citet{Serkowski1958,Serkowski1962} is often used for polarization
determinations based on optical extinction data.
Although investigated by \citet{Naghizadeh1993}, these classical
uncertainties still do not include asymmetrical terms and correlations
in the covariance matrix.  Here we extend the method to the general case,
by using the derivatives of $p$ and $\psi$
around the observed values of the $I$, $Q$, and $U$ parameters.
It should be noted that, since this approach is based on derivatives around the
observed values of ($I$, $Q$, $U$), it is only valid in the high
signal-to-noise regime.
The detailed derivation, provided in
Appendix~\ref{sec:classical_uncertainties}, leads to the expressions
\begin{eqnarray}
\sigma_{\rm p,C}^2 &= & \frac{1}{p^2 I^4} \times
 \Big( Q^2\sigma_{\rm Q}^2+U^2\sigma_{\rm U}^2+p^4I^2\sigma_{\rm I}^2
 \nonumber \\
 & & \quad\quad
 +2 QU\sigma_{\rm QU}-2IQp^2\sigma_{\rm IQ}-2IUp^2\sigma_{\rm IU} \Big) 
\end{eqnarray}
and
\begin{eqnarray}
\sigma_{\psi,{\rm C}}^2 & = & \frac{1}{4}
 \frac{Q^{2}\sigma_{\rm U}^2+U^{2}\sigma_{\rm Q}^2-2QU\sigma_{\rm QU}}
 {(Q^{2}+U^{2})^2} \, \, \mathrm{rad}^2,
\end{eqnarray}
where $I$, $Q$, $U$, and $p$ are the measured quantities, and
$\sigma_{XY}$ are the elements of the covariance matrix.
We recall that the maximum uncertainty on $\psi$ is equal to
$\pi/\sqrt{12} \, \mathrm{rad}$ 
(integral of the variance of the polarization angle over a flat distribution
between $-\pi/2$ and $\pi/2$).
When $\sigma_{\rm I}$ can be neglected, we obtain
\begin{equation}
\label{eq:classical_sigphi}
\sigma_{\psi,{\rm C}} =
 \sqrt{ \frac{Q^{2}\sigma_{\rm U}^2+U^{2}\sigma_{\rm Q}^2-2QU\sigma_{\rm QU}}
 {Q^{2}\sigma_{\rm Q}^2+U^{2}\sigma_{\rm U}^2+2QU\sigma_{\rm QU}} }
 \times \frac{\sigma_{\rm p,C}}{2 p} \, \, \mathrm{rad}.
\end{equation}
Because the uncertainty of $\psi$ is also often expressed in degrees,
we provide the associated conversions:
$\pi/\sqrt{12}\,\mathrm{rad}\,{=}\,51\pdeg96$;
and $1/2\,\mathrm{rad}\,{=}\,28\pdeg65$.
Moreover, under the canonical assumptions, 
we recover $\sigma_{\rm p,C}\,{=}\,\sigma_{\rm p,G}\,{=}\,\sigma_{\rm Q}/I_0
 \,{=}\,\sigma_{\rm U}/I_0$ and
$\sigma_{\psi,{\rm C}}\,{=}\,\sigma_{\rm p,C} / 2p \,\, \mathrm{rad}$.
 
Since the classical estimate of the uncertainty $\sigma_{\rm p,C}$ is equal
to $\sigma_{\rm p,G}$ under the standard simplifications of the covariance
matrix, it has the same deficiency at low S/N (see bottom left panel of
Fig.~\ref{fig:uncertainties_sigp}). 
The impact of the effective ellipticity 
of the covariance matrix (grey shaded area) tends to be negligible at high
S/N ($p_0/\sigma_{\rm p,G}>4$), and remains limited at low S/N. 
Thus this estimator of the polarization fraction uncertainty appears more
robust than the geometric and arithmetic estimators, while still being
easy to compute, and valid (even conservative) over a wide range of S/N.

The classical estimate of the polarization angle uncertainty,
$\sigma_{\psi,{\rm C}}$, is shown in Fig.~\ref{fig:uncertainties_sigpsi}
(right panel) in the {\it canonical}, {\it low}, and {\it extreme\/} regimes
of the covariance matrix.  It appears that $\sigma_{\psi,{\rm C}}$ is
strongly under-estimated at low S/N, mainly due to the
presence of the term $1/p$ in Eq.~\ref{eq:classical_sigphi}, 
where $p$ is strongly biased at low S/N.  For S/N$\,{>}\,4$,
the agreement between the probability $\mathcal{P}$ and the expected value
is good, while the impact of the ellipticity of the covariance matrix
becomes negligible only for S/N$\,{>}\,10$. 
Hence this estimator can cetainly be used at high S/N.

\begin{figure}
 \includegraphics[width=9cm]{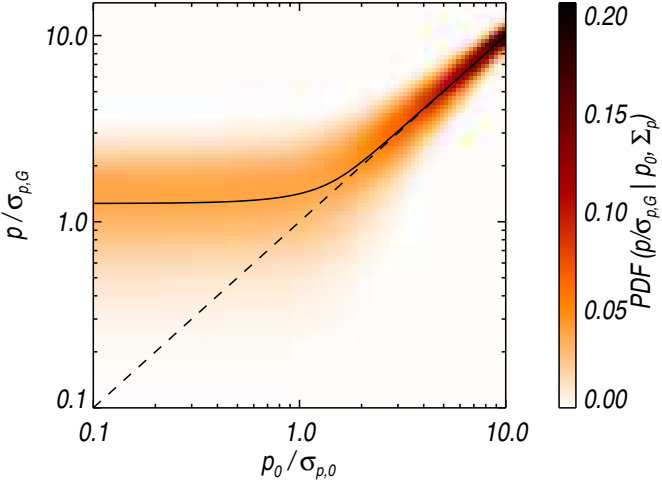} 
 \caption{Probability density function of the measured signal-to-noise
 ratio $p/\sigma_{\rm p,G}$ (where $\sigma_{\rm p,G}$ is the geometric
 estimate) as a function of the true signal-to-noise ratio
 $p_0/\sigma_{{\rm p},0}$, with no ellipticity and correlation in the
 covariance matrix $\tens{\Sigma}_{\rm p}$.
 The mean likelihood, $\overline{p}/\sigma_{\rm p,G}$ (full line),
 tends to $\sqrt{\pi/2}$
 at low S/N and to the 1:1 relation (dashed line) at high S/N
 ($p_0/\sigma_{{\rm p},0}\,{>}\,2$).}
 \label{fig:uncertainty_bias}
\end{figure}

\subsection{S/N estimates}
\label{sec:snr_estimates}

It is important to stress how any measurement of the signal-to-noise ratio,
$p/\sigma_{{\rm p,G}}$, is highly impacted by the bias on the measured
polarization fraction $p$, as shown in Fig.~\ref{fig:uncertainty_bias}.
We observe that at high S/N ($p_0/\sigma_{{\rm p},0}\,{>}\,2$) the measured 
S/N, here $p / \sigma_{\rm p,G}$, is very close to the true S/N (dashed line).
The mean likelihood of the measured S/N (solid line) 
flattens for lower true S/N, such that $\overline{p}/\sigma_{\rm p,G}$
tends to $\sqrt{\pi/2}$ for $p_0/\sigma_{{\rm p},0}\,{<}\,1$, 
which comes from the limit of the \citet{Rice1945} function when
$p_0/\sigma_{{\rm p},0}\rightarrow0$.
This should be taken into account carefully when dealing with polarization
measurements at intermediate S/N.  For any measurement with a S/N
$p_0/\sigma_{{\rm p},0}\,{<}\,2$, 
it is in fact impossible to obtain an estimate of the true S/N, 
because this is fully degenerate due to the bias of the polarization fraction.

\subsection{Confidence intervals}
\label{sec:confidence_intervals}

We have seen the limitations of the Gaussian assumption for computing
valid estimates of the polarization uncertainties.
To obtain a robust estimate of the uncertainty in $p$ and $\psi$
at low S/N, one has to construct the correct confidence regions or intervals. 
The $ \lambda$ \% confidence interval around a measurement $p$ 
is defined as the interval which has a probability of containing
the true value $p_0$ exactly equal to $\lambda/100$, where $(1-\lambda)$
is called ``critical parameter.''
This interval is constructed from the probability density function and does
not require any estimate of the true polarization parameters.
\citet{Mood1974}, \citet{Simmons1985} and \citet{Vaillancourt2006} 
provided a simple way to construct such 
confidence intervals for the polarization fraction $p$ when the
usual simplifications of the covariance matrix are assumed.
\citet{Naghizadeh1993} have provided estimates 
of the confidence intervals for the polarization angle $\psi$ under
similar assumptions, and this is even simpler,
because in that case $f_{\psi}(\psi\,|\,p_0,\psi_0,\tens{\Sigma}_{\rm p})$ 
only depends on the signal-to-noise ratio $p_0/\sigma_{{\rm p},0}$.

\begin{figure}
\center
\includegraphics[width=8cm]{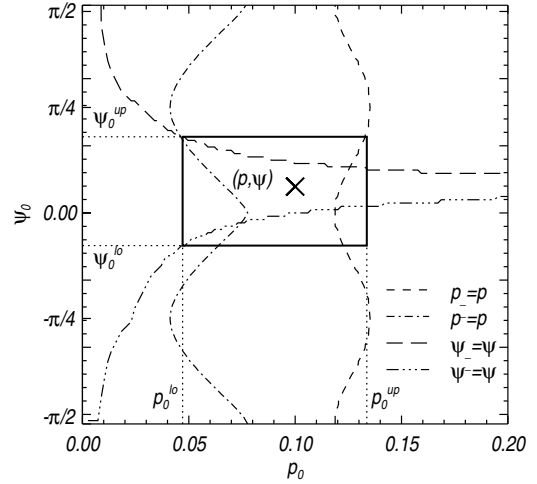} 
\caption{Construction of 68\% confidence intervals
$[p_0^{\rm low},p_0^{\rm up}]$ and $[\psi_0^{\rm low},\psi_0^{\rm up}]$
(full line box) of $p_0$ and $\psi_0$, based on the upper and lower
loci $p\,{=}\,p_{-}$, $p\,{=}\,p^{-}$, $\psi\,{=}\,\psi_-$,
and $\psi\,{=}\,\psi^-$, built from probability 
density functions $f_{\rm 2D}$ and a given measurement ($p,\psi$) (indicated
by the cross).}
\label{fig:confidence_intervals_summary}
\end{figure}

Once the covariance matrix is allowed to include ellipticity and correlations,
we have seen in Sect.~\ref{sec:marginal_p_psi_pdf} and
Appendix~\ref{sec:f_1d_full} how
the marginalized probability density functions
$f_{\rm p}(p\,|\,p_0,\psi_0,\tens{\Sigma}_{\rm p})$
and $f_{\psi}(\psi\,|\,p_0,\psi_0,\tens{\Sigma}_{\rm p})$ 
depend on the true polarization fraction $p_0$ and the true polarization angle
$\psi_0$.  This leads is to consider $\psi_0$ as a ``nuisance parameter''
when building confidence intervals of $p_0$, and vice-versa.
We propose below an extension of the \citet{Simmons1985} technique,
using an iterative method to build the confidence intervals of $p_0$ and
$\psi_0$ simultaneously. 

For each possible value of $p_0$ and $\psi_0$ (spanning the range 0 to 1, 
and $-\pi/2$ to $\pi/2$, respectively),
we compute the quantities $p_{-}$, $p^{-}$, 
$\psi_-$ and $\psi^-$, which provide the lower and upper limits in
$p$ and $\psi$ of the region $\Omega (\lambda, p_0,\psi_0)$ defined by
 \begin{equation}
 \label{eq:omega}
\iint_{\Omega(\lambda,p_0,\psi_0)}
 f_{\rm 2D}(p,\psi\,|\,p_0,\psi_0,\tens{\Sigma}_{\rm p}) \, dpd\psi
 = \frac{\lambda}{100}
\end{equation}
and such that the contour of the region $\Omega$ is an iso-probability
contour of the probability density function $f_{\rm 2D}$.
We stress that the choice of a confidence interval is still subjective
and may be shifted by any arbitrary value of $p$ or $\psi$, provided that
the integral over the newly defined region is also $\lambda/100$.
The definition we have chosen ensures that the region
$\Omega(\lambda, p_0,\psi_0)$ is the smallest possible.
We also note that 
\begin{equation}
\int_{p_-}^{p^-} \int_{\psi_-}^{\psi^-} f_{\rm 2D} \, dpd\psi
 > \iint_{\Omega(\lambda, p_0,\psi_0)} f_{\rm 2D} \, dpd\psi, 
\end{equation}
which implies that the rectangular region bounded by $p_{-}$, $p^{-}$,
$\psi_-$ and $\psi^-$ is a conservative choice.  For a given $\lambda$
and covariance matrix $\tens{\Sigma}_{\rm p}$, 
we can finally obtain a set of four upper and lower limits on $p$ and $\psi$:
$p_{-}(p_0,\psi_0)$; $p^{-}(p_0,\psi_0)$; $\psi_-(p_0,\psi_0)$;
and $\psi^-(p_0,\psi_0)$.  We illustrate this with the example
of ($p$, $\psi$) set to (0.1, $\pi$/8) in
Fig.~\ref{fig:confidence_intervals_summary} (this point shown by the cross).
For given polarization measurements ($p$, $\psi$),
we trace the loci $p_-(p_0,\psi_0)\,{=}\,p$ (dashed line), 
$p^-(p_0,\psi_0)\,{=}\,p$ (dot-dash line), 
$\psi_-(p_0,\psi_0)\,{=}\,\psi$ (long dashed line), 
and $\psi^-(p_0,\psi_0)\,{=}\,\psi$ (dash-dot-dot-dot line). 
Finally, the 68\% confidence intervals $[p_0^{\rm low},p_0^{\rm up}]$
of $p_0$ and [$\psi_0^{\rm low},\psi_0^{\rm up}]$ of $\psi_0$ are defined by
building the smallest rectangular region
(solid line in Fig.~\ref{fig:confidence_intervals_summary}) that
simultaneously covers the domain in $p_0$ and $\psi_0$ between the upper and
lower limits defined above, and which satisfies the conditions:
\begin{eqnarray}
p_0^{\rm low} &=& \mathrm{min}_{p_0}
 \left(\: p=p^{-} \left\{ p_0\: , \: \psi_0 \in
 [\psi_0^{\rm low},\psi_0^{\rm up}] \right\} \: \right ); \nonumber \\
p_0^{\rm up} &=& \mathrm{max}_{p_0} \left( \: p = p_{-}
 \left\{ p_0\: , \: \psi_0 \in [\psi_0^{\rm low} ,\psi_0^{\rm up}]
 \right\} \: \right ); \nonumber \\
\psi_0^{\rm low}&=& \mathrm{min}_{\psi_0}
 \left( \: \psi = \psi^{-} \left\{ p_0 \in [p_0^{\rm low},
 p_0^{\rm up} ] \: , \: \psi_0\right\} \: \right ); \nonumber \\
\psi_0^{\rm up}&=& \mathrm{max}_{\psi_0}
 \left( \: \psi = \psi_{-} \left\{ p_0 \in
 [p_0^{\rm low}, p_0^{\rm up} ] \: , \: \psi_0\right\} \: \right ).
\end{eqnarray}
Using these conditions, the confidence interval of $p_0$ takes into account
the nuisance parameter $\psi_0$ over its own confidence interval,
and vice-versa.  This has to be constructed iteratively, starting with 
$\psi_0^{\rm low}\,{=}\,-\pi/2$ and $\psi_0^{\rm up}\,{=}\,\pi/2$, to build
first guesses for $p_0^{\rm low}$ and $p_0^{\rm up}$, which
are then used to build a new estimate of the confidence intervals of $\psi_0$,
and so on until convergence.  In practice, it converges very quickly.
We emphasize that these confidence intervals are conservative, because they
include the impact of the nuisance parameters, implying that
\begin{equation}
\mathrm{Pr}\, \Big ( p_0^{\rm low} \le p_0 \le p_0^{\rm up}\, ;
 \, \psi_0^{\rm low} \le \psi_0 \le \psi_0^{\rm up} \, \Big )
 \big |_{p,\psi,\tens{\Sigma}_{\rm p}} \ge \frac{\lambda}{100}
\end{equation}
whatever the true values $p_0$, $\psi_0$.

\section{Conclusions}
\label{sec:conclusion}

This paper represents the first step in an extensive study of polarization
analysis methods.  We focused here on the impact of the full covariance matrix
on naive polarization measurements, and especially the impact on the bias. 
We have derived analytical expressions for the probability density function
of the polarization parameters ($I, p, \psi$)
in the 3D and 2D cases, taking into account the full covariance matrix
$\tens{\Sigma}$ of the Stokes parameters $I$, $Q$, and $U$. 

The asymmetries of the covariance matrix can be characterized by the effective
ellipticity $\varepsilon_{\rm eff}$, expressed as a function of the ellipticity
$\varepsilon$ and the correlation $\rho$ between $Q$ and $U$ in a given
reference frame, and by the correlation parameters $\rho_{\rm Q}$ and
$\rho_{\rm U}$ between the intensity $I$ and the $Q$ and $U$ parameters.
We have quantified departures from the canonical case
($\varepsilon_{\rm eff}\,{=}\,1$), usually assumed in earlier works on
polarization.  We explored this effect for three regimes of the covariance
matrix: the {\it canonical\/} case, $\varepsilon_{\rm eff}\,{=}\,1$);
the {\it low\/} regime, $1\,{<}\,\varepsilon_{\rm eff}\,{<}\,1.1$;
and the {\it extreme\/} regime $1\,{<}\,\varepsilon_{\rm eff}\,{<}\,2$. 
We first emphasized the impact of the true polarization angle $\psi_0$,
which can produce variations in the polarization fraction bias of up to
30\% of the dispersion of $p$, in the {\it extreme\/} regime, and
up to 5\% in the {\it low\/} regime.  We then estimated
the statistical bias on the polarization angle measurement $\psi$.
This can reach up to $9^\circ$ when the ellipticity or the correlation
between the $Q$ and $U$ Stokes components becomes important
($\varepsilon_{\rm eff}\,{\sim}\,2$), and the S/N is low. 
However, when values of the effective ellipticity are in the {\it low\/}
regime (i.e., less than 10\% greater than the
canonical values) the bias on $\psi$ remains limited (i.e., $<1^\circ$),
and well below (by a factor of 5--25) the level of the measurement uncertainty.
Thus the bias on $\psi$ can be neglected, to first order, for small departures
of the covariance matrix from the canonical case. 

On the other hand, we have quantified the impact of the uncertainty of the
intensity on the relative and absolute statistical bias of the polarization
fraction and angle.
We provided the modified probability density function in ($p,\psi$) arising
from a finite signal-to-noise ratio of the intensity, $I_0/\sigma_{\rm I}$. 
We have shown that, above an intensity S/N of 5, the relative bias on 
the polarization fraction $p$ remains globally unchanged at polarization S/N
$p_0/\sigma_{\rm p,G}\,{<}\,2$, while
it is slightly enhanced when the intensity and the polarization S/N lie in
the intermediate range, $p_0/\sigma_{\rm p,G}\,{>}\,2$.
For S/N on the intensity $I_0/\sigma_{\rm I}$ below 5, the relative bias on
$p$ drops suddenly to 0, because of the
increasing dispersion.  Indeed, the absolute bias
can be higher by a factor as large as 5 when the S/N on $I$ drops below 2 to 3;
this is associated with a dramatical increase in the dispersion of the
polarization fraction, which diverges and strongly overwhelms 
the increase of the bias at low S/N.
Hence the uncertainty of the intensity has to be properly taken into account
when analysing polarization data for faint objects, in order to 
derive the correct polarization fraction bias and uncertainty.
Similarly, the case of faint polarized objects on top of a varying but
unpolarized background can lead to a question about the correct intensity
offset to subtract, yielding an effective additional uncertainty on the
intensity.

The impact of correlations between the intensity and the $Q$ and $U$
components has also been quantified in the case of a finite S/N on the
intensity.  It has been shown that the bias on $p$ is only slightly affected
(below 10\% difference compared with the canonical case) even at low S/N on
$I$, when the correlations $\rho_{\rm Q}$ and $\rho_{\rm U}$ span the range
$-0.2$ to $0.2$.

We have additionally addressed the question of how to obtain a robust estimate
of the uncertainties on polarization measurements ($p,\psi$).
We extended the often used procedure of \citet{Simmons1985} by building
confidence intervals for polarization fraction and angle simultaneously,
taking into account the full properties of the covariance matrix.
This method makes it possible to build conservative 
confidence intervals around polarization measurements.

We have explored the domain of validity for the commonly used polarization
uncertainty estimators based on the variance of the 
probability density function (assuming a Gaussian distribution).
The true dispersion of the polarization fraction has been shown to provide
robust estimates only at high S/N (above 3), while the true dispersion of
the polarization angle yields conservative estimates for S/N$\,{>}\,0.5$.
Simple estimators, such as the geometric and arithmetic polarization fraction
uncertainties, appear sensitive to the effective ellipticity of the covariance
matrix at high S/N, while they provide conservative estimates over a wide
range of S/N (above 0.5) in the canonical case.  The classical method, usually
adopted to analyse optical extinction polarization data, provides
the most robust estimates of $\sigma_{\rm p}$ for S/N above 0.5,
with respect to the ellipticity of the covariance matrix, 
but poor estimates of $\sigma_{\psi}$, which are valid only at very high S/N
(above 5). 

We have seen how much the naive polarization estimates provide poor
determinations of the true polarization parameters, and how it can be difficult
to recover the true S/N of a measurement. 
In a companion paper (Montier et al.\ in preparation), we review
different estimators of the true polarization from experimental measurements
that partially correct this bias in $p$ and $\psi$, using full knowledge of
the polarization covariance matrix.

\begin{acknowledgements}
This paper was developed to support the analysis of data from the
\Planck\ satellite.  The development of \Planck\ has been supported by: ESA; CNES and
CNRS/INSU-IN2P3-INP (France); ASI, CNR, and INAF (Italy); NASA and DoE
(USA); STFC and UKSA (UK); CSIC, MICINN, JA, and RES (Spain); Tekes,
AoF, and CSC (Finland); DLR and MPG (Germany); CSA (Canada); DTU Space
(Denmark); SER/SSO (Switzerland); RCN (Norway); SFI (Ireland);
FCT/MCTES (Portugal); and PRACE (EU). A description of the Planck
Collaboration and a list of its members, including the technical or
scientific activities in which they have been involved, can be found
at \url{http://www.sciops.esa.int/index.php?project=planck&page=}\\ \url{Planck_Collaboration}.
We acknowledge the use of the Legacy Archive for Microwave Background
Data Analysis (LAMBDA), part of the High Energy Astrophysics Science
Archive Center (HEASARC). HEASARC/LAMBDA is a service of the
Astrophysics Science Division at the NASA Goddard Space Flight Center.
Some of the results in this paper have been derived using the
{\healpix} package.   
We would also like to thank P. Leahy, S. Prunet and D. Scott for their very useful comments.
\end{acknowledgements}

\bibliographystyle{aa}
\bibliography{Planck_bib,biblio_v1.7}

\begin{thebibliography}{31}
\expandafter\ifx\csname natexlab\endcsname\relax\def\natexlab#1{#1}\fi

\bibitem[{{Aalo} {et~al.}(2007){Aalo}, {Efthymoglou}, \& {Chayawan}}]{Aalo2007}
{Aalo}, V.~A., {Efthymoglou}, G.~P., \& {Chayawan}, C. 2007, IEEE
  Communications letlers, 11, 985

\bibitem[{{Abramowitz} \& {Stegun}(1964)}]{AbramowitzStegun1964}
{Abramowitz}, M. \& {Stegun}, I. 1964, {Handbook of Mathematical Functions}

\bibitem[{{Beno{\^i}t} {et~al.}(2004){Beno{\^i}t}, {Ade}, {Amblard}, {Ansari},
  {Aubourg}, {Bargot}, {Bartlett}, {Bernard}, {Bhatia}, {Blanchard}, {Bock},
  {Boscaleri}, {Bouchet}, {Bourrachot}, {Camus}, {Couchot}, {de Bernardis},
  {Delabrouille}, {D{\'e}sert}, {Dor{\'e}}, {Douspis}, {Dumoulin}, {Dupac},
  {Filliatre}, {Fosalba}, {Ganga}, {Gannaway}, {Gautier}, {Giard},
  {Giraud-H{\'e}raud}, {Gispert}, {Guglielmi}, {Hamilton}, {Hanany},
  {Henrot-Versill{\'e}}, {Kaplan}, {Lagache}, {Lamarre}, {Lange},
  {Mac{\'{\i}}as-P{\'e}rez}, {Madet}, {Maffei}, {Magneville}, {Marrone},
  {Masi}, {Mayet}, {Murphy}, {Naraghi}, {Nati}, {Patanchon}, {Perrin}, {Piat},
  {Ponthieu}, {Prunet}, {Puget}, {Renault}, {Rosset}, {Santos}, {Starobinsky},
  {Strukov}, {Sudiwala}, {Teyssier}, {Tristram}, {Tucker}, {Vanel}, {Vibert},
  {Wakui}, \& {Yvon}}]{Benoit2004}
{Beno{\^i}t}, A., {Ade}, P., {Amblard}, A., {et~al.} 2004, \aap, 424, 571

\bibitem[{{Bernard} {et~al.}(2007){Bernard}, {Ade}, {De Bernardis}, {Giard},
  {Griffin}, {Hargrave}, {Laurens}, {Leriche}, {Leroy}, {Longval}, {Marty},
  {Madden}, {Maffei}, {Masi}, {Meny}, {Miville-Desch{\^e}nes}, {Narbonne},
  {Nati}, {Pajot}, {Pisano}, {Pointecouteau}, {Ponthieu}, {Ristorcelli},
  {Rodriguez}, {Roudil}, {Salatino}, \& {Savini}}]{Bernard2007}
{Bernard}, J.-P., {Ade}, P., {De Bernardis}, P., {et~al.} 2007, in EAS
  Publications Series, Vol.~23, EAS Publications Series, ed. M.-A.
  {Miville-Desch{\^e}nes} \& F.~{Boulanger}, 189--203

\bibitem[{{Boissier} {et~al.}(2009){Boissier}, {Bockel{\'e}e-Morvan}, {Biver},
  {Crovisier}, {Moreno}, {Lellouch}, \& {Neri}}]{Boissier2009}
{Boissier}, J., {Bockel{\'e}e-Morvan}, D., {Biver}, N., {et~al.} 2009, Earth
  Moon and Planets, 105, 89

\bibitem[{{Clarke} {et~al.}(1993){Clarke}, {Naghizadeh-Khouei}, {Simmons}, \&
  {Stewart}}]{Clarke1993}
{Clarke}, D., {Naghizadeh-Khouei}, J., {Simmons}, J.~F.~L., \& {Stewart}, B.~G.
  1993, \aap, 269, 617

\bibitem[{{Dotson} {et~al.}(2010){Dotson}, {Vaillancourt}, {Kirby}, {Dowell},
  {Hildebrand}, \& {Davidson}}]{Dotson2010}
{Dotson}, J.~L., {Vaillancourt}, J.~E., {Kirby}, L., {et~al.} 2010, \apjs, 186,
  406

\bibitem[{{Dowell} {et~al.}(1998){Dowell}, {Hildebrand}, {Schleuning},
  {Vaillancourt}, {Dotson}, {Novak}, {Renbarger}, \& {Houde}}]{Dowell1998}
{Dowell}, C.~D., {Hildebrand}, R.~H., {Schleuning}, D.~A., {et~al.} 1998, \apj,
  504, 588

\bibitem[{{Fissel} {et~al.}(2010){Fissel}, {Ade}, {Angil{\`e}}, {Benton},
  {Chapin}, {Devlin}, {Gandilo}, {Gundersen}, {Hargrave}, {Hughes}, {Klein},
  {Korotkov}, {Marsden}, {Matthews}, {Moncelsi}, {Mroczkowski}, {Netterfield},
  {Novak}, {Olmi}, {Pascale}, {Savini}, {Scott}, {Shariff}, {Soler}, {Thomas},
  {Truch}, {Tucker}, {Tucker}, {Ward-Thompson}, \& {Wiebe}}]{Fissel2010}
{Fissel}, L.~M., {Ade}, P.~A.~R., {Angil{\`e}}, F.~E., {et~al.} 2010, in
  Society of Photo-Optical Instrumentation Engineers (SPIE) Conference Series,
  Vol. 7741, Society of Photo-Optical Instrumentation Engineers (SPIE)
  Conference Series

\bibitem[{{Girart} {et~al.}(2006){Girart}, {Rao}, \& {Marrone}}]{Girart2006}
{Girart}, J.~M., {Rao}, R., \& {Marrone}, D.~P. 2006, Science, 313, 812

\bibitem[{{Gradshteyn} \& {Ryzhik}(2007)}]{GradshteynRyzhik2007}
{Gradshteyn}, I.~S. \& {Ryzhik}, I.~M. 2007, {Table of Integrals, Series, and
  Products}

\bibitem[{{Greaves} {et~al.}(2003){Greaves}, {Holland}, {Jenness},
  {Chrysostomou}, {Berry}, {Murray}, {Tamura}, {Robson}, {Ade}, {Nartallo},
  {Stevens}, {Momose}, {Morino}, {Moriarty-Schieven}, {Gannaway}, \&
  {Haynes}}]{Greaves2003}
{Greaves}, J.~S., {Holland}, W.~S., {Jenness}, T., {et~al.} 2003, \mnras, 340,
  353

\bibitem[{{Li} {et~al.}(2008){Li}, {Dowell}, {Kirby}, {Novak}, \&
  {Vaillancourt}}]{Li2008}
{Li}, H., {Dowell}, C.~D., {Kirby}, L., {Novak}, G., \& {Vaillancourt}, J.~E.
  2008, \ao, 47, 422

\bibitem[{{Mood} \& {Graybill}(1974)}]{Mood1974}
{Mood}, A.~M. \& {Graybill}, A.~F. 1974, {Introduction to the Sheory of
  Statistics, 3rd ed. McGraw-Hill, New-York}

\bibitem[{{Naghizadeh-Khouei} \& {Clarke}(1993)}]{Naghizadeh1993}
{Naghizadeh-Khouei}, J. \& {Clarke}, D. 1993, \aap, 274, 968

\bibitem[{{Page} {et~al.}(2007){Page}, {Hinshaw}, {Komatsu}, {Nolta},
  {Spergel}, {Bennett}, {Barnes}, {Bean}, {Dor{\'e}}, {Dunkley}, {Halpern},
  {Hill}, {Jarosik}, {Kogut}, {Limon}, {Meyer}, {Odegard}, {Peiris}, {Tucker},
  {Verde}, {Weiland}, {Wollack}, \& {Wright}}]{page2007}
{Page}, L., {Hinshaw}, G., {Komatsu}, E., {et~al.} 2007, \apjs, 170, 335

\bibitem[{{P{\'e}rez-S{\'a}nchez} \& {Vlemmings}(2013)}]{Perez2013}
{P{\'e}rez-S{\'a}nchez}, A.~F. \& {Vlemmings}, W.~H.~T. 2013, \aap, 551, A15

\bibitem[{{Platt} {et~al.}(1991){Platt}, {Hildebrand}, {Pernic}, {Davidson}, \&
  {Novak}}]{Platt1991}
{Platt}, S.~R., {Hildebrand}, R.~H., {Pernic}, R.~J., {Davidson}, J.~A., \&
  {Novak}, G. 1991, \pasp, 103, 1193

\bibitem[{{Quinn}(2012)}]{Quinn2012}
{Quinn}, J.~L. 2012, \aap, 538, A65

\bibitem[{{Renbarger} {et~al.}(2004){Renbarger}, {Chuss}, {Dotson}, {Griffin},
  {Hanna}, {Loewenstein}, {Malhotra}, {Marshall}, {Novak}, \&
  {Pernic}}]{Renbarger2004}
{Renbarger}, T., {Chuss}, D.~T., {Dotson}, J.~L., {et~al.} 2004, \pasp, 116,
  415

\bibitem[{{Rice}(1945)}]{Rice1945}
{Rice}, S.~O. 1945, Bell Systems Tech.~J., Volume 24, p.~46-156, 24, 46

\bibitem[{{Schleuning} {et~al.}(1997){Schleuning}, {Dowell}, {Hildebrand},
  {Platt}, \& {Novak}}]{Schleuning1997}
{Schleuning}, D.~A., {Dowell}, C.~D., {Hildebrand}, R.~H., {Platt}, S.~R., \&
  {Novak}, G. 1997, \pasp, 109, 307

\bibitem[{{Serkowski}(1958)}]{Serkowski1958}
{Serkowski}, K. 1958, \actaa, 8, 135

\bibitem[{{Serkowski}(1962)}]{Serkowski1962}
{Serkowski}, K. 1962, Advances in Astronomy and Astrophysics, 0, 290

\bibitem[{{Simmons} \& {Stewart}(1985)}]{Simmons1985}
{Simmons}, J.~F.~L. \& {Stewart}, B.~G. 1985, \aap, 142, 100

\bibitem[{{Tauber} {et~al.}(2010){Tauber}, {Mandolesi}, {Puget}, {Banos},
  {Bersanelli}, {Bouchet}, {Butler}, {Charra}, {Crone}, {Dodsworth}, \&
  et~al.}]{Tauber2010}
{Tauber}, J.~A., {Mandolesi}, N., {Puget}, J., {et~al.} 2010, \aap, 520, A1

\bibitem[{{Thum} {et~al.}(2008){Thum}, {Wiesemeyer}, {Paubert}, {Navarro}, \&
  {Morris}}]{Thum2008}
{Thum}, C., {Wiesemeyer}, H., {Paubert}, G., {Navarro}, S., \& {Morris}, D.
  2008, \pasp, 120, 777

\bibitem[{{Vaillancourt}(2006)}]{Vaillancourt2006}
{Vaillancourt}, J.~E. 2006, \pasp, 118, 1340

\bibitem[{{Vaillancourt} \& {Matthews}(2012)}]{Vaillancourt2012}
{Vaillancourt}, J.~E. \& {Matthews}, B.~C. 2012, \apjs, 201, 13

\bibitem[{{Vinokur}(1965)}]{Vinokur1965}
{Vinokur}, M. 1965, Annales d'Astrophysique, 28, 412

\bibitem[{{Wardle} \& {Kronberg}(1974)}]{Wardle1974}
{Wardle}, J.~F.~C. \& {Kronberg}, P.~P. 1974, \apj, 194, 249

\end{thebibliography}

\onecolumn
\appendix

\section{Expressions for PDFs}
\label{sec:PDFexpressions}
Here we present expressions for the 2D probability density functions,
which are discussed in Section~\ref{sec:pdf}:

\begin{align}
f(I,p,\psi\,|\,I_0,p_0,\psi_0,\tens{\Sigma})
 &= \frac{2|p|\,I^2} {\sqrt{(2\pi)^3} \sigma^3} \, 
\exp \left \lgroup - \frac{1}{2} 
\left[ \begin{array}{c} 
I -I_0 \\
p \, I\, \cos(2\psi)-p_0\,I_0\cos(2\psi_0) \\
p \, I\, \sin(2\psi)-p_0\,I_0\sin(2\psi_0) \\\end{array}
\right] ^{\rm T}
\tens{\Sigma}^{-1}
\left[ \begin{array}{c} 
I-I_0 \\
p\,I\,\cos(2\psi)-p_0\,I_0\,\cos(2\psi_0)\\
p\,I\,\sin(2\psi)-p_0\,I_0\,\sin(2\psi_0)\\\end{array}
\right]
\right \rgroup ;
\label{eq:f_ipphi}\\
\nonumber\\
f_{\rm 2D}(p,\psi\,|\,I_0,p_0,\psi_0,\tens{\Sigma}) 
 &= \frac{|p|}{2\pi\sigma^3}\exp{\left(-\frac{I_0^2}{2}\gamma\right)}
 \left\{\sqrt{\frac{2}{\pi}}\frac{\beta I_0}{\alpha^2}
 +\frac{1}{\alpha^{3/2}}\left[1+\frac{\beta^2I_0^2}{\alpha}\right]
 \exp{\left(\frac{\beta^2I_0^2}{2\alpha}\right)\left[1+\mathrm{erf}
 \left(\frac{\beta I_0}{\sqrt{2\alpha}}\right)\right]}\right\}
 \quad\ \ \ \ \textrm{for} \quad p\geqslant 0 ;
\label{eq:f_2d_ppos}\\
\nonumber\\
f_{\rm 2D}(p,\psi\,|\,I_0,p_0,\psi_0,\tens{\Sigma})
 &= \frac{|p|}{2\pi\sigma^3}\exp{\left(-\frac{I_0^2}{2}\gamma\right)}
 \left\{-\sqrt{\frac{2}{\pi}}\frac{\beta I_0}{\alpha^2}
 +\frac{1}{\alpha^{3/2}}\left[1+\frac{\beta^2I_0^2}{\alpha}\right]
 \exp{\left(\frac{\beta^2I_0^2}{2\alpha}\right)\left[1-\mathrm{erf}
 \left(\frac{\beta I_0}{\sqrt{2\alpha}}\right)\right]}\right\}
 \quad\ \textrm{for} \quad p\leqslant 0 ;
\label{eq:f_2d_pneg}\\
\nonumber\\
f_{\rm 2D}(p,\psi\,|\,p_0,\psi_0, \tens{\Sigma}_{\rm p})
 &= \frac{p} {\pi \sigma_{\rm p,G}^2} \, 
 \exp \left \lgroup - \frac{1}{2} 
 \left[ \begin{array}{c} 
 p \, \cos(2\psi)-p_0 \cos(2\psi_0) \\
 p \, \sin(2\psi)-p_0 \sin(2\psi_0) \\\end{array}
 \right]^{\rm T}
 \tens{\Sigma}_{\rm p}^{-1}
 \left[ \begin{array}{c} 
 p\,\cos(2\psi)-p_0\,\cos(2\psi_0)\\
 p\,\sin(2\psi)-p_0\,\sin(2\psi_0)\\\end{array}
 \right]
 \right \rgroup \quad \textrm{for} \quad \sigma_{\rm I} = 0 .
 \label{eq:f_2d_polar}
\end{align}

\section{Computation of $f_{\rm 2D}$}
\label{appendix:feta}
The 3D PDF of $(I,p,\psi)$ is given by 
\begin{equation}
f(I,p,\psi)=2\,|p|\,I^2\,F\left(I,pI\cos{2\psi},pI\sin{2\psi}\right).
\end{equation}
To compute the 2D PDF of $(p,\psi)$ we marginalize over total intensity.
However, some care is required here, because the above expression for
$f(I,p,\psi)$ is only valid for $pI\geqslant 0$ (i.e., we cannot measure
negative $p$ unless $I$ happens to be negative due to noise) and $f$
must be taken to be null otherwise.  This means that the marginalization
is performed over $I\geqslant 0$ for positive $p$ and over $I\leqslant 0$
for negative $p$:
\begin{align}
f_{\rm 2D}&=\int_0^{+\infty}2\,|p|\,I^2\,
 F\left(I,pI\cos{2\psi},pI\sin{2\psi}\right)dI,
 \quad \textrm{for} \quad p\geqslant 0; \\
f_{\rm 2D}&=\int_{-\infty}^02\,|p|\,I^2\,
 F\left(I,pI\cos{2\psi},pI\sin{2\psi}\right)dI,
 \quad\ \ \textrm{for} \quad p\leqslant 0.
\end{align}
The integrand may be written so as to exhibit the dependence on total
intensity,
\begin{equation}
f=\frac{2\,|p|\,I^2}{(2\pi)^{3/2}\sigma^3}
 \exp{\left[-\frac{1}{2}\left(I^2\alpha-2II_0\beta+I_0^2\gamma\right)\right]}
\end{equation}
and then we make use of the functions \citep{GradshteynRyzhik2007}:
\begin{align}
G_-(x,y) &= \int_{-\infty}^0I^2e^{-x I^2+2y I}dI\ \
 = -\frac{y}{2x^2}+\sqrt{\frac{\pi}{x^5}}
 \frac{2y^2+x}{4}\exp{\left(\frac{y^2}{x}\right)}
 \left[1-\mathrm{erf}\left(\frac{y}{\sqrt{x}}\right)\right]; \\
G_+(x,y) &= \int_0^{+\infty}I^2e^{-x I^2+2y I}dI
 = \frac{y}{2x^2}+\sqrt{\frac{\pi}{x^5}}
 \phantom{0}\frac{2y^2+x}{4}\exp{\left(\frac{y^2}{x}\right)}
 \left[1+\mathrm{erf}\left(\frac{y}{\sqrt{x}}\right)\right].
\end{align}
Elementary replacement of $(x,y)$ by $(\alpha/2,I_0\beta/2)$ yields the
PDF of Eqs.~\ref{eq:f_2d_ppos} and \ref{eq:f_2d_pneg}, given in the main
body of the text.

\section{Illustrations of $f_{\rm 2D}$}
\label{sec:illustration_pdf}
 
\begin{figure*}[tp]
\begin{tabular}{cccccc}
& \begin{minipage}[c]{.15\linewidth}
$\quad$
 \begin{tabular}{l}
$ \varepsilon = 1$ \\
$ \rho = 0$ \\
\end{tabular} \\
$\Bigg($
 \begin{tabular}{l}
$ \varepsilon_{\rm eff} = 1$ \\
$ \theta = 0$ \\
\end{tabular} 
$\Bigg)$
\end{minipage} & 
 \begin{minipage}[c]{.15\linewidth}
$\quad$
 \begin{tabular}{l}
$ \varepsilon = 1/2$ \\
$ \rho = 0$ \\
\end{tabular} \\ 
$\Bigg($
 \begin{tabular}{l}
$ \varepsilon_{\rm eff} = 2$ \\
$ \theta = \pi$ \\
\end{tabular} 
$\Bigg)$
\end{minipage} & 
\begin{minipage}[c]{.15\linewidth}
$\quad$
 \begin{tabular}{l}
$ \varepsilon = 2$ \\
$ \rho = 0$ 
\end{tabular} \\ 
$\Bigg($
 \begin{tabular}{l}
$ \varepsilon_{\rm eff} = 2$ \\
$ \theta = 0$ \\
\end{tabular} 
$\Bigg)$
\end{minipage} &
\begin{minipage}[c]{.15\linewidth}
$\quad$
 \begin{tabular}{l}
$ \varepsilon = 1$ \\
$ \rho = -1/2$ 
\end{tabular} \\ 
$\Bigg($
 \begin{tabular}{l}
$ \varepsilon_{\rm eff} \sim 1.73$ \\
$ \theta = -\pi/4$ \\
\end{tabular} 
$\Bigg)$
\end{minipage} &
\begin{minipage}[c]{.15\linewidth}
$\quad$
 \begin{tabular}{l}
$ \varepsilon = 1$ \\
$ \rho = 1/2$
\end{tabular} \\ 
$\Bigg($
 \begin{tabular}{l}
$ \varepsilon_{\rm eff} \sim 1.73$ \\
$ \theta = \pi/4$ \\
\end{tabular} 
$\Bigg)$
\end{minipage} 
\\ \\
\begin{minipage}[c]{.13\linewidth}
 \begin{tabular}{l}
 $p_0/\sigma_{\rm p,G}=0.01$ 
 \end{tabular} 
\end{minipage} &
\begin{minipage}[c]{.15\linewidth}
 \includegraphics[width=3cm, viewport=200 0 600 400]
 {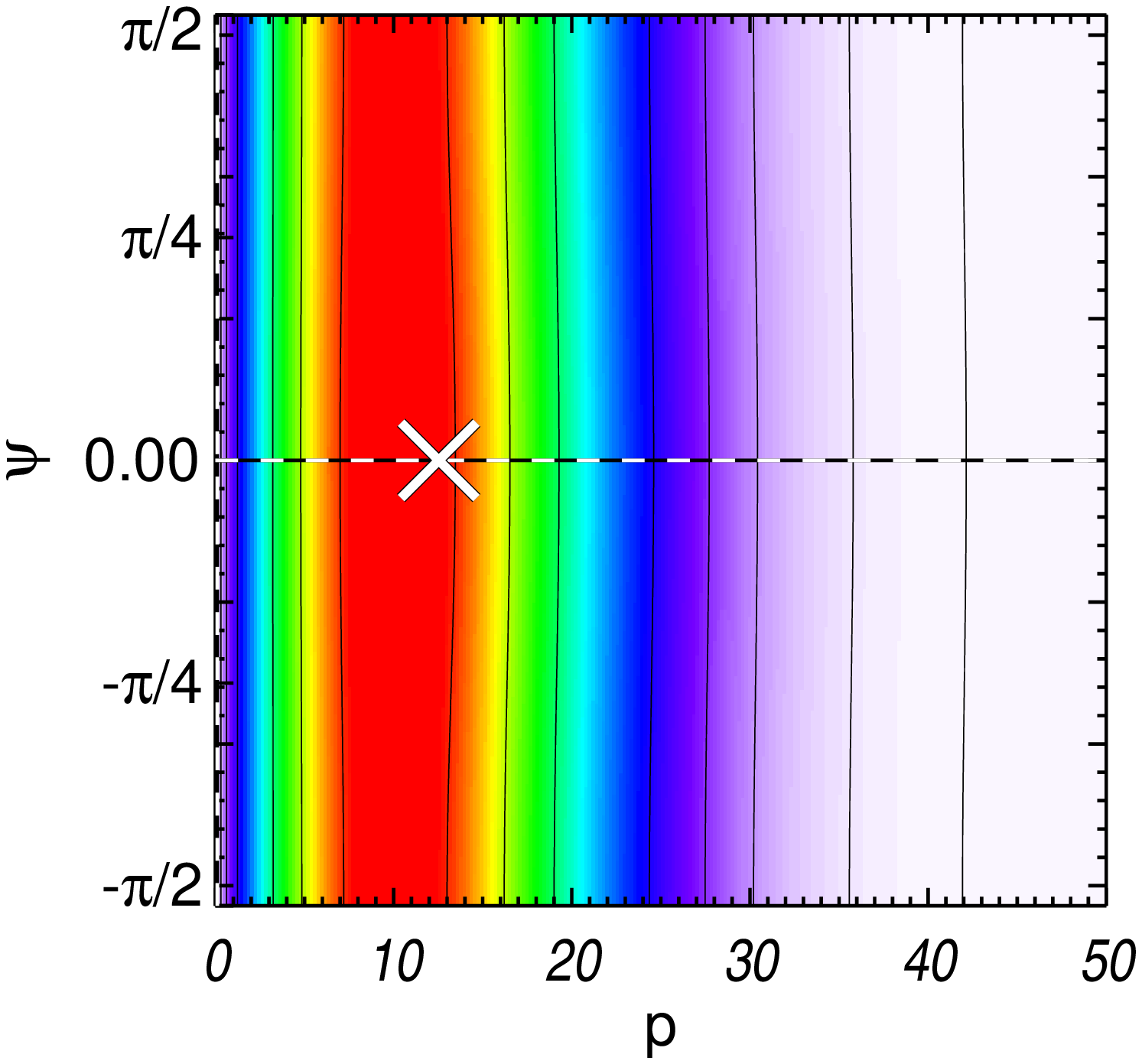}
\end{minipage} & 
\begin{minipage}[c]{.15\linewidth}
 \includegraphics[width=3cm, viewport=200 0 600 400]
 {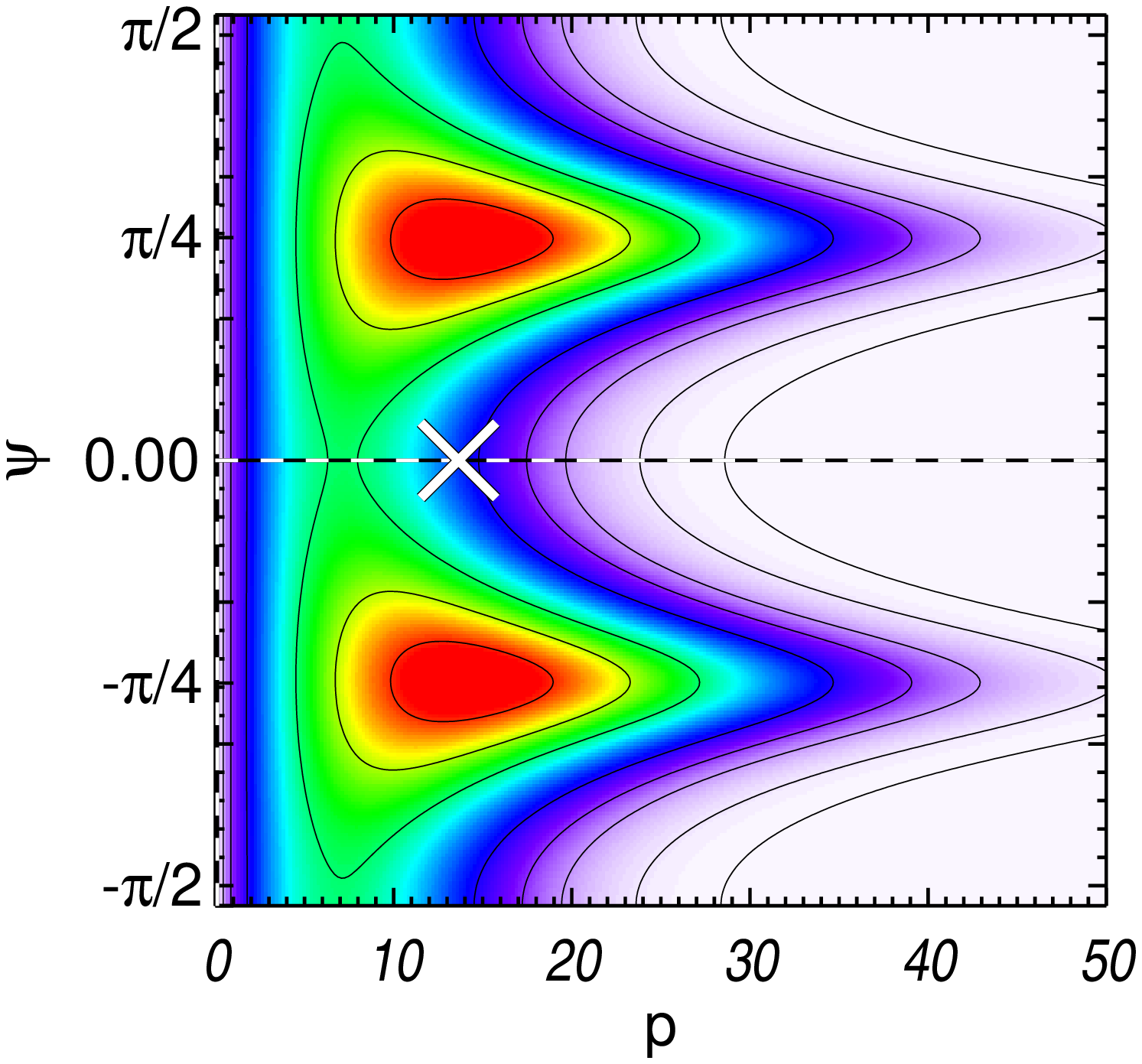}
\end{minipage} &
\begin{minipage}[c]{.15\linewidth}
 \includegraphics[width=3cm, viewport=200 0 600 400]
 {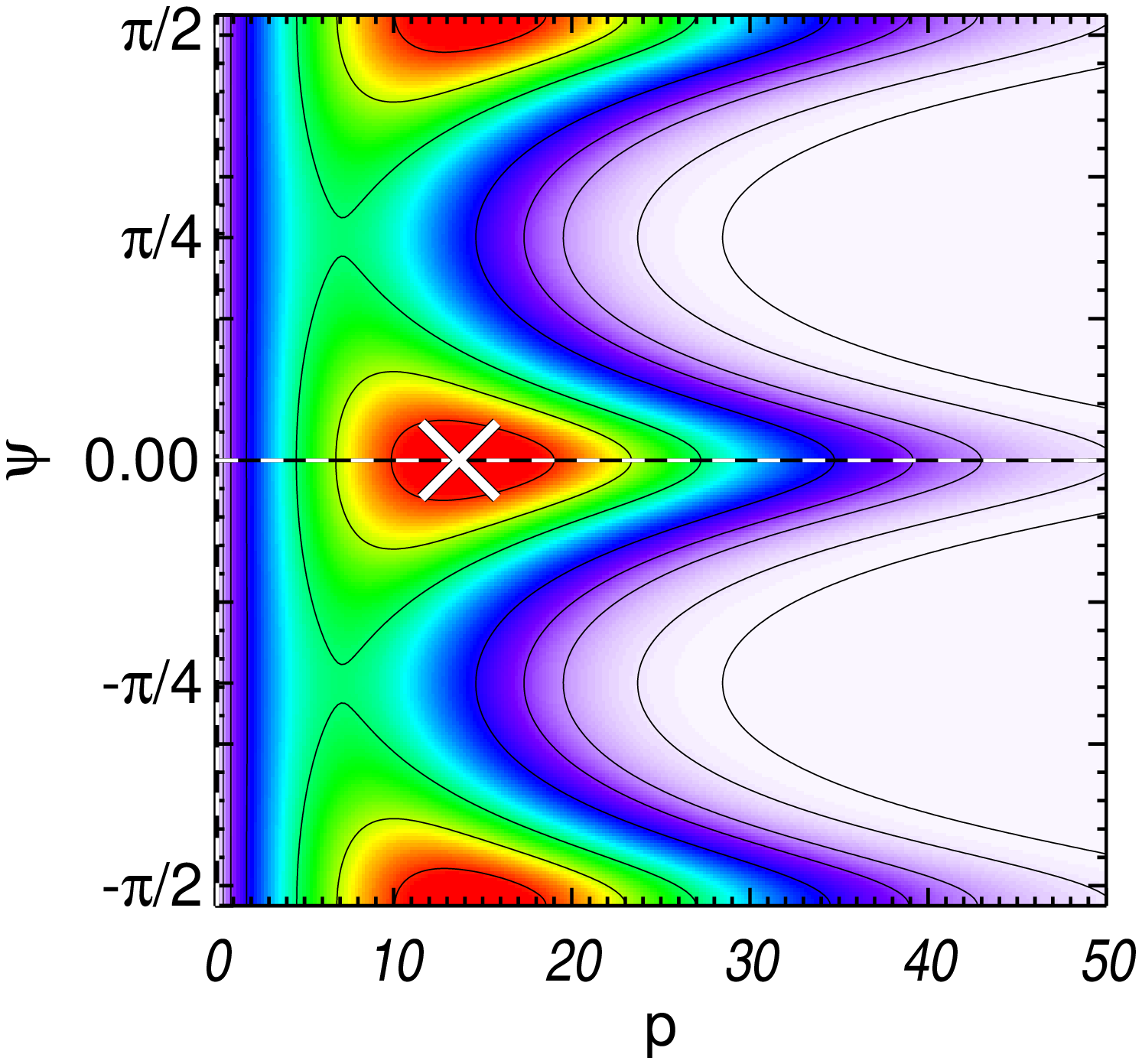}
\end{minipage} &
\begin{minipage}[c]{.15\linewidth}
 \includegraphics[width=3cm, viewport=200 0 600 400]
 {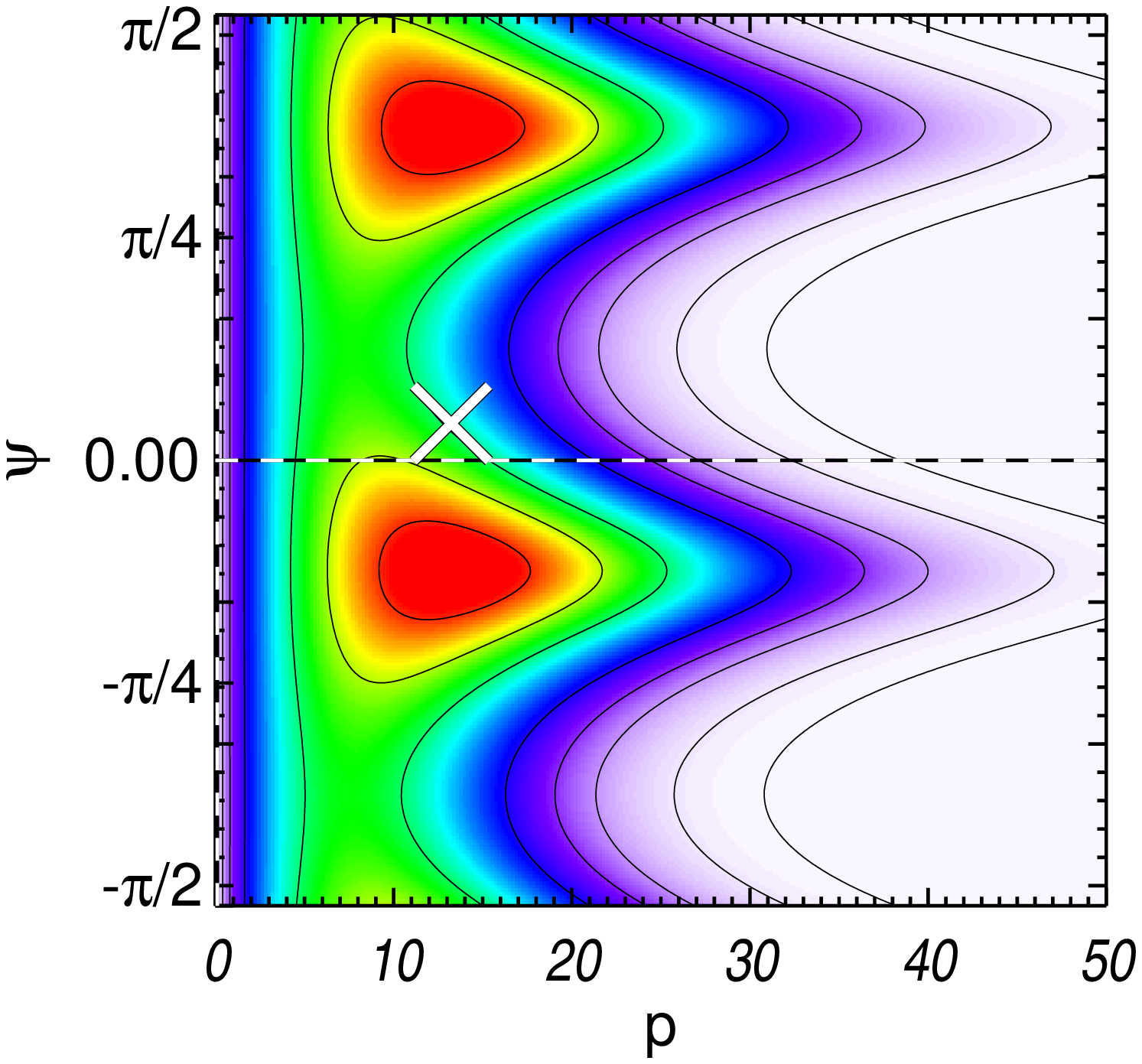}
\end{minipage} &
\begin{minipage}[c]{.15\linewidth}
 \includegraphics[width=3cm, viewport=200 0 600 400]
 {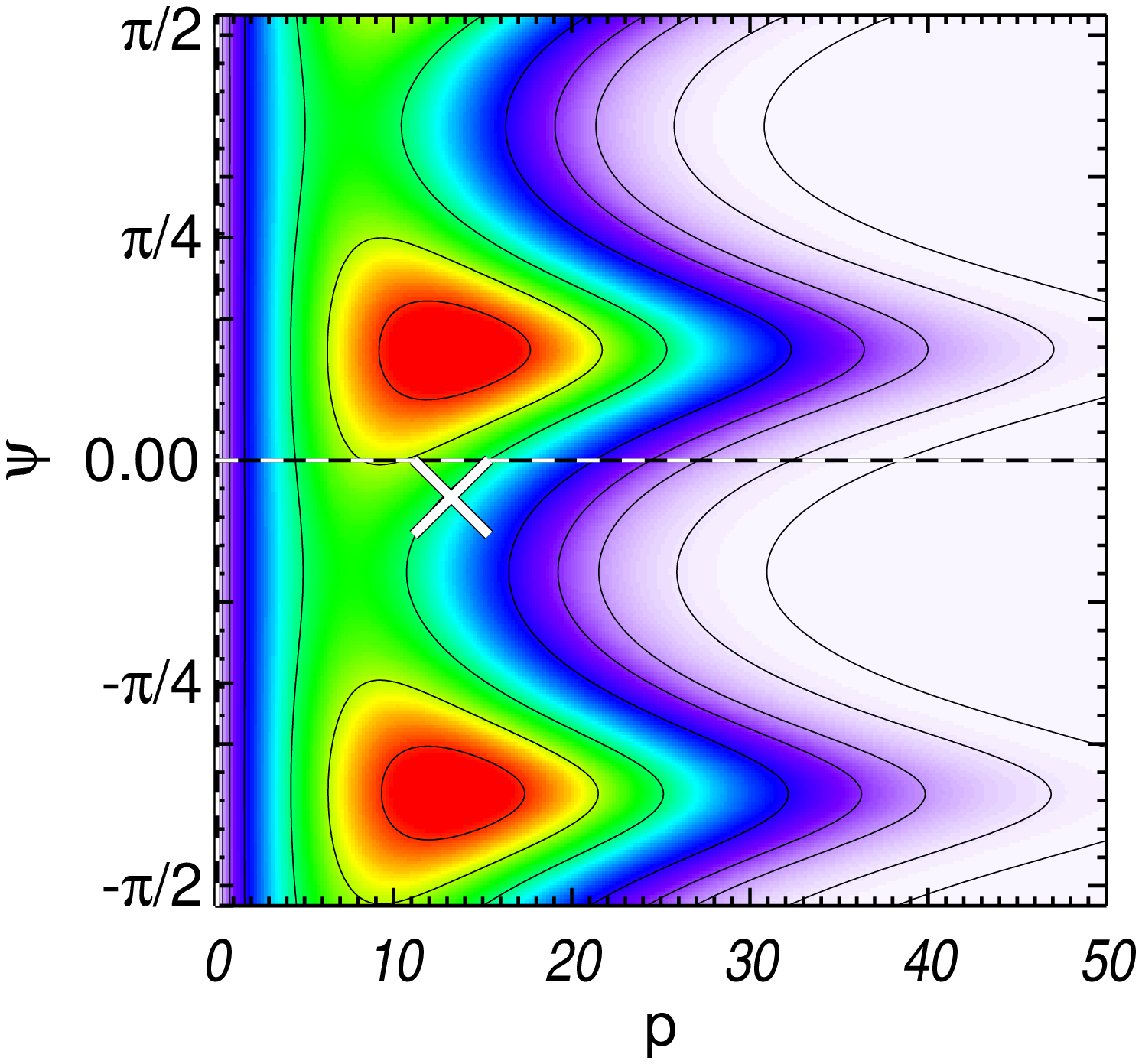}
\end{minipage} \\

\begin{minipage}[c]{.13\linewidth}
 \begin{tabular}{l}
 $p_0/\sigma_{\rm p,G}=0.5$ 
 \end{tabular} 
\end{minipage} &
\begin{minipage}[c]{.15\linewidth}
 \includegraphics[width=3cm, viewport=200 0 600 400]
 {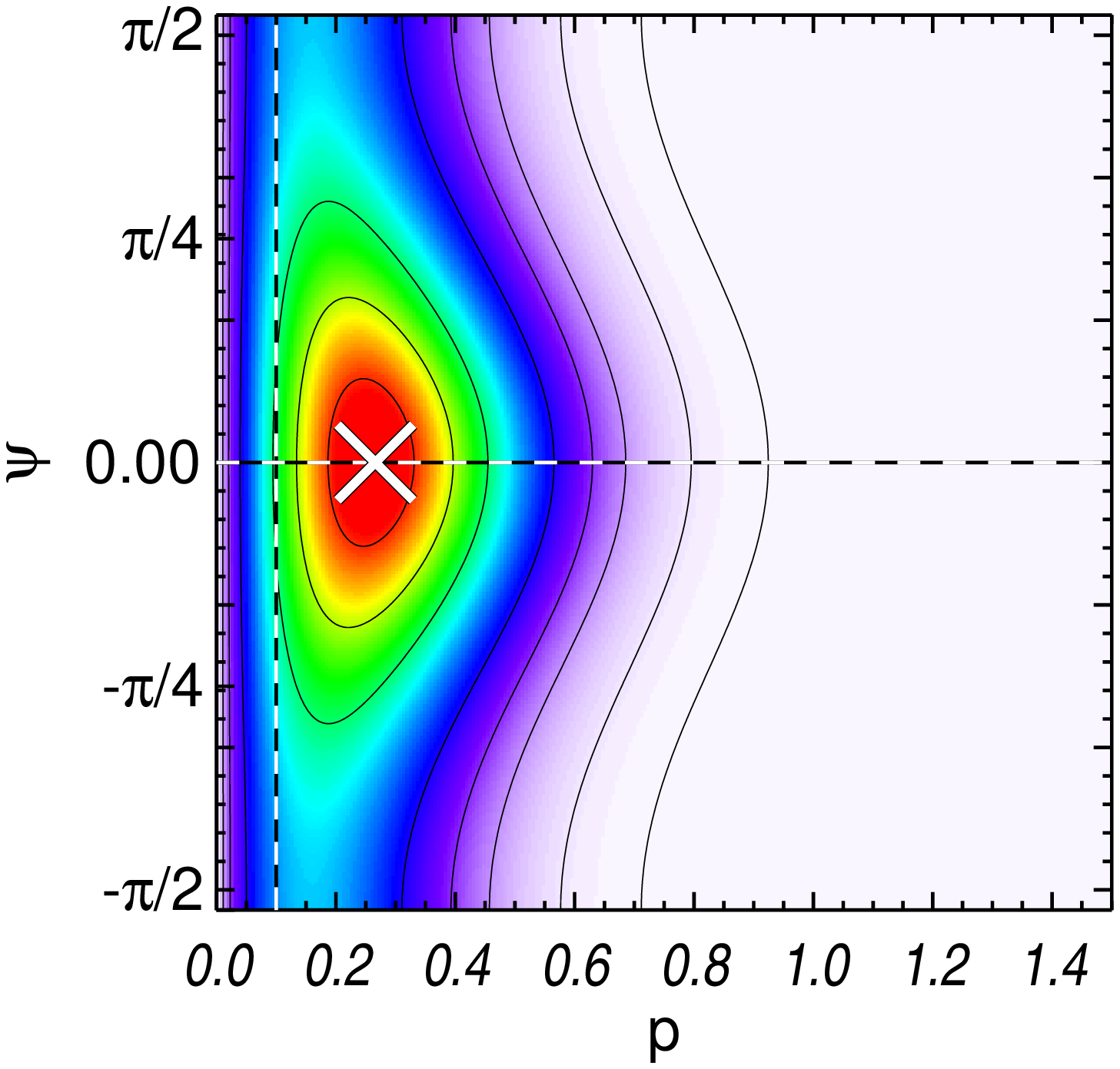}
\end{minipage} &
\begin{minipage}[c]{.15\linewidth}
 \includegraphics[width=3cm, viewport=200 0 600 400]
 {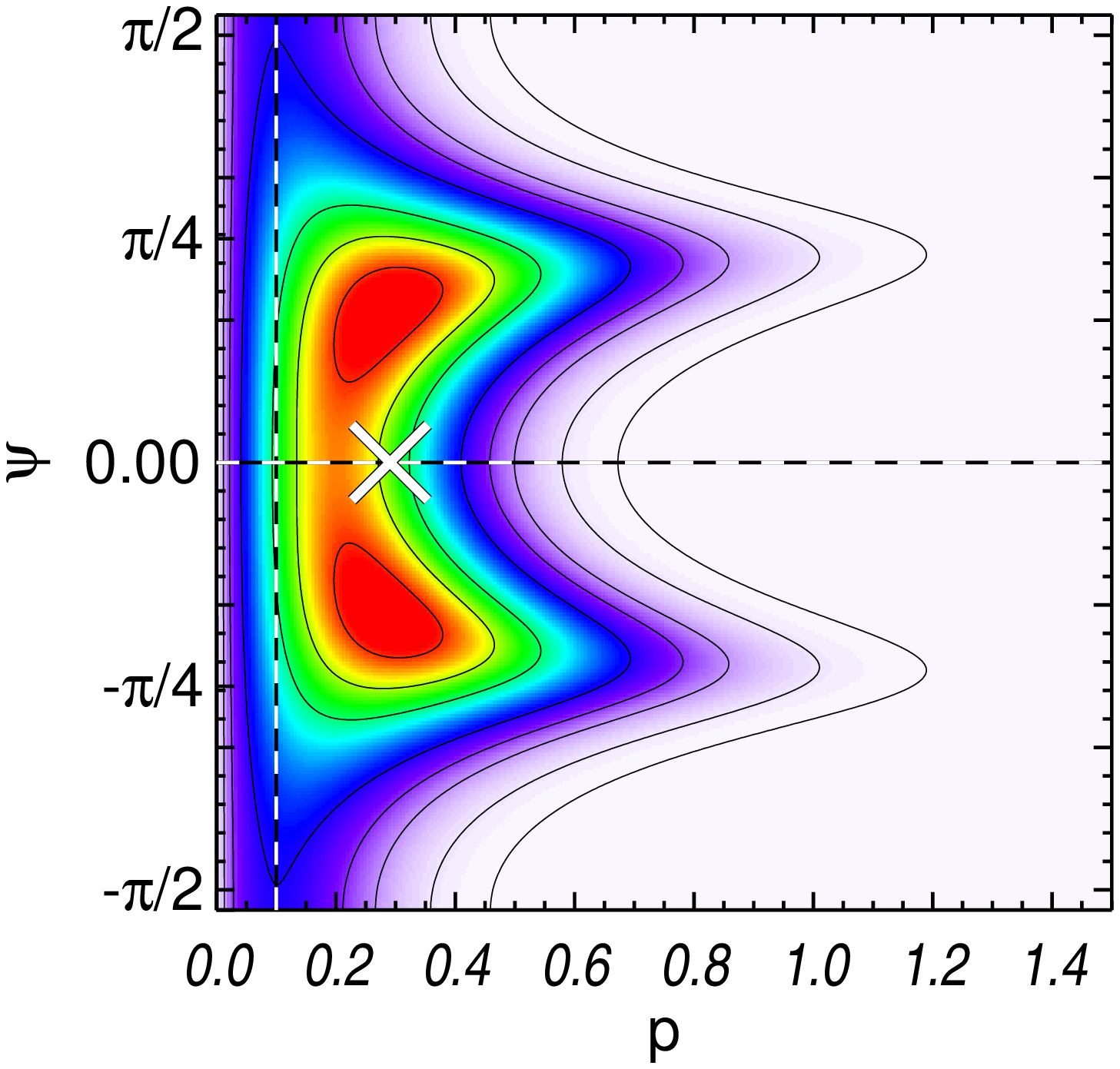}
\end{minipage} &
\begin{minipage}[c]{.15\linewidth}
 \includegraphics[width=3cm, viewport=200 0 600 400]
 {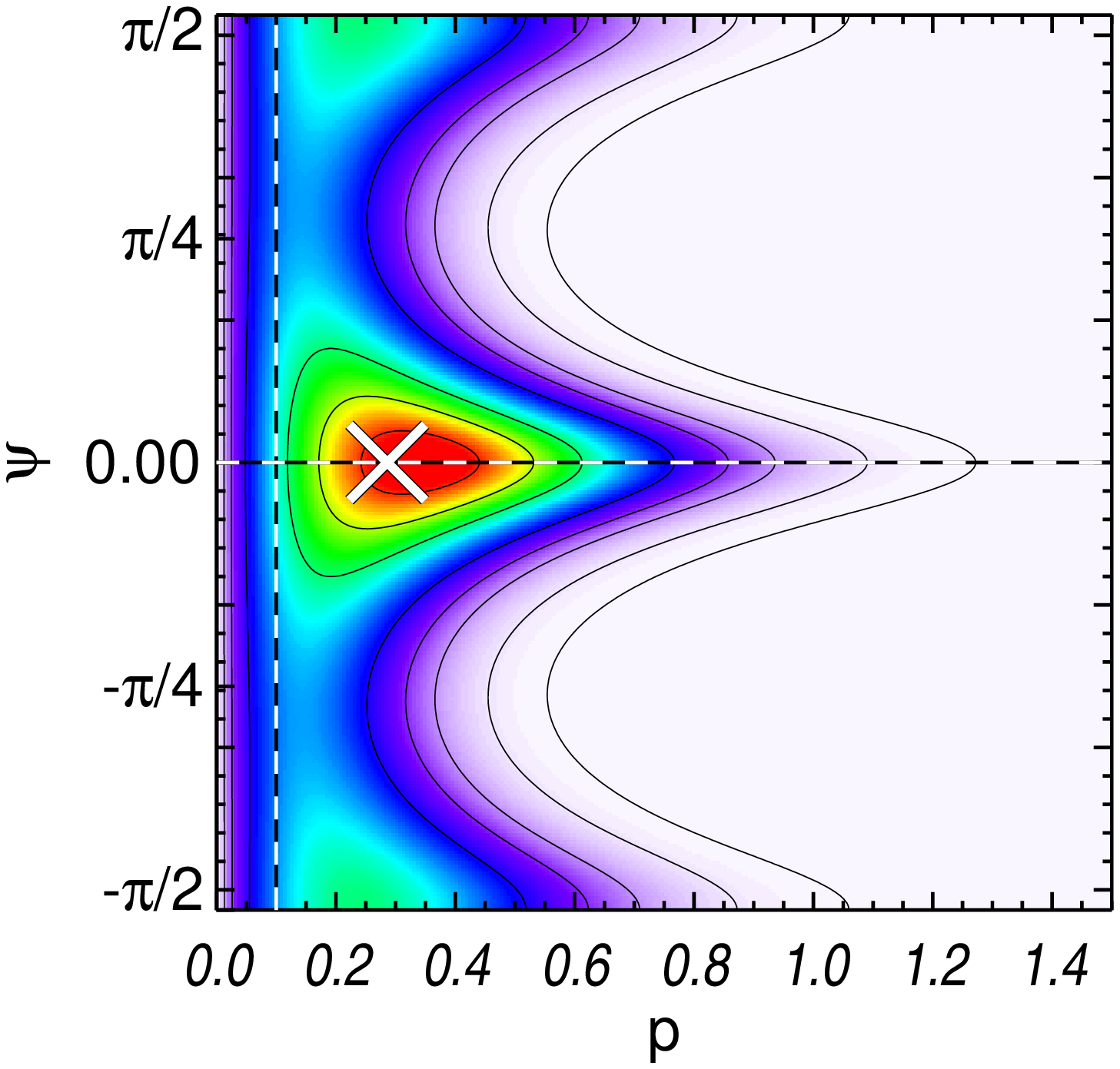}
\end{minipage} &
\begin{minipage}[c]{.15\linewidth}
 \includegraphics[width=3cm, viewport=200 0 600 400]
 {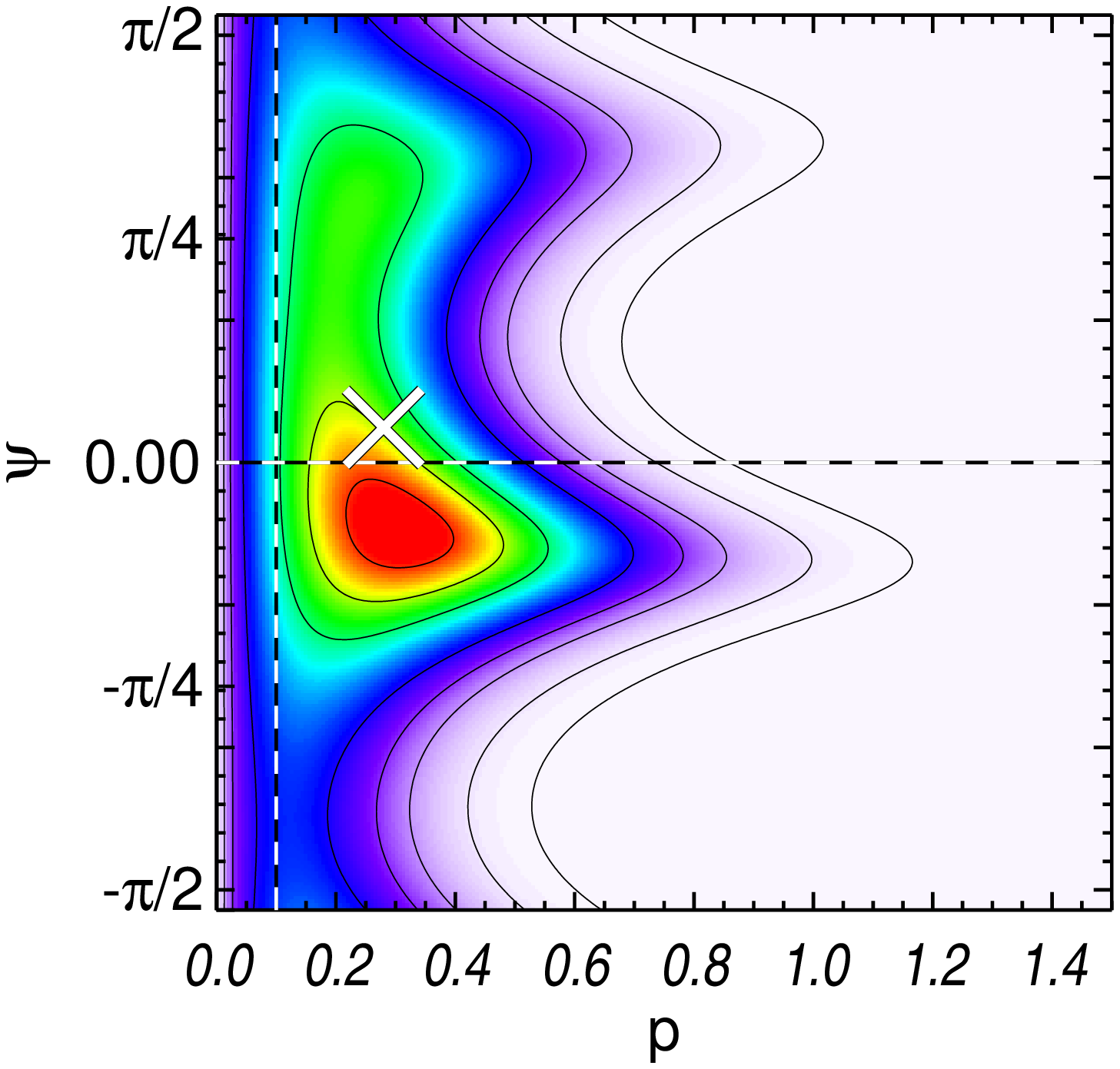}
\end{minipage} &
\begin{minipage}[c]{.15\linewidth}
 \includegraphics[width=3cm, viewport=200 0 600 400]
 {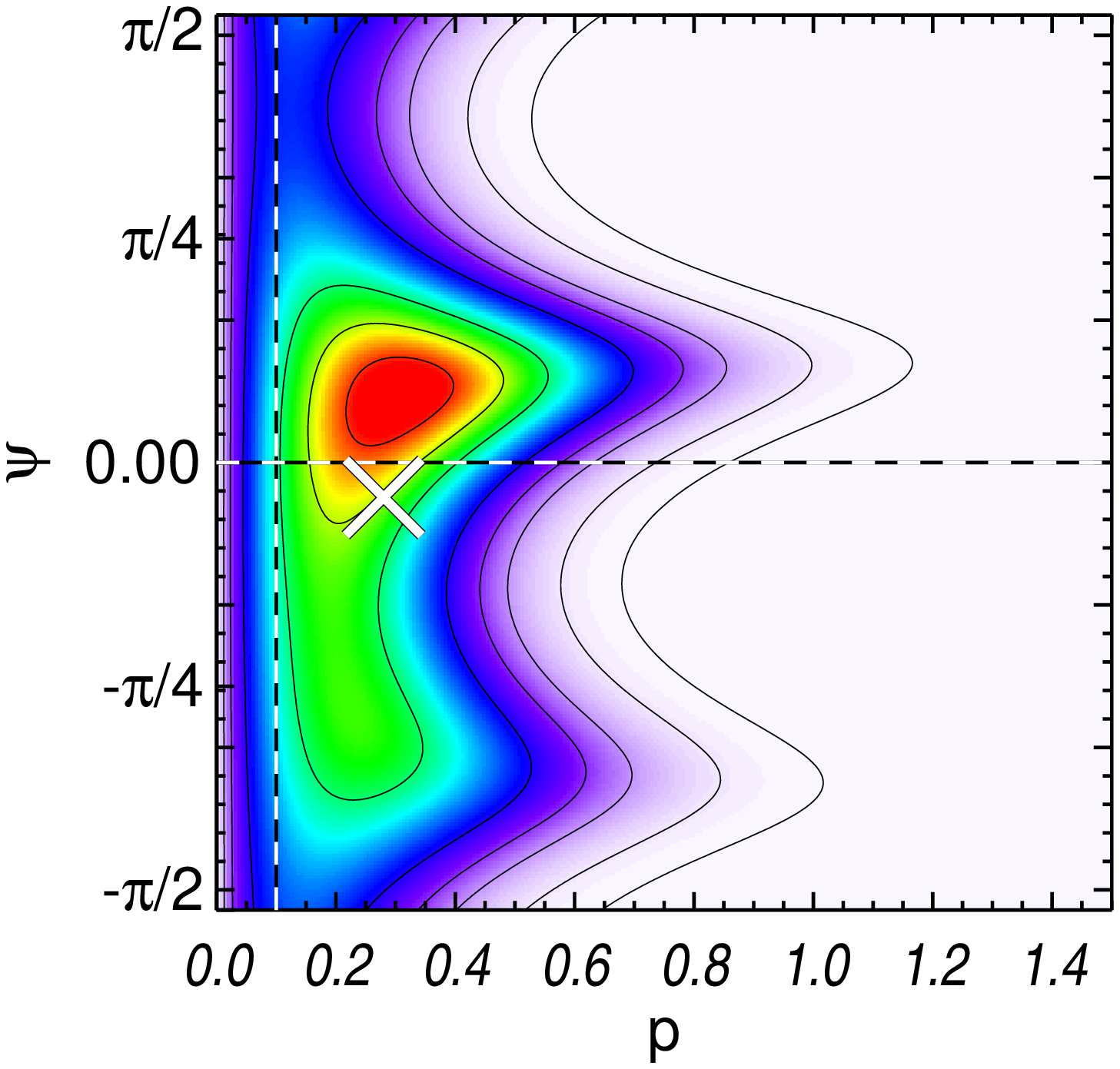}
\end{minipage} \\

\begin{minipage}[c]{.13\linewidth}
 \begin{tabular}{l}
 $p_0/\sigma_{\rm p,G}=1$ 
 \end{tabular} 
\end{minipage} &
\begin{minipage}[c]{.15\linewidth}
 \includegraphics[width=3cm, viewport=200 0 600 400]
 {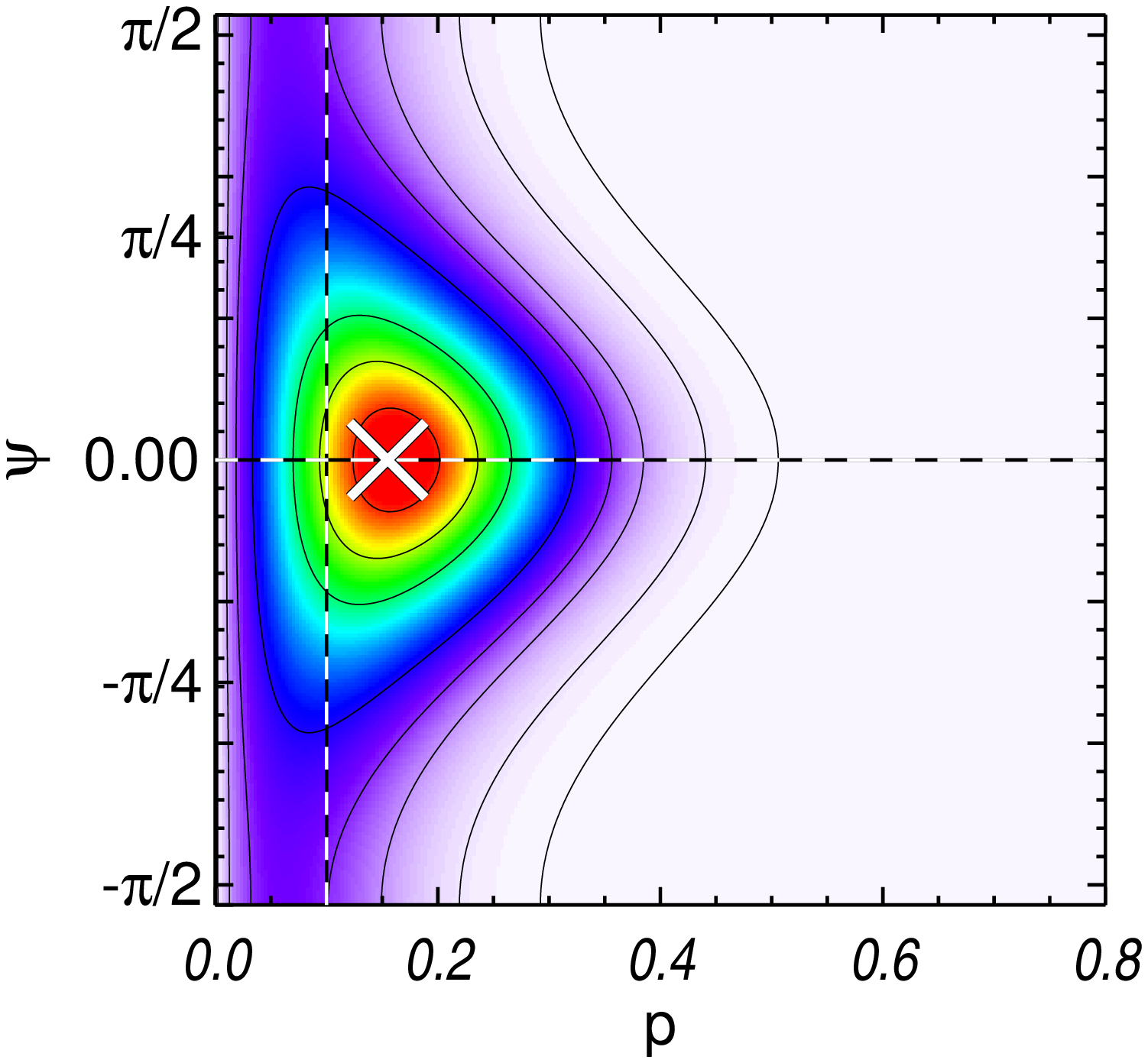}
\end{minipage} &
\begin{minipage}[c]{.15\linewidth}
 \includegraphics[width=3cm, viewport=200 0 600 400]
 {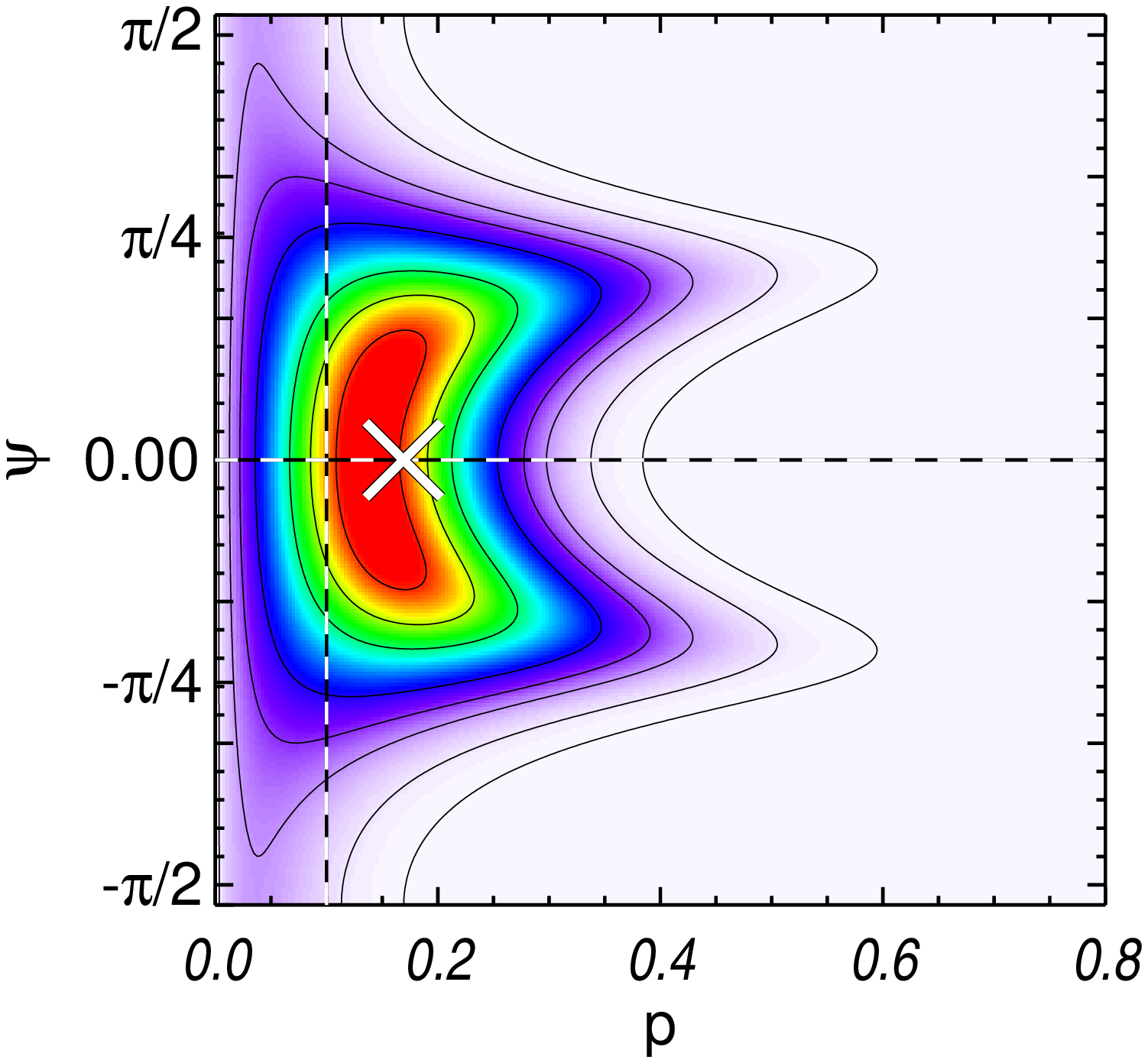}
\end{minipage} &
\begin{minipage}[c]{.15\linewidth}
 \includegraphics[width=3cm, viewport=200 0 600 400]
 {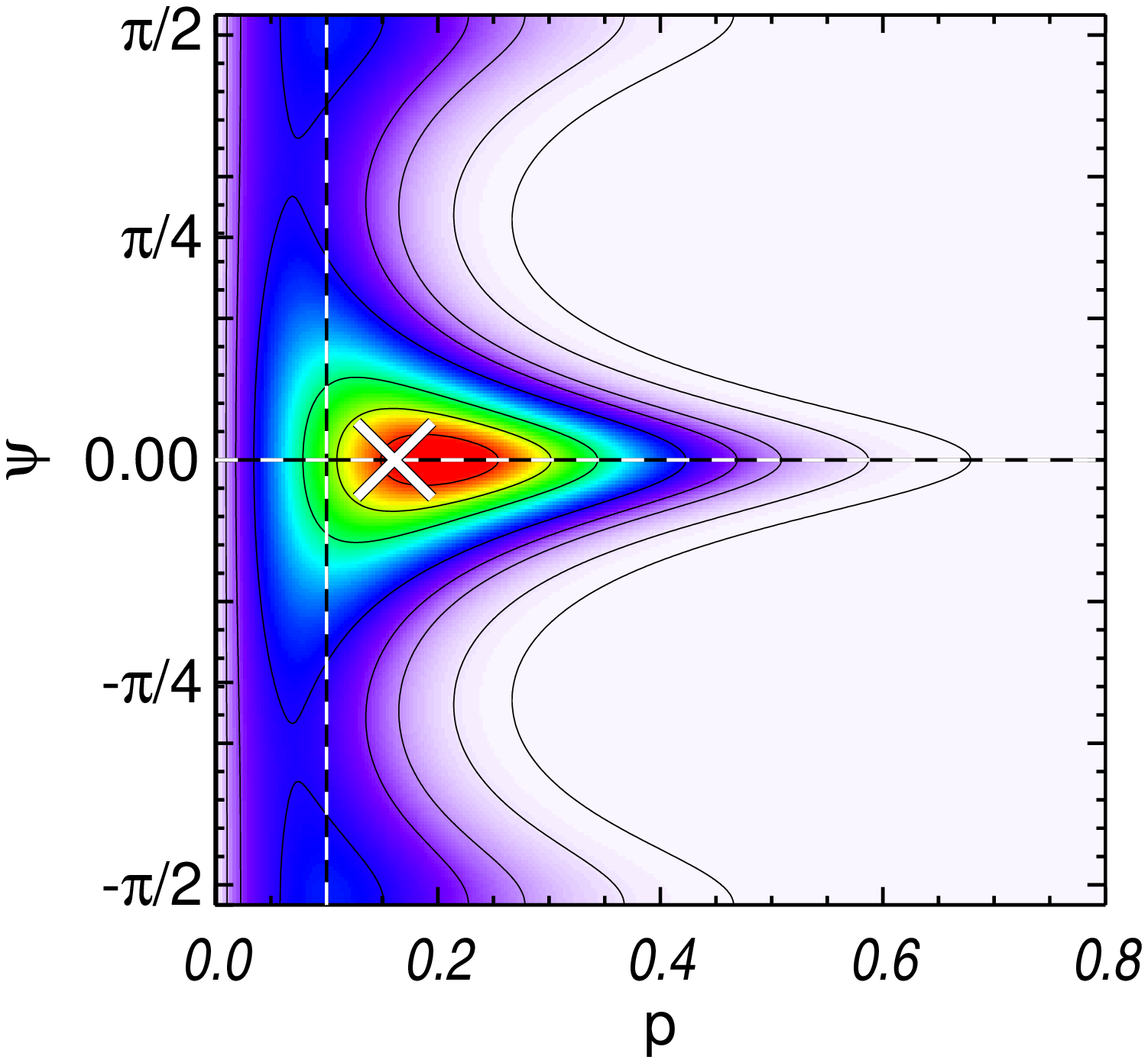}
\end{minipage} &
\begin{minipage}[c]{.15\linewidth}
 \includegraphics[width=3cm, viewport=200 0 600 400]
 {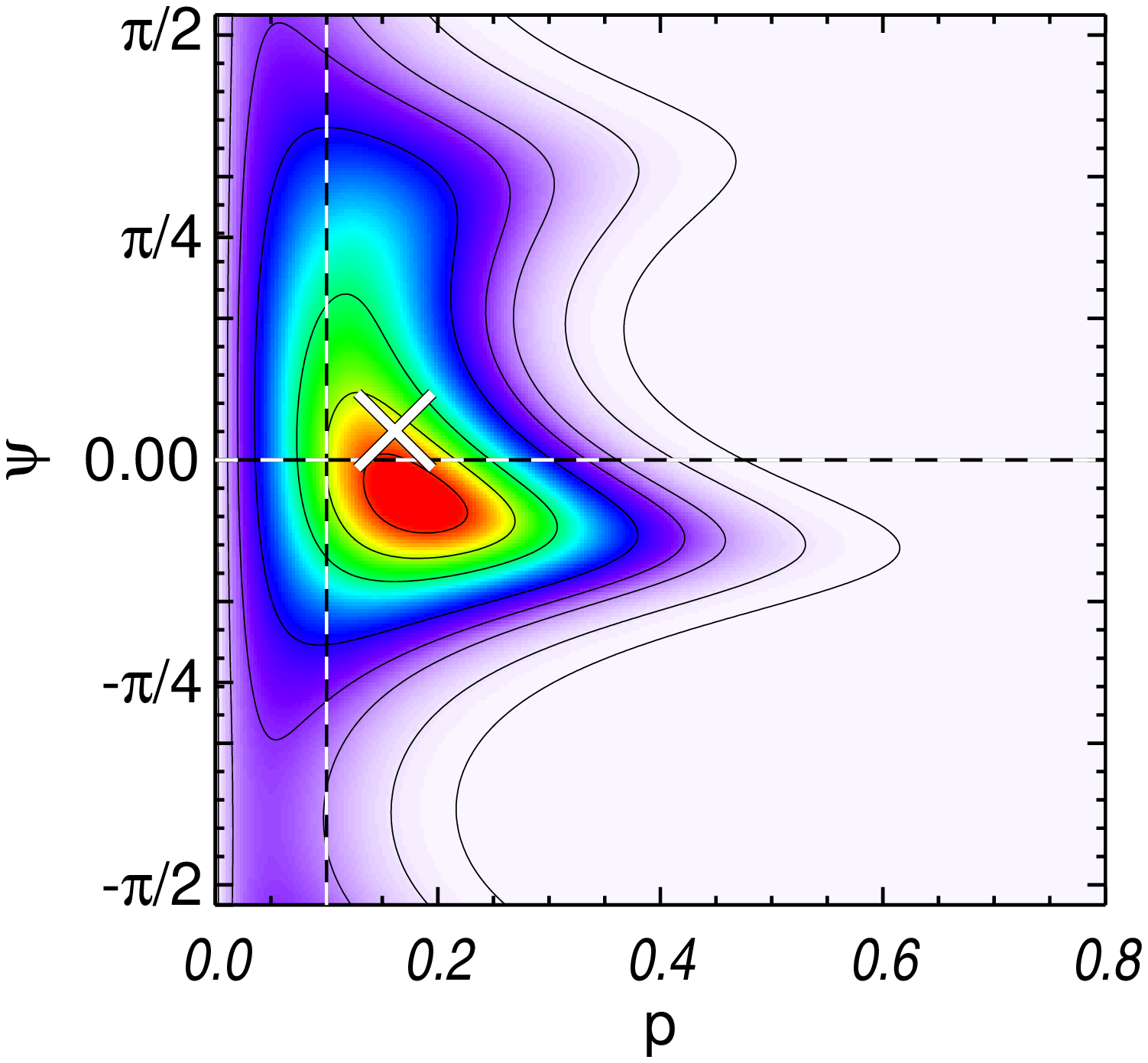}
\end{minipage} &
\begin{minipage}[c]{.15\linewidth}
 \includegraphics[width=3cm, viewport=200 0 600 400]
 {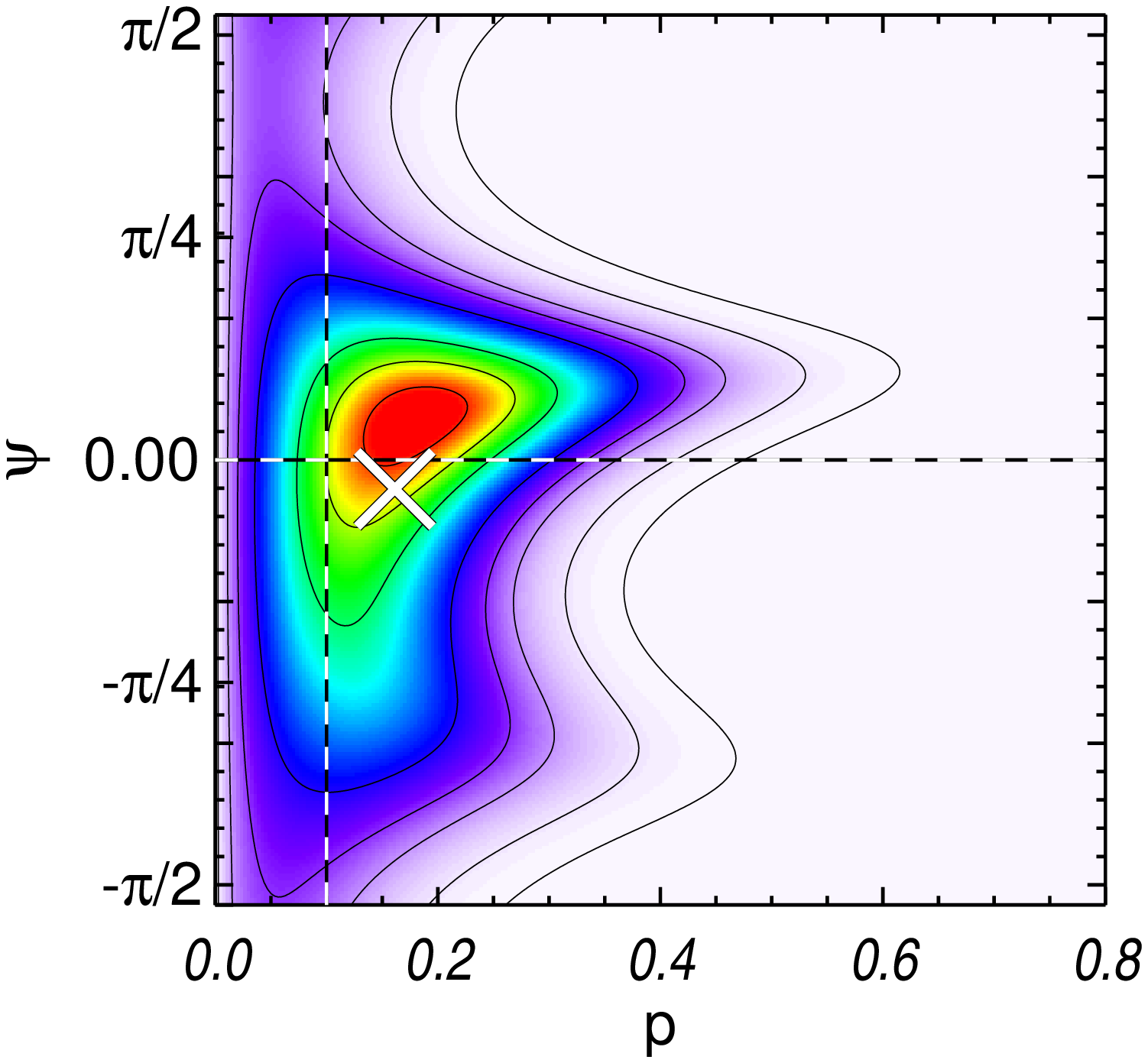}
\end{minipage} \\

\begin{minipage}[c]{.13\linewidth}
 \begin{tabular}{l}
 $p_0/\sigma_{\rm p,G}=5$ 
 \end{tabular} 
\end{minipage} &
\begin{minipage}[c]{.15\linewidth}
 \includegraphics[width=3cm, viewport=200 0 600 400]
 {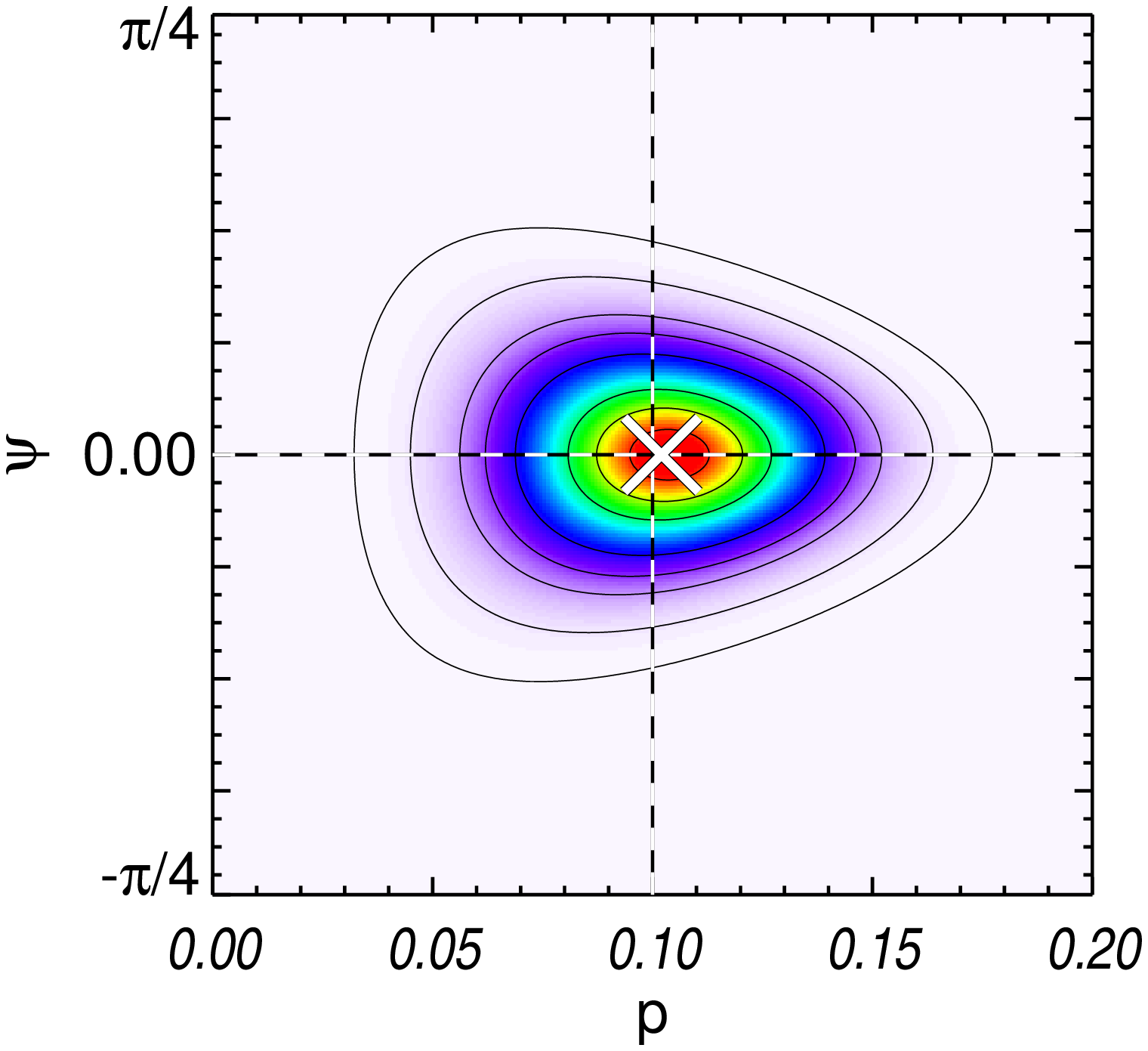}
\end{minipage} &
\begin{minipage}[c]{.15\linewidth}
 \includegraphics[width=3cm, viewport=200 0 600 400]
 {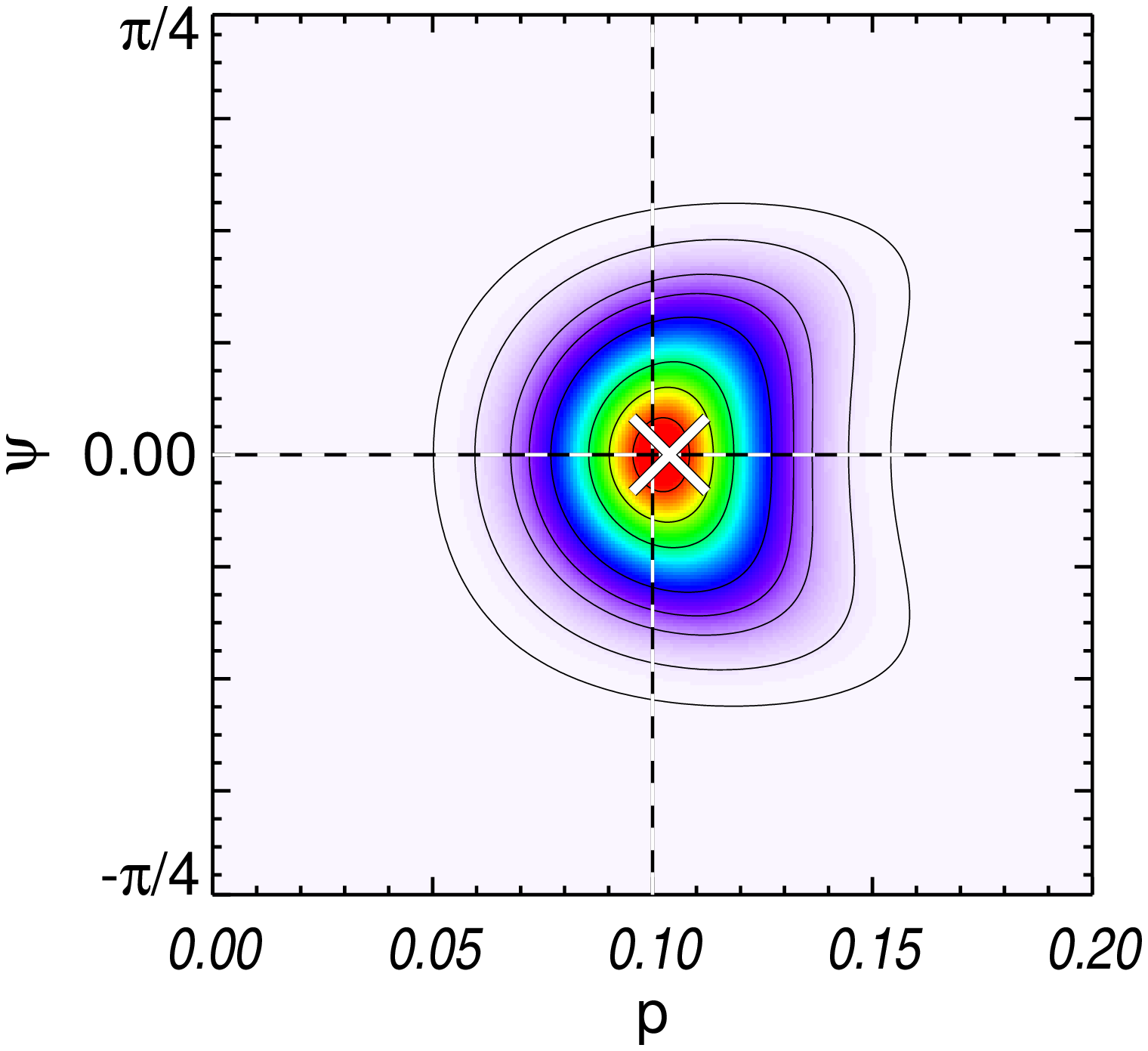}
\end{minipage} &
\begin{minipage}[c]{.15\linewidth}
 \includegraphics[width=3cm, viewport=200 0 600 400]
 {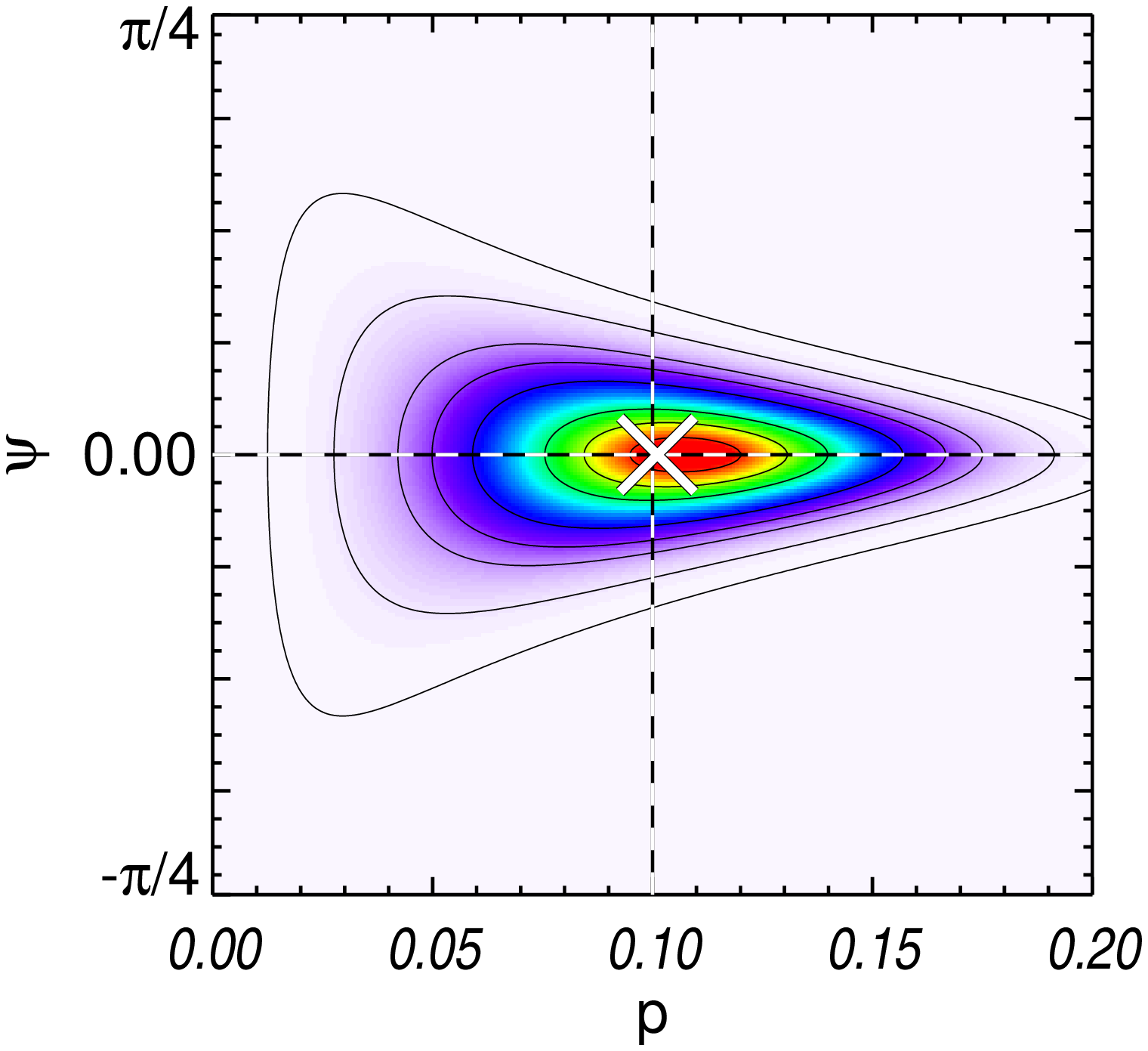}
\end{minipage} &
\begin{minipage}[c]{.15\linewidth}
 \includegraphics[width=3cm, viewport=200 0 600 400]
 {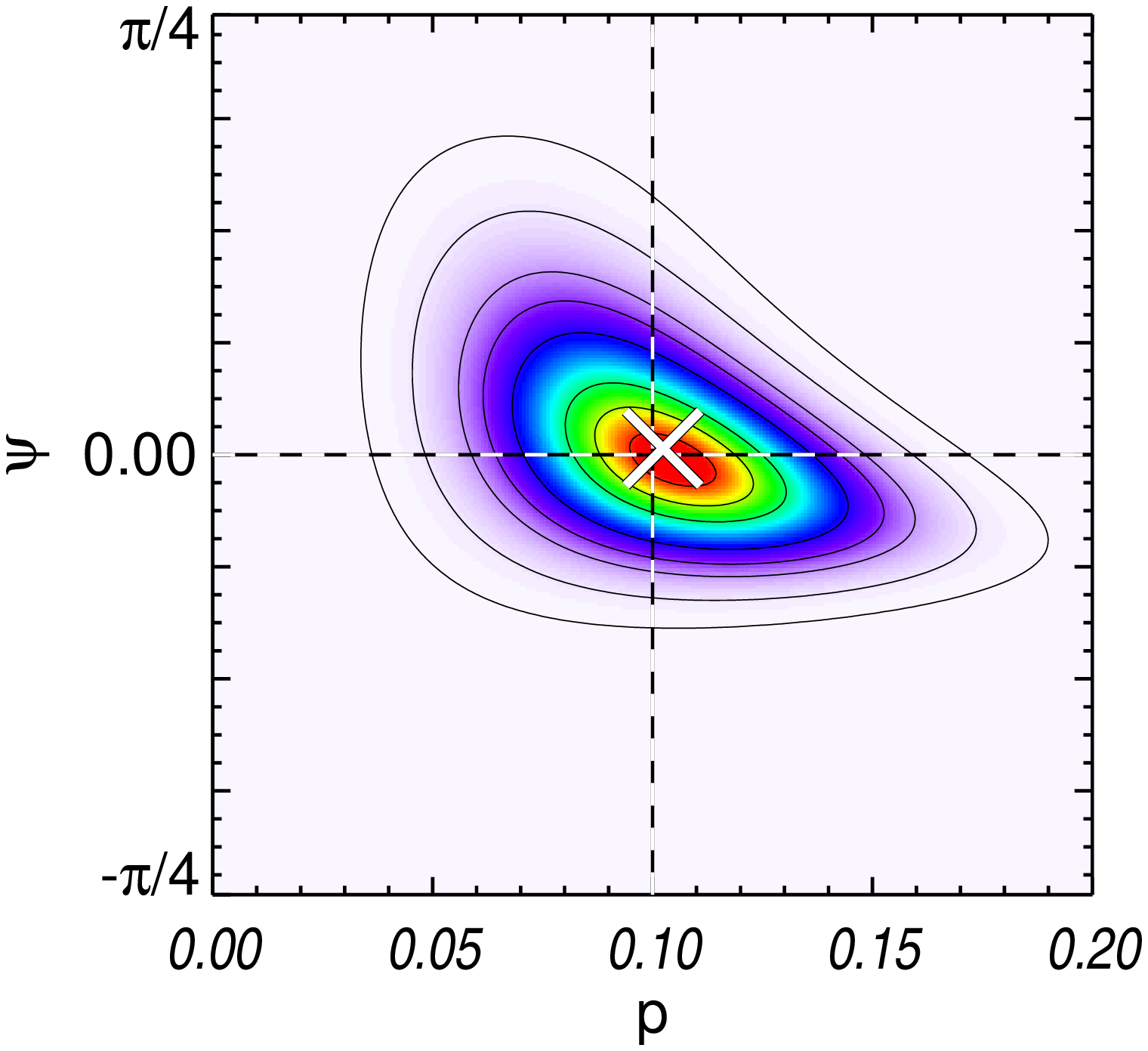}
\end{minipage} &
\begin{minipage}[c]{.15\linewidth}
 \includegraphics[width=3cm, viewport=200 0 600 400]
 {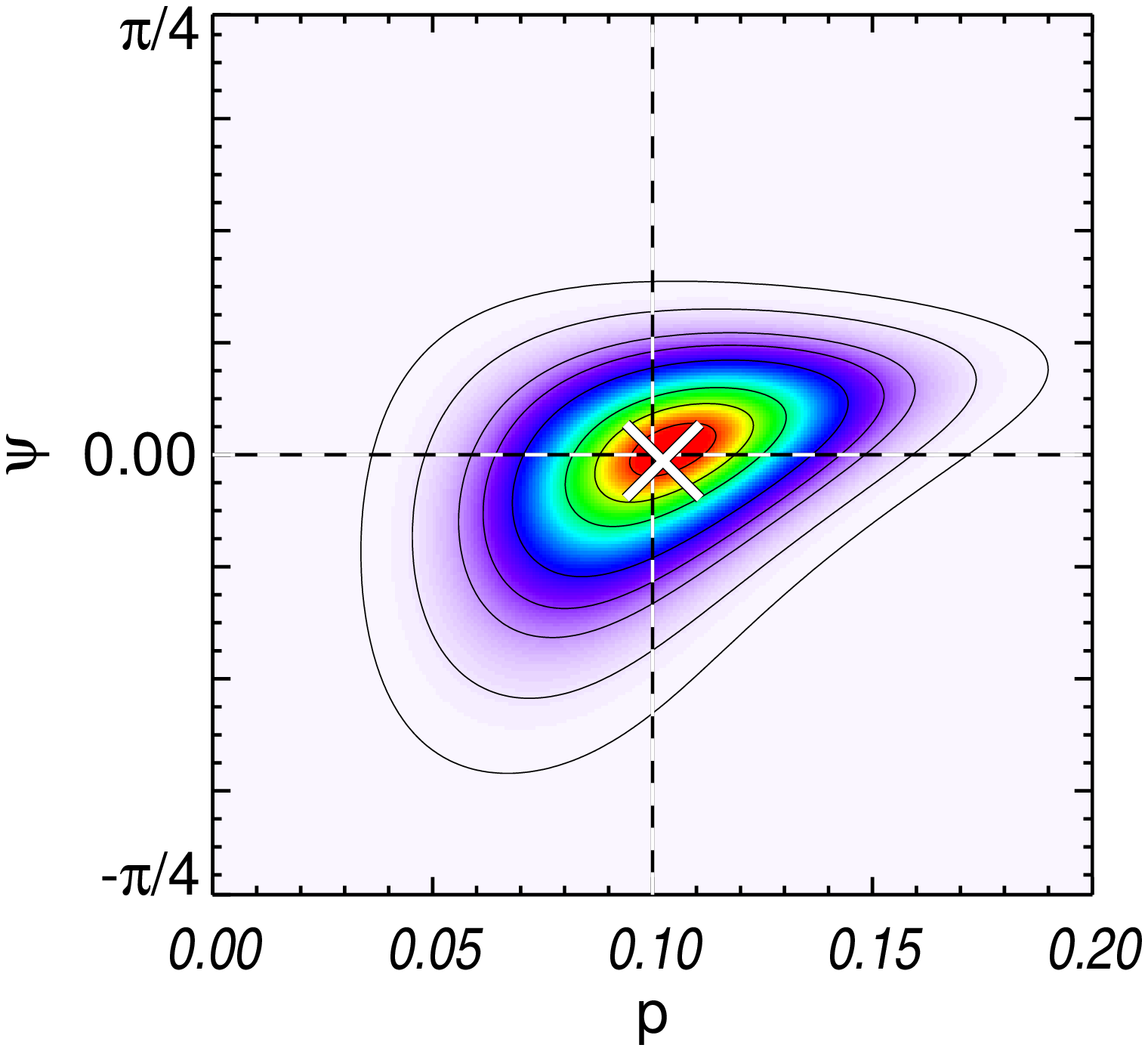}
\end{minipage} \\
\end{tabular} 
\caption{Probability density functions,
$f_{\rm 2D}(p,\psi\,|\,p_0,\psi_0,\tens{\Sigma}_{\rm p})$,
with infinite S/N on intensity, computed for a given set of polarization
parameters, namely $\psi_0\,{=}\,0^\circ$ and $p_0\,{=}\,0.1$ (dashed lines).
Each row corresponds to a specific level of the signal-to-noise ratio
$p_0/\sigma_{\rm p,G}\,{=}\,0.01, 0.5, 1$, and $5$, from top to bottom. 
Various configurations of the covariance matrix are shown (in the
different columns).
Furthest left is the standard case: no ellipticity and no correlation.
The next two columns show the impact of ellipticities $\varepsilon\,{=}\,1/2$
and $2$. The last two columns deal 
with correlations $\rho\,{=}\,-1/2$ and $+1/2$. White crosses indicate the
mean likelihood estimates of the PDF ($\overline{p},\overline{\psi}$).
The contour levels are shown at 0.1, 1, 5, 10, 20, 50, 70, and 90\% of
the maximum of the distribution.}
\label{fig:pdf_impact_epsirho}
\end{figure*}

We illustrate the shape of the 2D probability density function
$f_{\rm 2D}(p,\psi\,|\,I_0, p_0, \psi_0, \tens{\Sigma})$ in
Fig.~\ref{fig:pdf_impact_epsirho}, for the case of a perfectly known
intensity having no correlation with the polarization.
Starting from a given couple of true polarization parameters
$\psi_0\,{=}\,0^\circ$ and $p_0\,{=}\,0.1$, 
the PDF is computed for various signal-to-noise ratios
$p_0/\sigma_{\rm p,G}$ and settings of the covariance matrix.
The signal-to-noise ratio $p_0/\sigma_{\rm p,G}$ is varied from
0.01 to 0.5, 1, and 5 (top to bottom).  The dashed crossing lines show
the location of the initial true polarization values.
The leftmost column shows the results obtained when the covariance matrix
is assumed to be diagonal and symmetric, (i.e.,
$\varepsilon\,{=}\,1$ and $\rho\,{=}\,0$), as was usually done in
previous works on polarization data.
The distribution along the $\psi$ axis is fully symmetric around 0,
implying the absence of bias on the polarization angle.
When varying the ellipticity $\varepsilon$ from 1/2 to 2 (columns 2 and 3),
we still observe symmetrical PDFs in this configuration, but
multiple peaks appear at low signal-to-noise ratio. 
In the presence of correlation, i.e., $\rho\,{=}\,-1/2$ and $1/2$
(columns 4 and 5), the maximum peak is now slightly shifted in $p$ and $\psi$, 
with an asymmetric PDF around the initial $\psi_0$ value.

\begin{figure*}
\begin{tabular}{ccccc}
&
 \begin{minipage}[c]{.15\linewidth}
$\quad$
 \begin{tabular}{l}
 $\varepsilon = 1/2$ \\
 $\rho = 0$ \\
\end{tabular} \\ 
$\Bigg($
 \begin{tabular}{l}
 $\varepsilon_{\rm eff} = 2$ \\
 $\theta = \pi$ \\
\end{tabular} 
$\Bigg)$
\end{minipage} & 
\begin{minipage}[c]{.15\linewidth}
$\quad$
 \begin{tabular}{l}
 $\varepsilon = 2$ \\
 $\rho = 0$ 
\end{tabular} \\ 
$\Bigg($
 \begin{tabular}{l}
 $\varepsilon_{\rm eff} = 2$ \\
 $\theta = 0$ \\
\end{tabular} 
$\Bigg)$
\end{minipage} &
\begin{minipage}[c]{.15\linewidth}
$\quad$
 \begin{tabular}{l}
 $\varepsilon = 1$ \\
 $\rho = -1/2$ 
\end{tabular} \\ 
$\Bigg($
 \begin{tabular}{l}
 $\varepsilon_{\rm eff} \sim 1.73$ \\
 $\theta = -\pi/4$ \\
 \end{tabular} 
$\Bigg)$
\end{minipage} &
\begin{minipage}[c]{.15\linewidth}
$\quad$
 \begin{tabular}{l}
 $\varepsilon = 1$ \\
 $\rho = 1/2$
 \end{tabular} \\ 
$\Bigg($
 \begin{tabular}{l}
 $\varepsilon_{\rm eff} \sim 1.73$ \\
 $\theta = \pi/4$ \\
 \end{tabular} 
$\Bigg)$
\end{minipage} \\ \\
\begin{minipage}[c]{.15\linewidth}
 \begin{tabular}{l}
 $\psi_0=3\pi/8$ 
 \end{tabular} 
\end{minipage} &
\begin{minipage}[c]{.15\linewidth}
 \includegraphics[width=3cm, viewport=200 0 600 400]
 {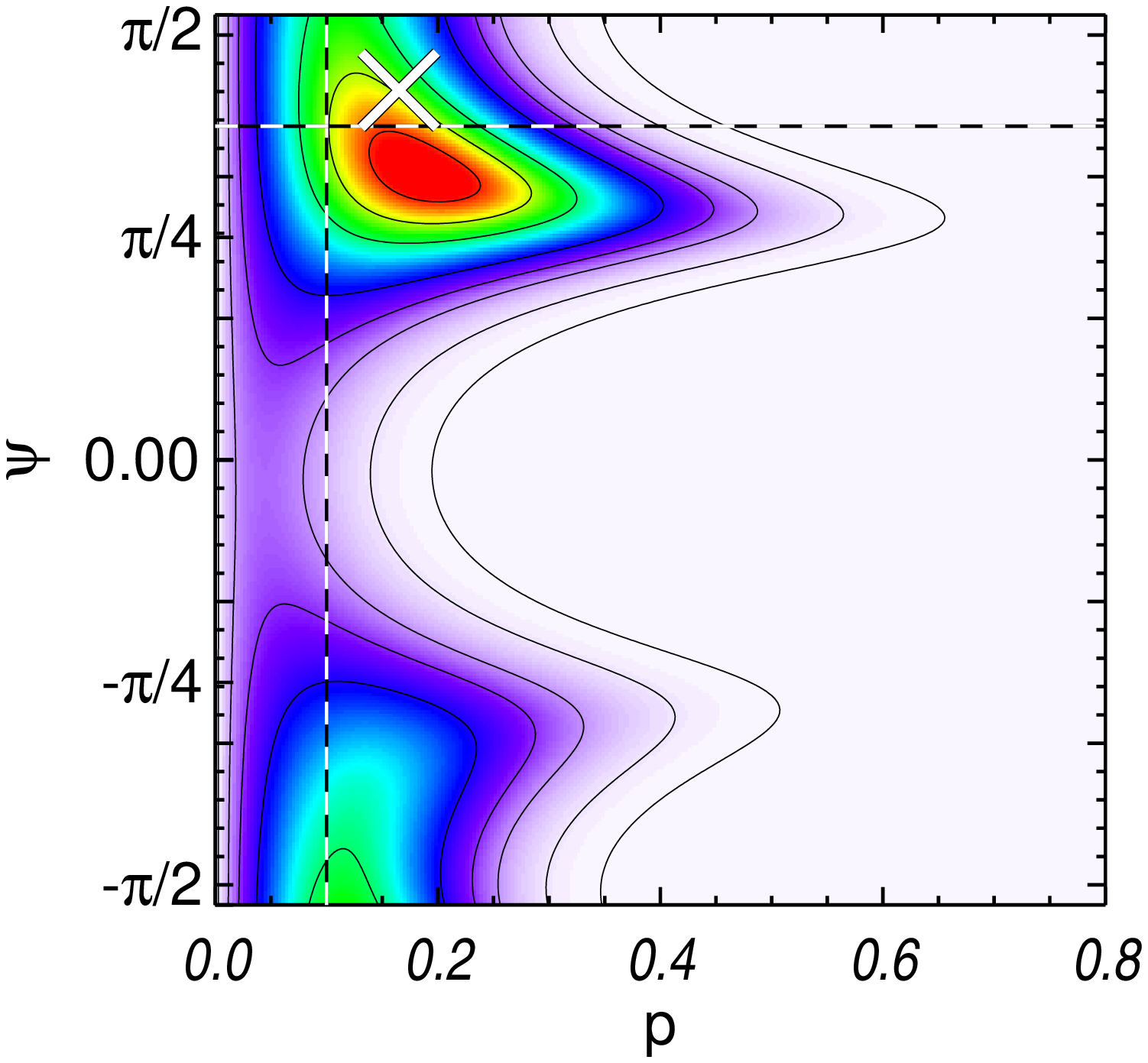}
\end{minipage}&
\begin{minipage}[c]{.15\linewidth}
 \includegraphics[width=3cm, viewport=200 0 600 400]
 {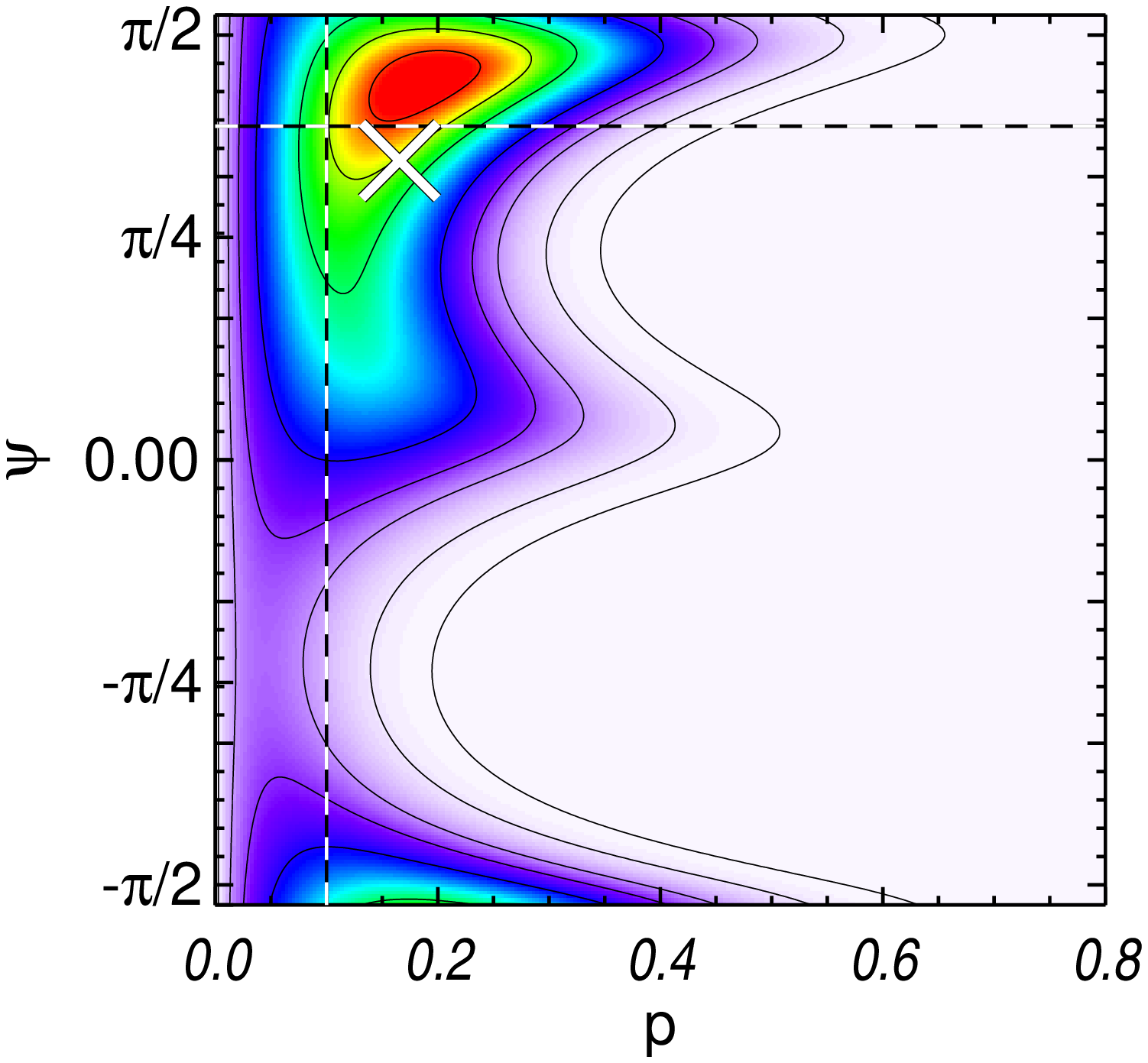}
\end{minipage}&
\begin{minipage}[c]{.15\linewidth}
 \includegraphics[width=3cm, viewport=200 0 600 400]
 {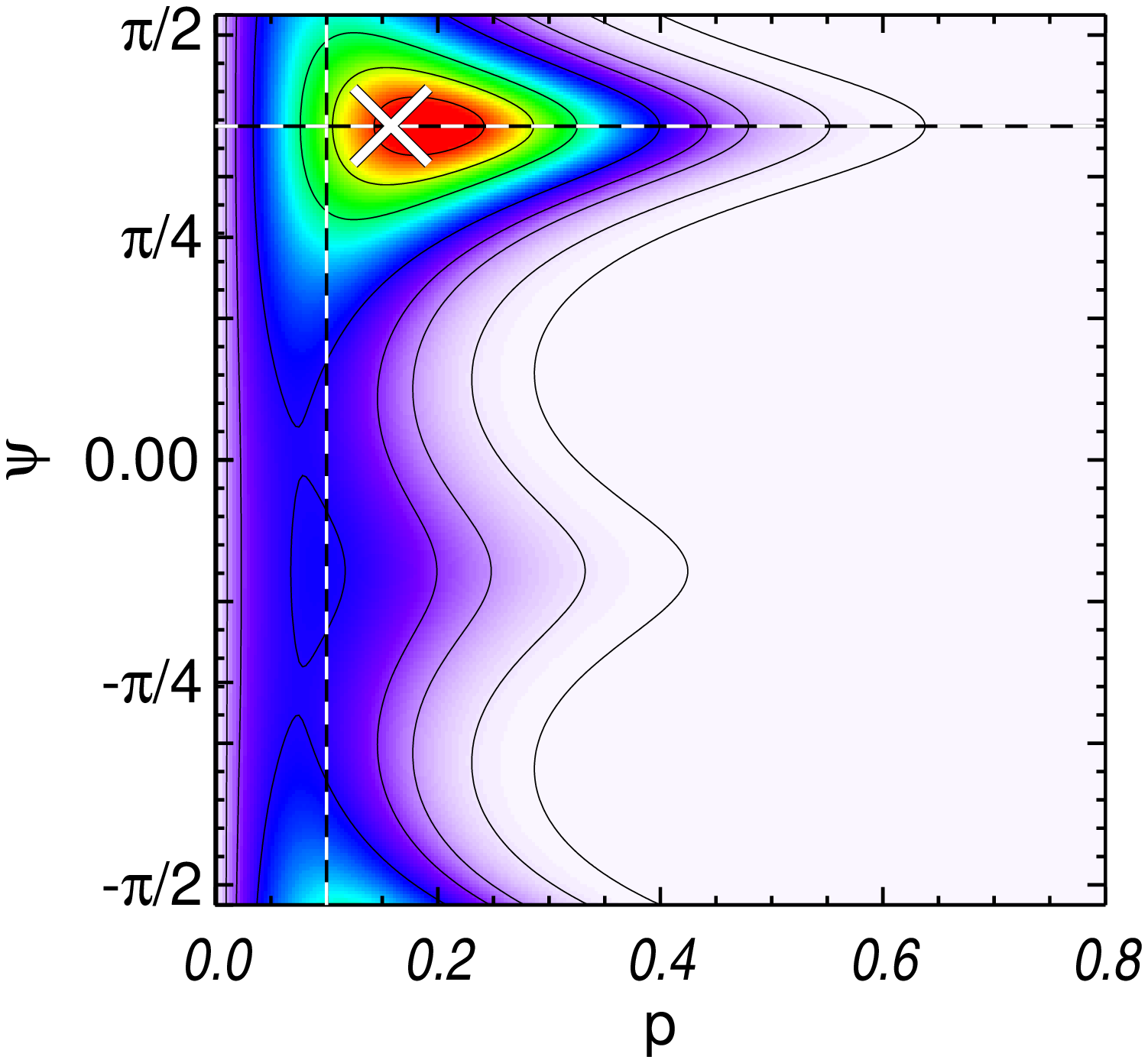}
\end{minipage}&
\begin{minipage}[c]{.15\linewidth}
 \includegraphics[width=3cm, viewport=200 0 600 400]
 {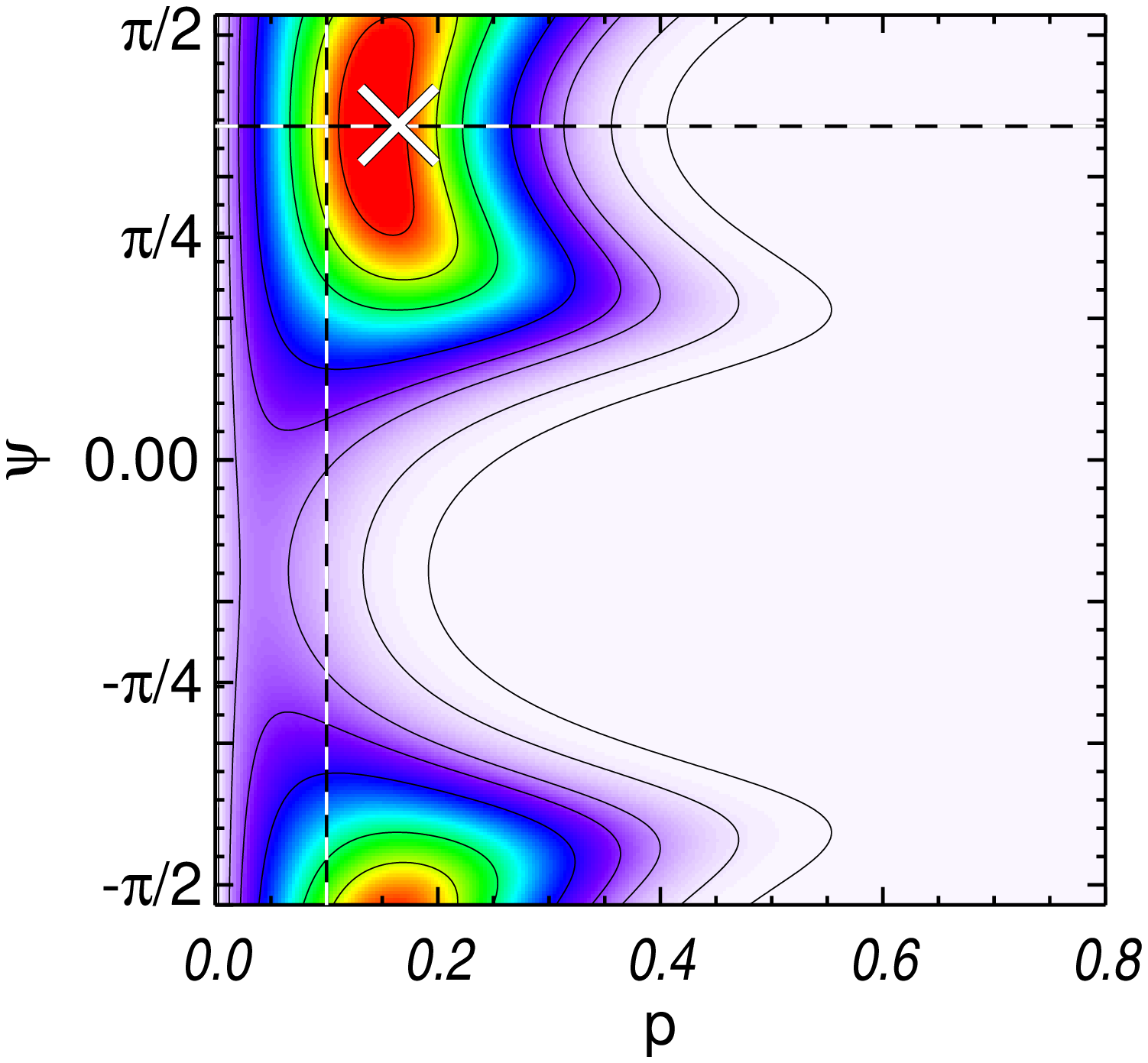}
\end{minipage} \\

\begin{minipage}[c]{.15\linewidth}
 \begin{tabular}{l}
 $\psi_0=\pi/4$ 
 \end{tabular} 
\end{minipage} &
\begin{minipage}[c]{.15\linewidth}
 \includegraphics[width=3cm, viewport=200 0 600 400]
 {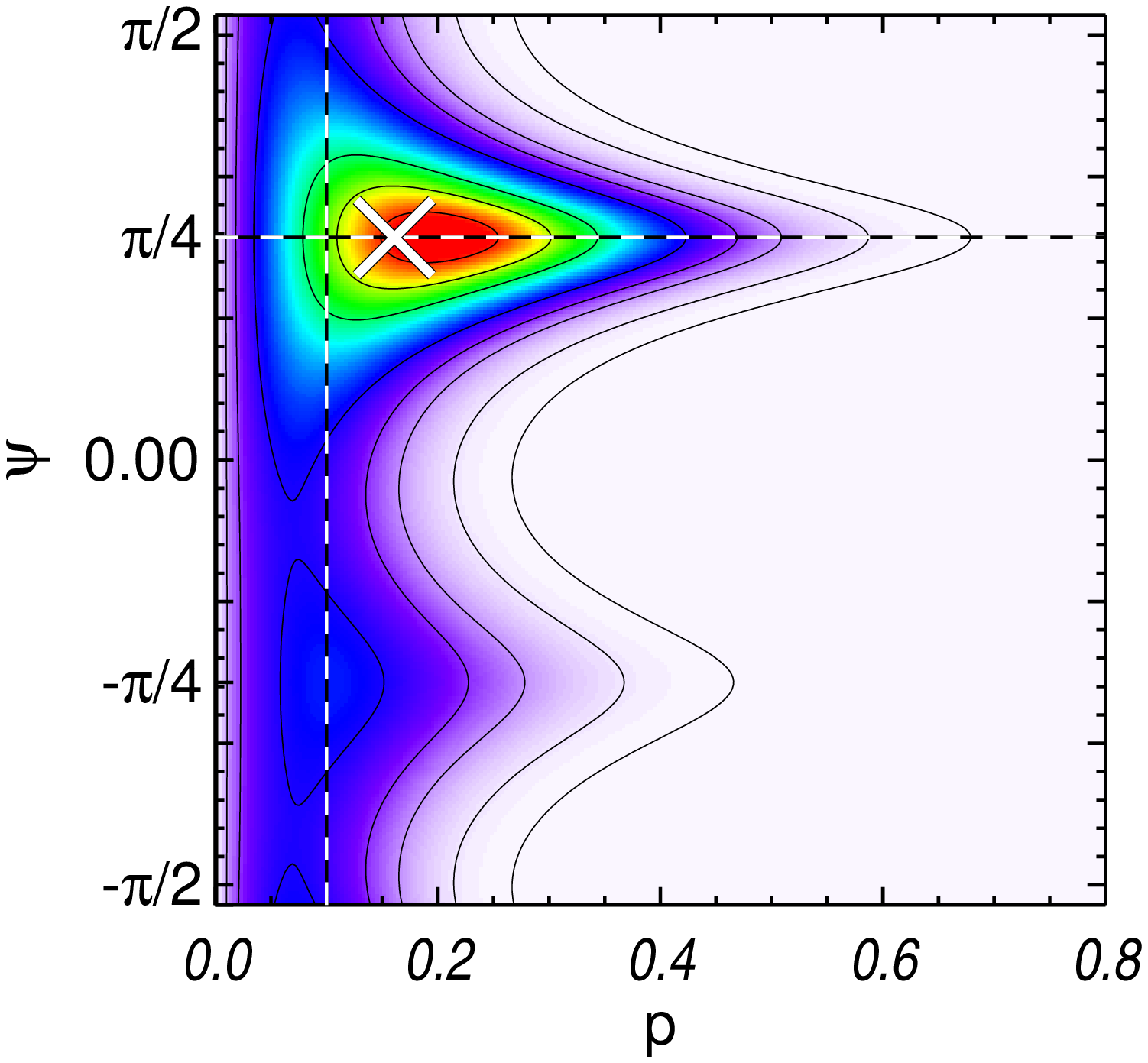}
\end{minipage}&
\begin{minipage}[c]{.15\linewidth}
 \includegraphics[width=3cm, viewport=200 0 600 400]
 {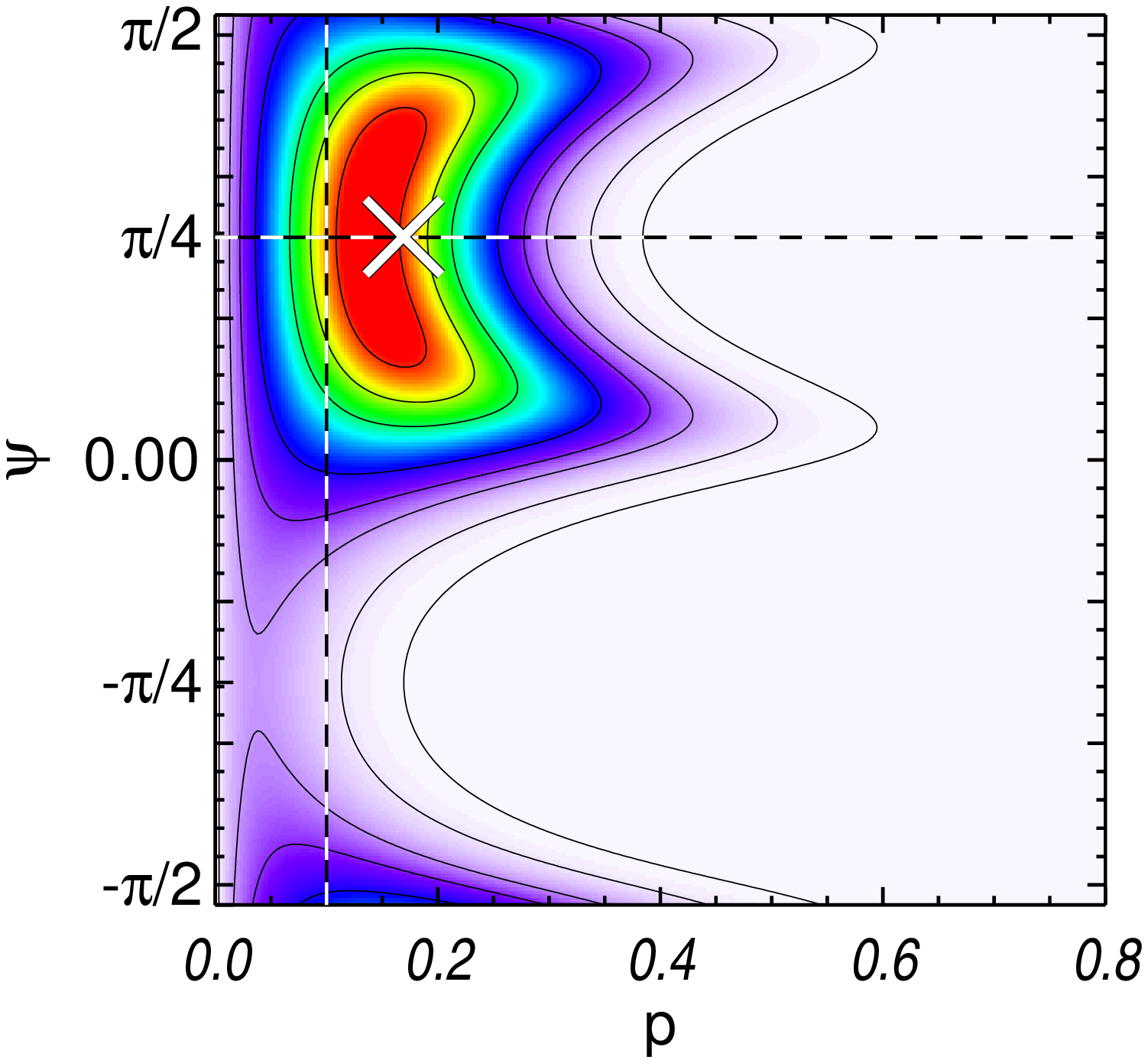}
\end{minipage}&
\begin{minipage}[c]{.15\linewidth}
 \includegraphics[width=3cm, viewport=200 0 600 400]
 {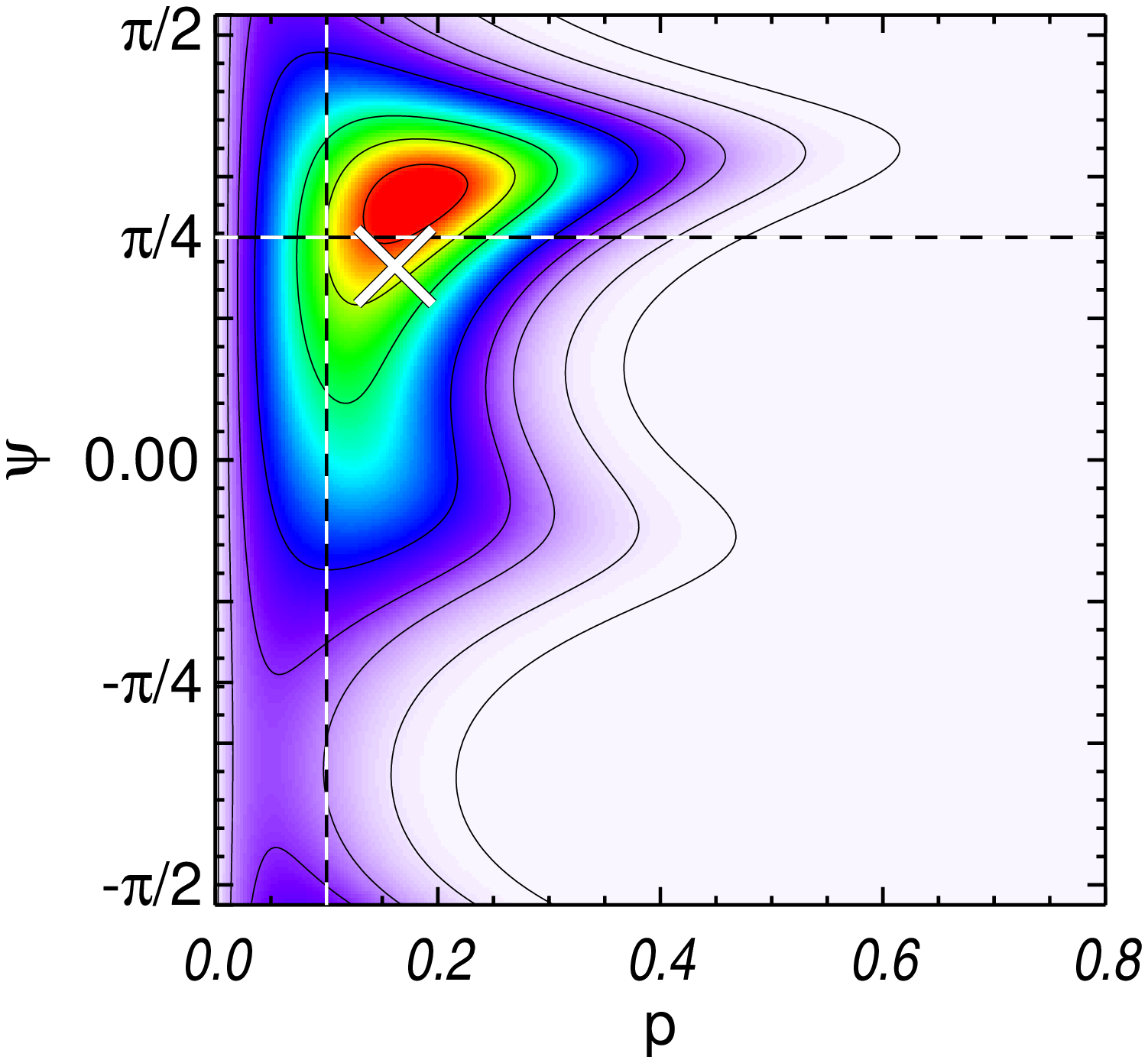}
\end{minipage}&
\begin{minipage}[c]{.15\linewidth}
 \includegraphics[width=3cm, viewport=200 0 600 400]
 {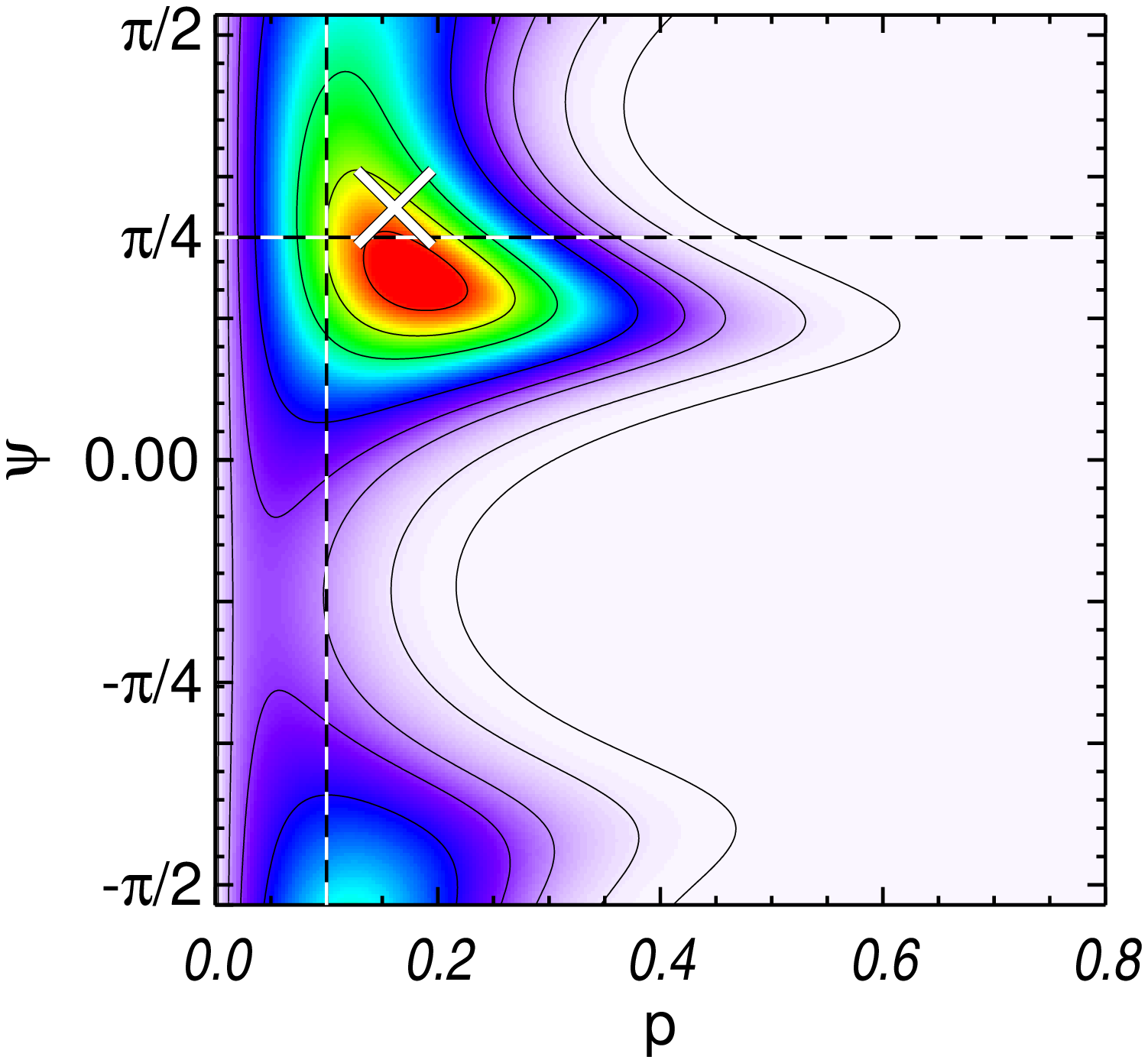}
\end{minipage} \\

\begin{minipage}[c]{.15\linewidth}
 \begin{tabular}{l}
 $\psi_0=\pi/8$ 
 \end{tabular} 
\end{minipage} &
\begin{minipage}[c]{.15\linewidth}
 \includegraphics[width=3cm, viewport=200 0 600 400]
 {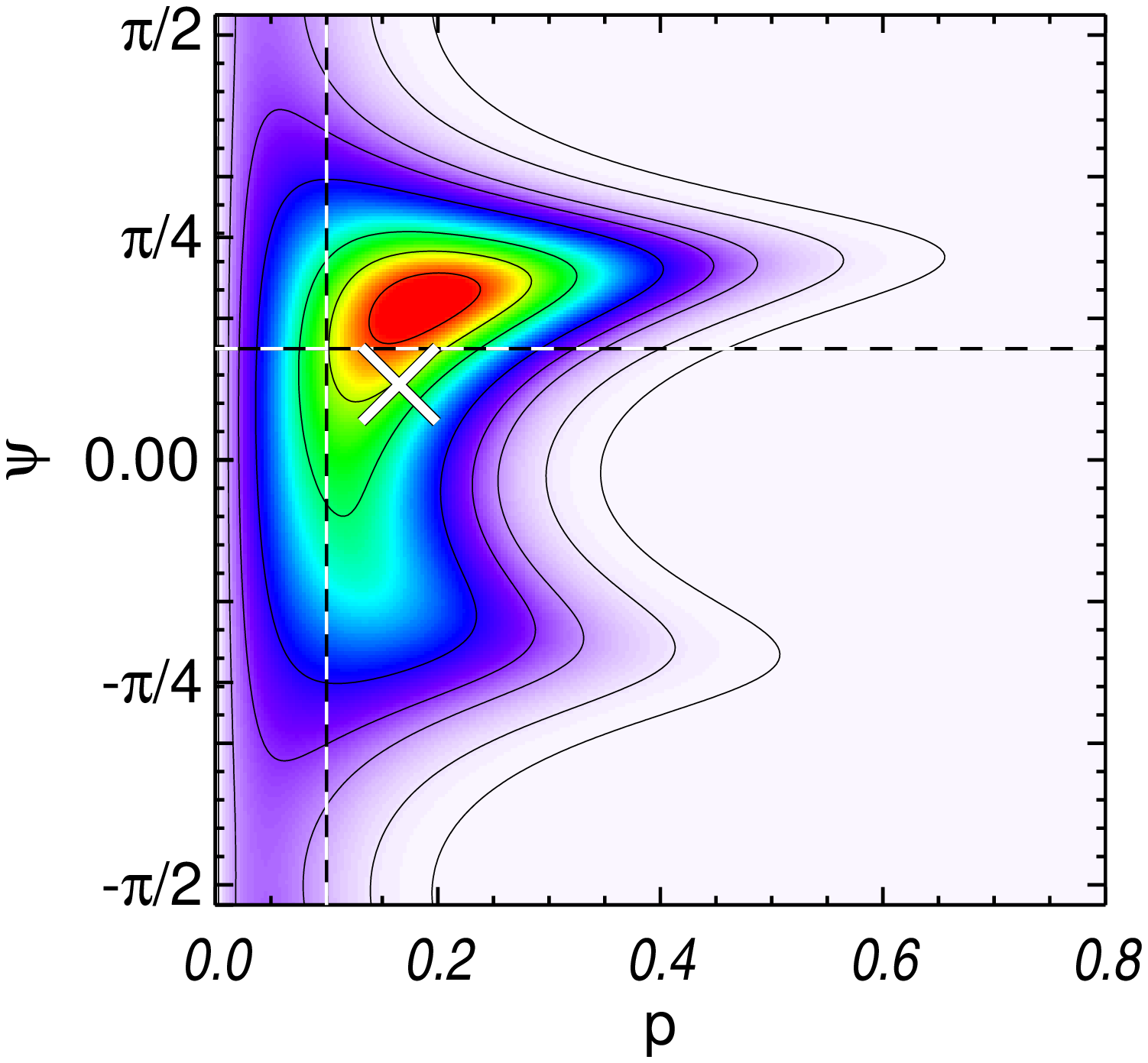}
\end{minipage}&
\begin{minipage}[c]{.15\linewidth}
 \includegraphics[width=3cm, viewport=200 0 600 400]
 {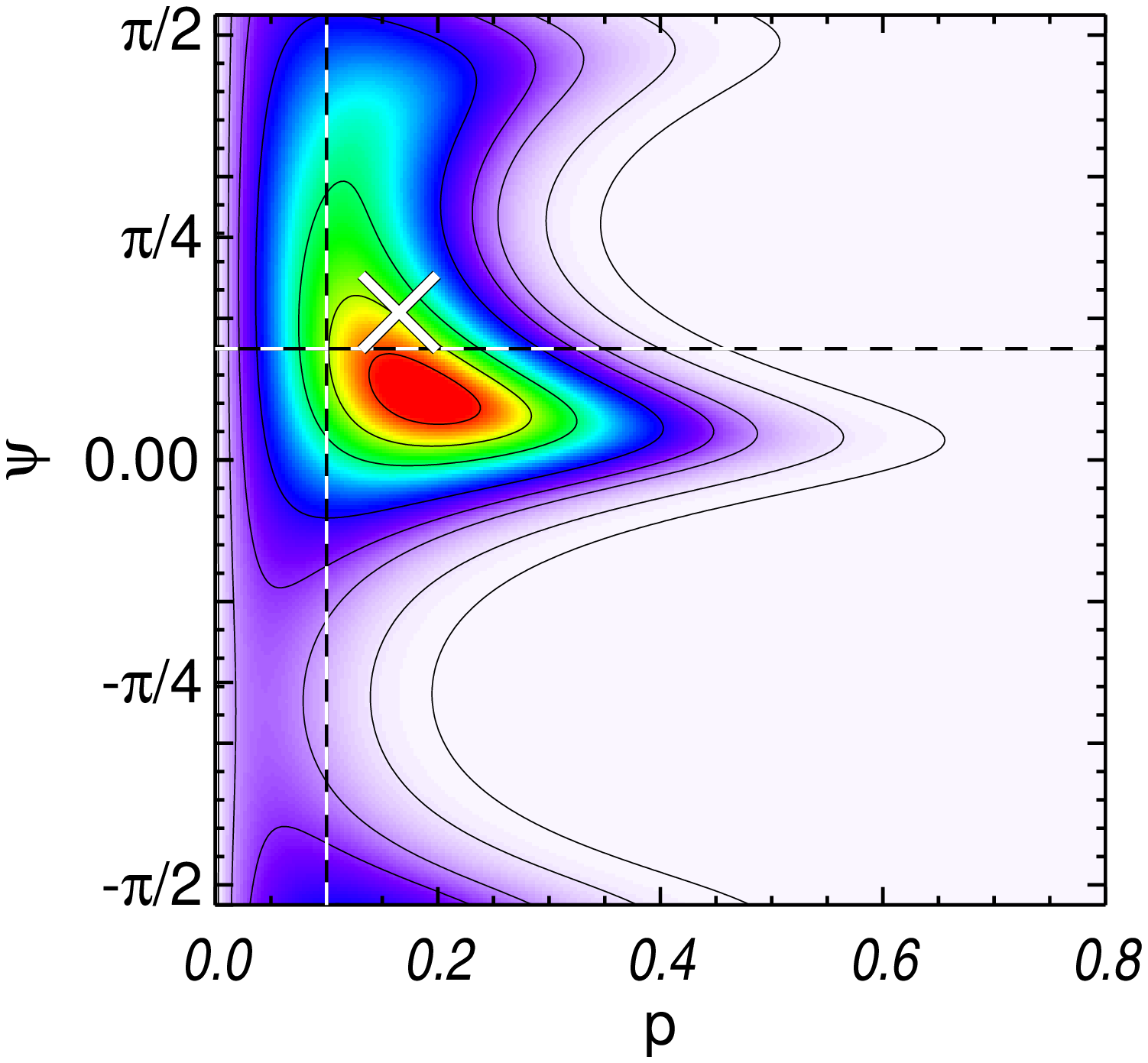}
\end{minipage}&
\begin{minipage}[c]{.15\linewidth}
 \includegraphics[width=3cm, viewport=200 0 600 400]
 {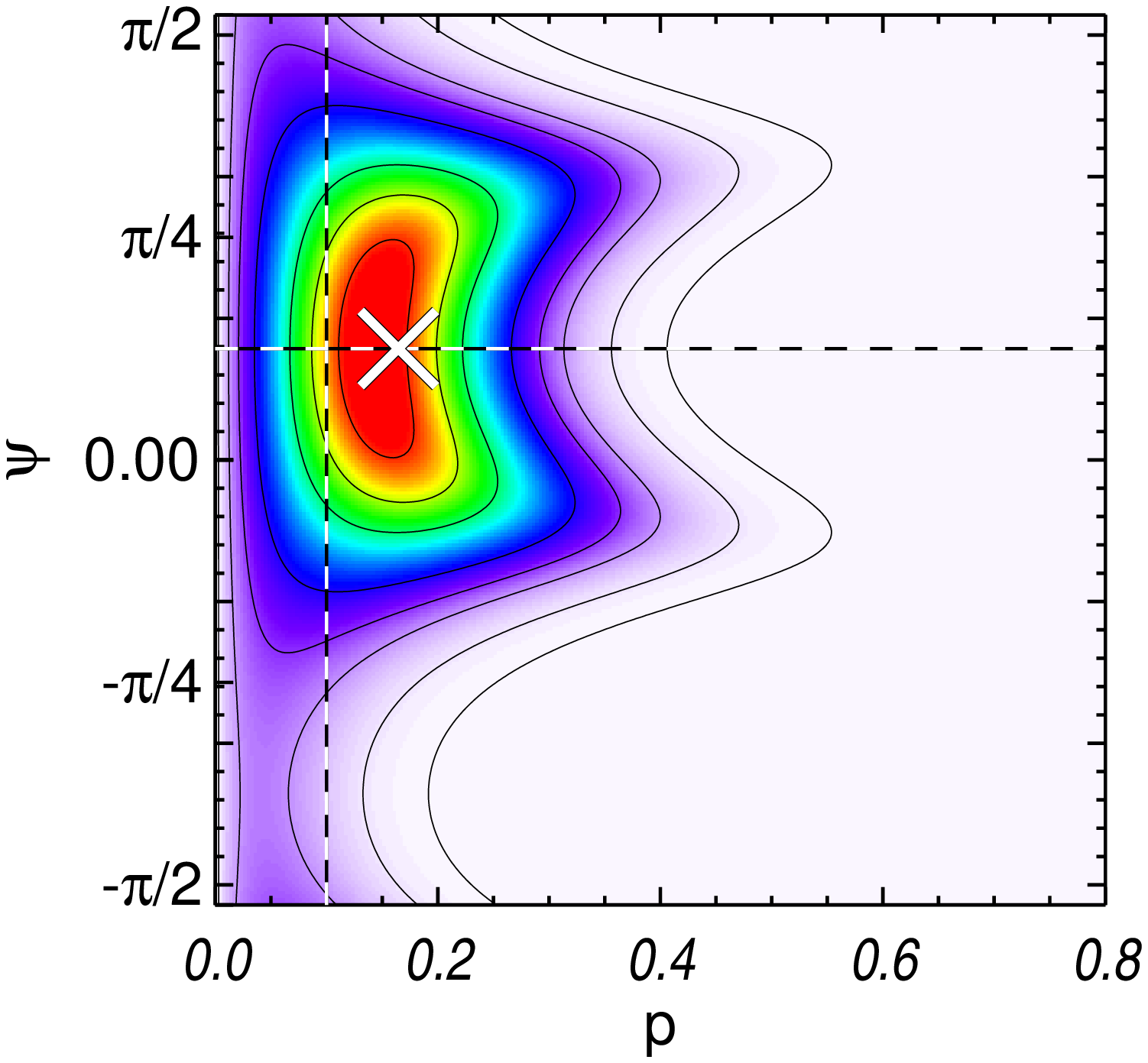}
\end{minipage}&
\begin{minipage}[c]{.15\linewidth}
 \includegraphics[width=3cm, viewport=200 0 600 400]
 {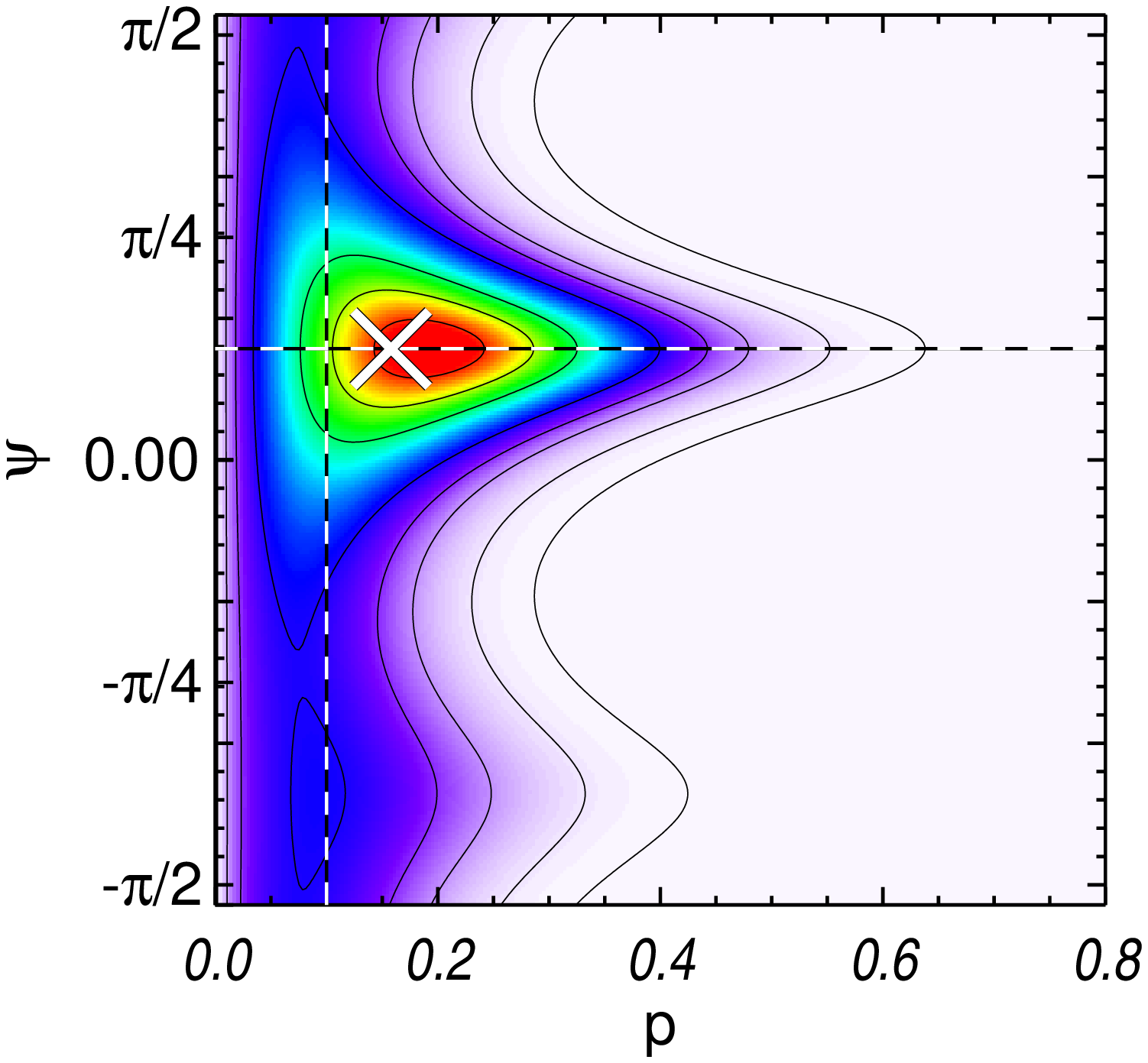}
\end{minipage} \\

\begin{minipage}[c]{.15\linewidth}
 \begin{tabular}{l}
 $\psi_0=0$ 
 \end{tabular} 
\end{minipage} &
\begin{minipage}[c]{.15\linewidth}
 \includegraphics[width=3cm, viewport=200 0 600 400]
 {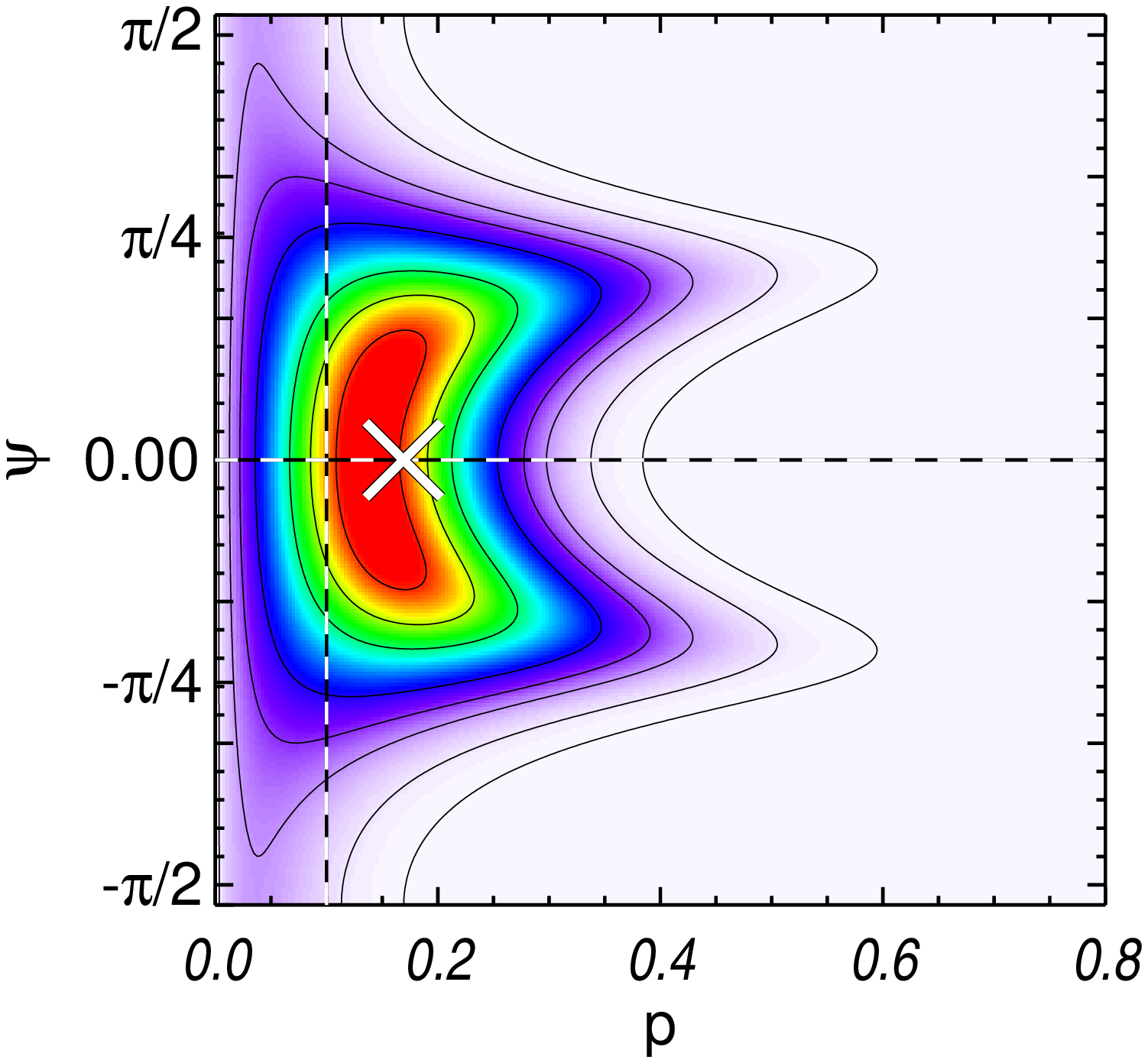}
\end{minipage}&
\begin{minipage}[c]{.15\linewidth}
 \includegraphics[width=3cm, viewport=200 0 600 400]
 {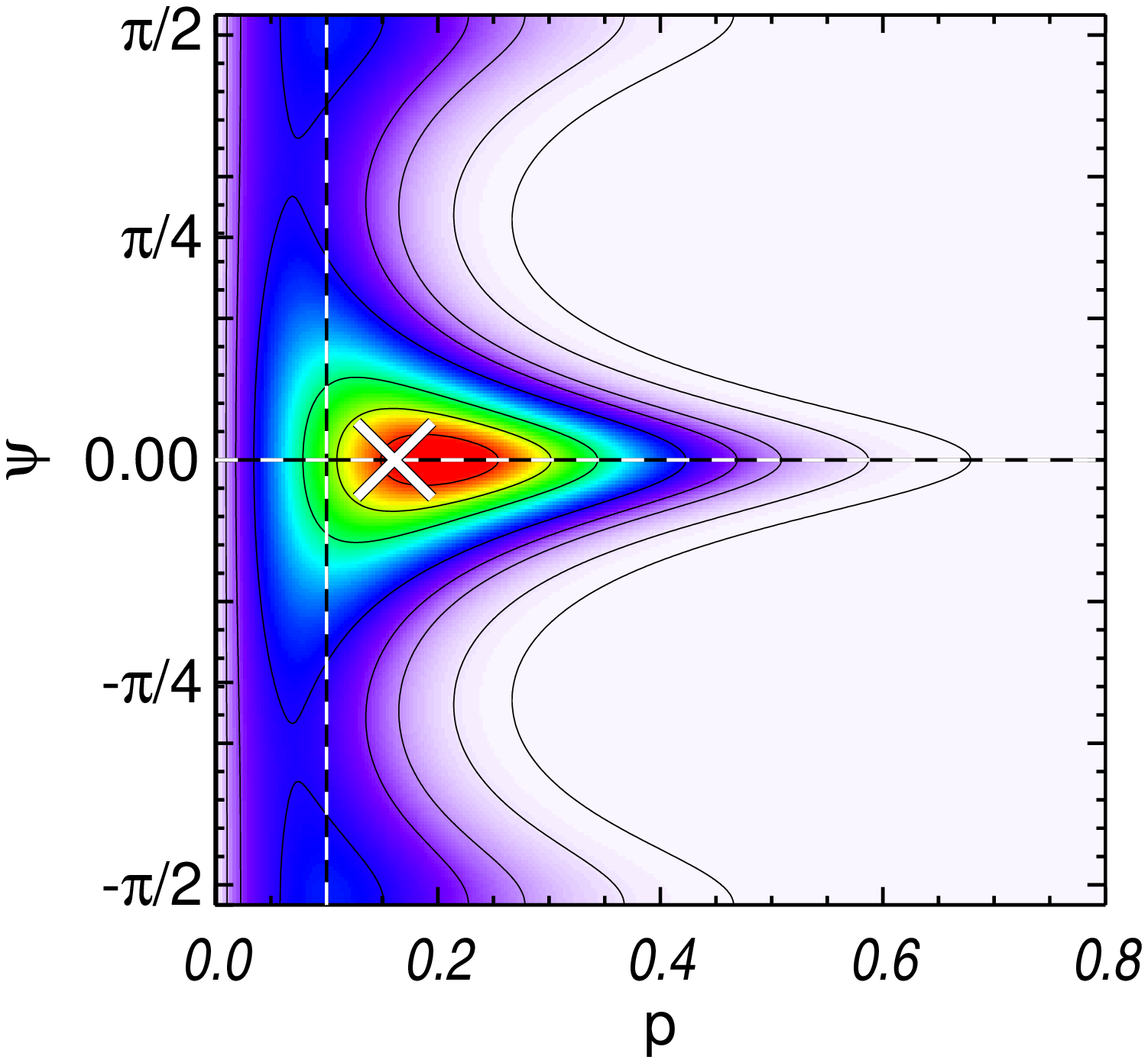}
\end{minipage}&
\begin{minipage}[c]{.15\linewidth}
 \includegraphics[width=3cm, viewport=200 0 600 400]
 {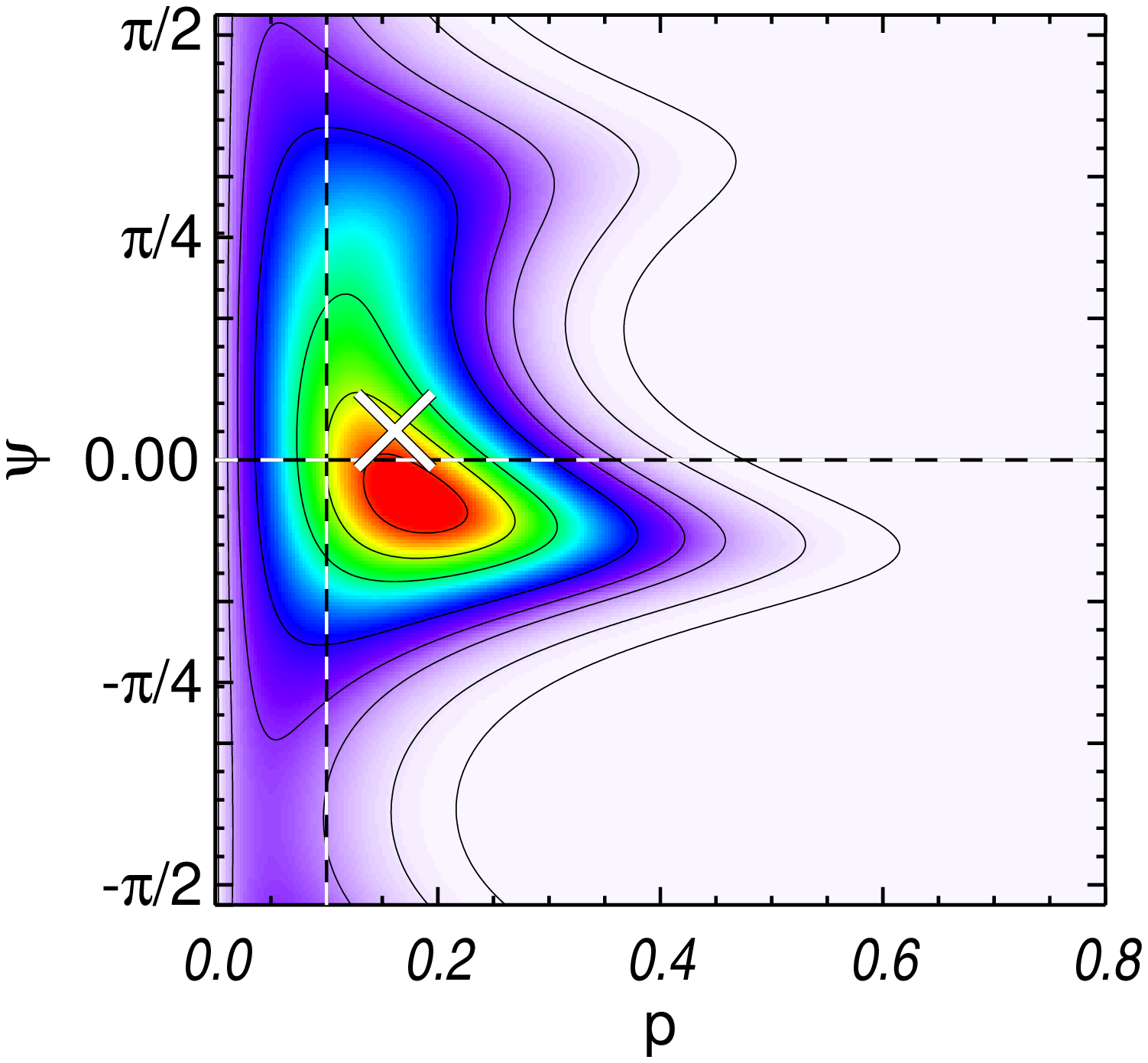}
\end{minipage}&
\begin{minipage}[c]{.15\linewidth}
 \includegraphics[width=3cm, viewport=200 0 600 400]
 {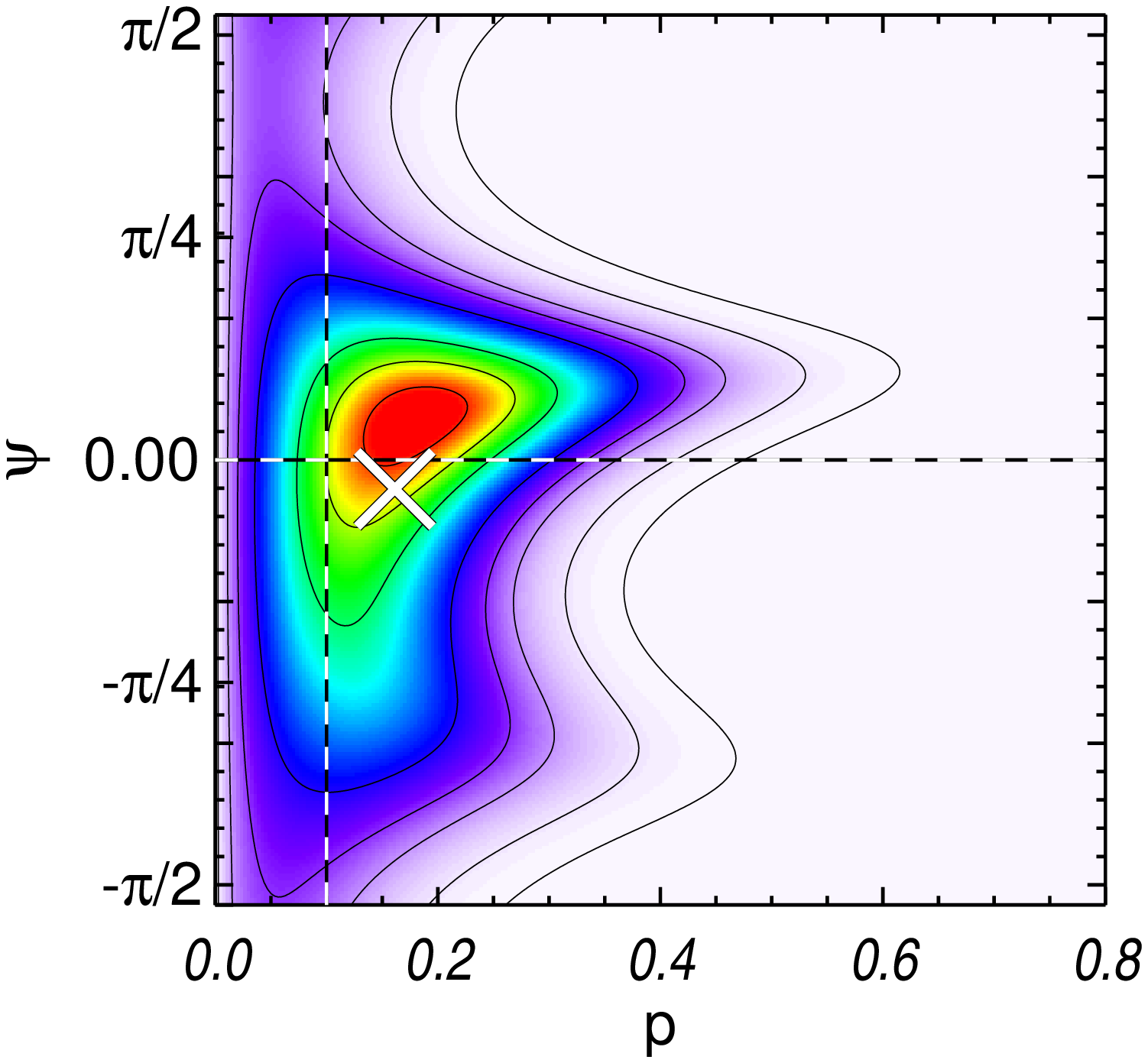}
\end{minipage} \\

\begin{minipage}[c]{.15\linewidth}
 \begin{tabular}{l}
 $\psi_0=-\pi/8$ 
 \end{tabular} 
\end{minipage} &
\begin{minipage}[c]{.15\linewidth}
 \includegraphics[width=3cm, viewport=200 0 600 400]
 {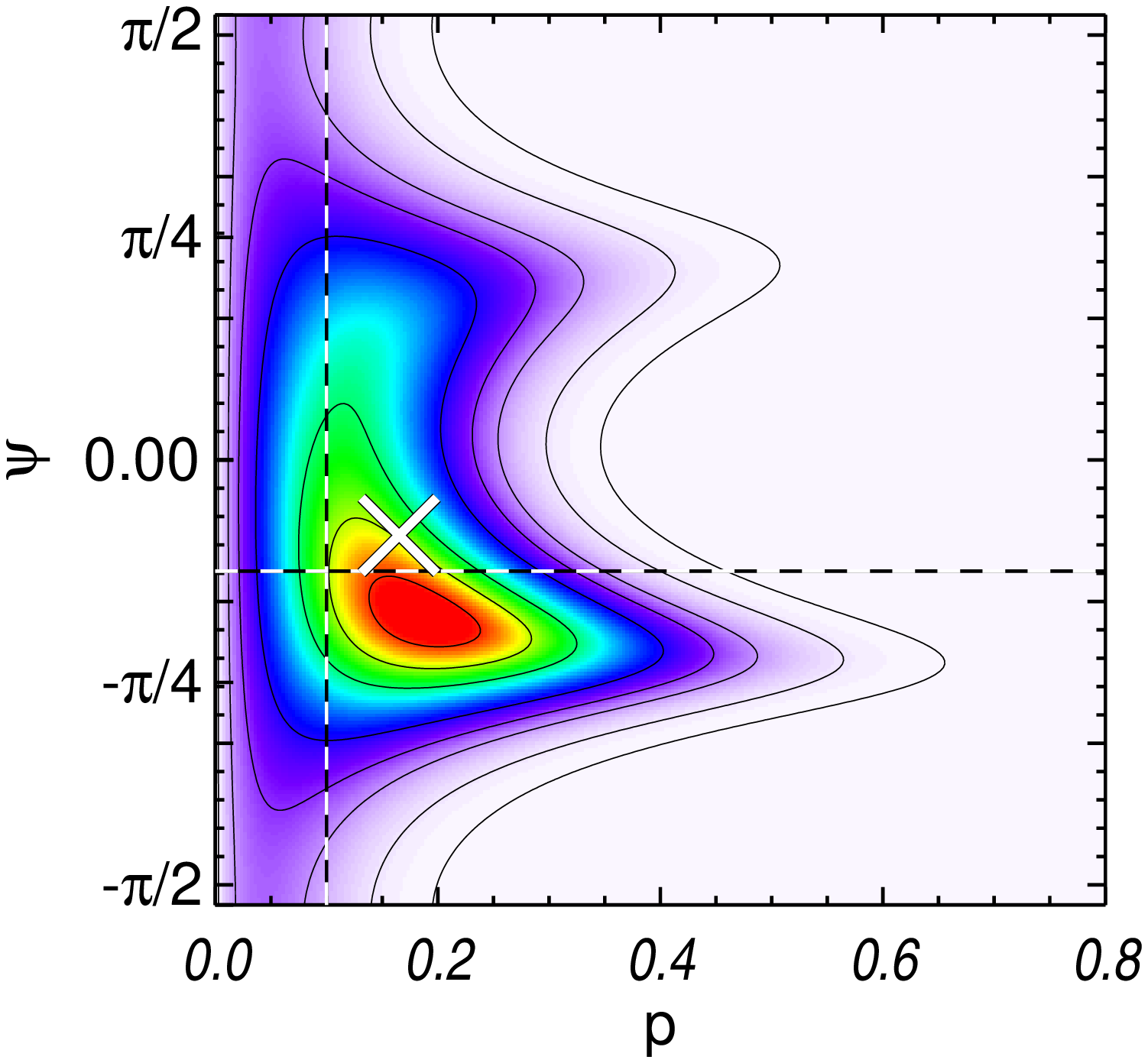}
\end{minipage}&
\begin{minipage}[c]{.15\linewidth}
 \includegraphics[width=3cm, viewport=200 0 600 400]
 {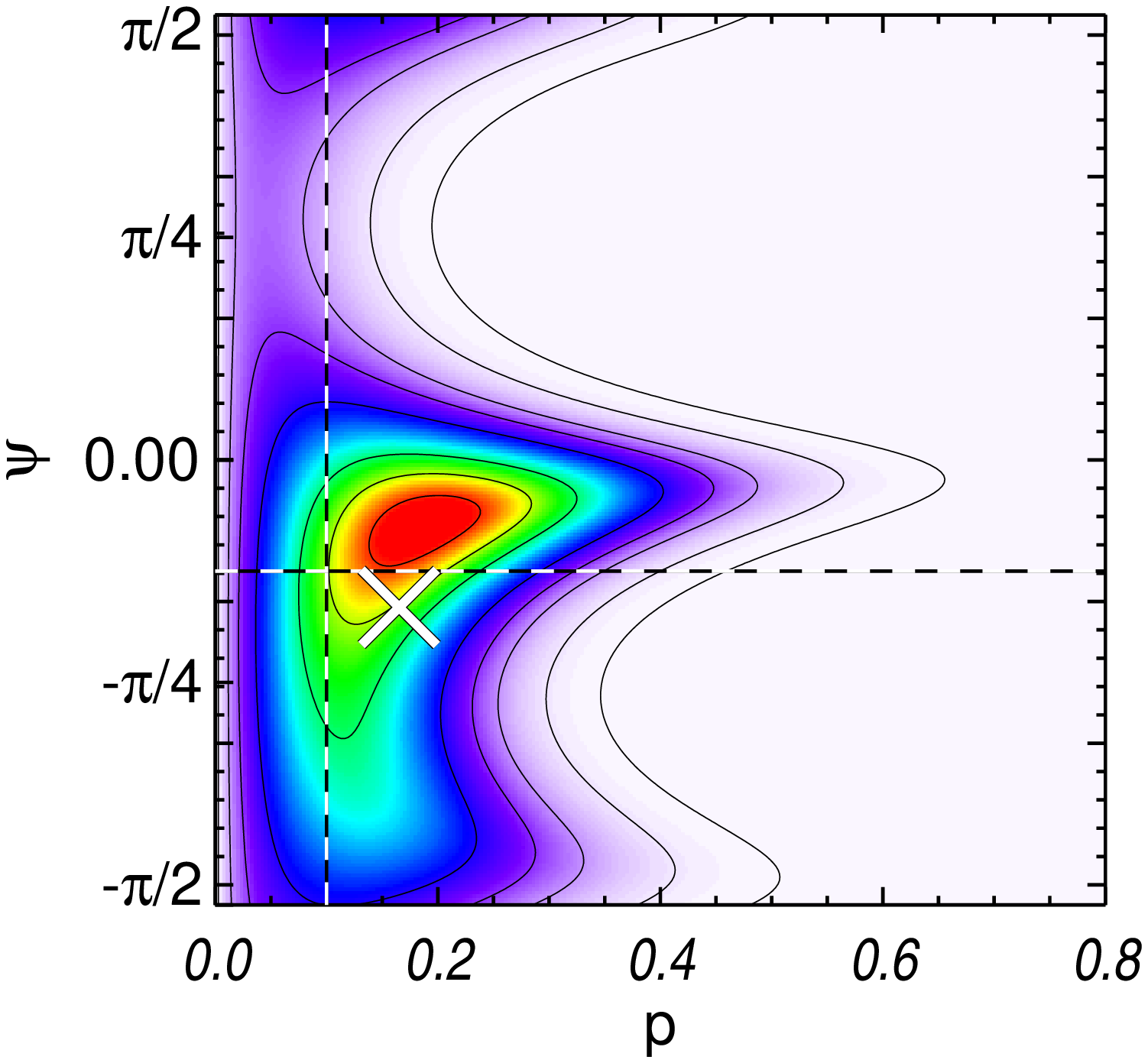}
\end{minipage}&
\begin{minipage}[c]{.15\linewidth}
 \includegraphics[width=3cm, viewport=200 0 600 400]
 {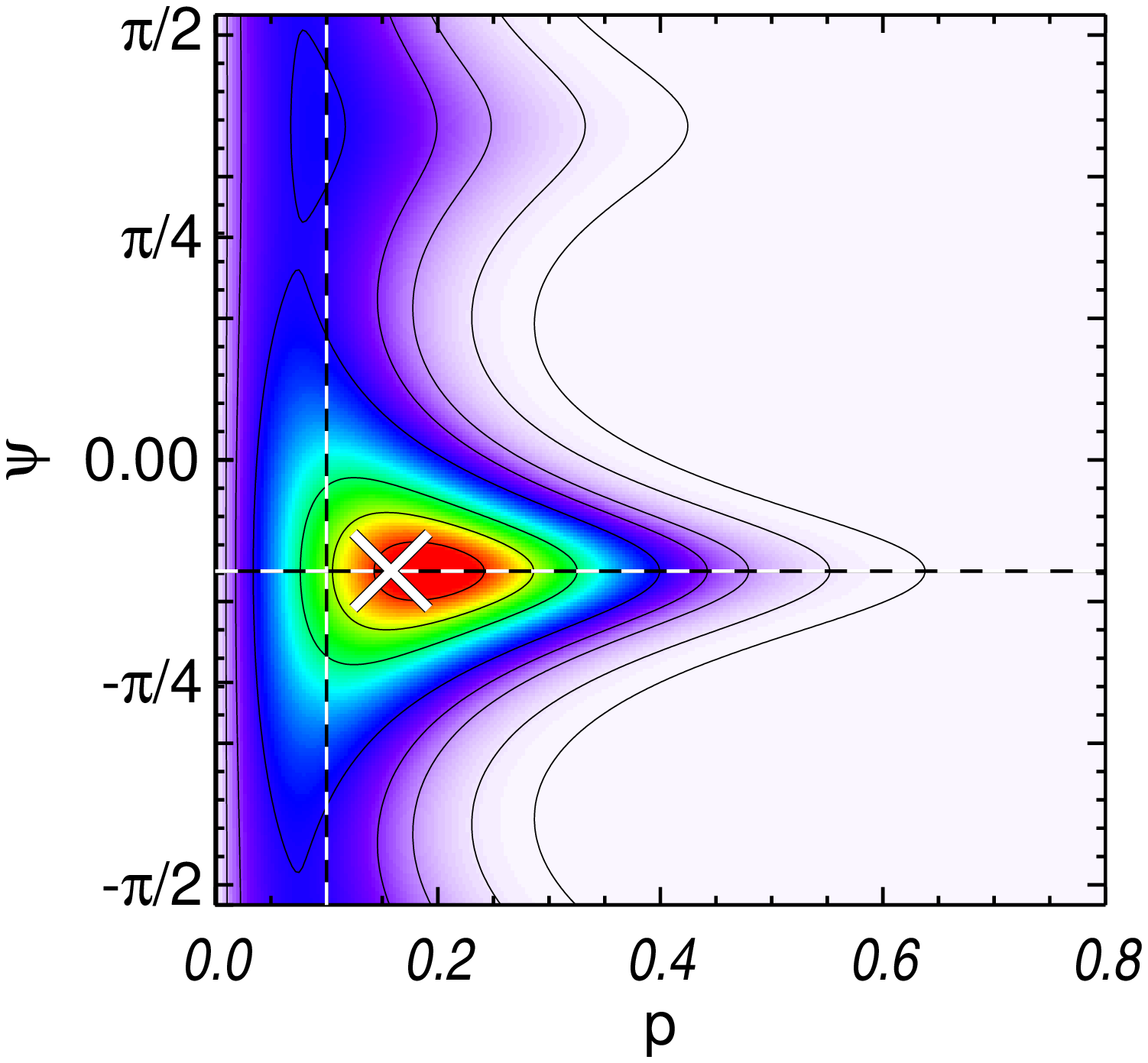}
\end{minipage}&
\begin{minipage}[c]{.15\linewidth}
 \includegraphics[width=3cm, viewport=200 0 600 400]
 {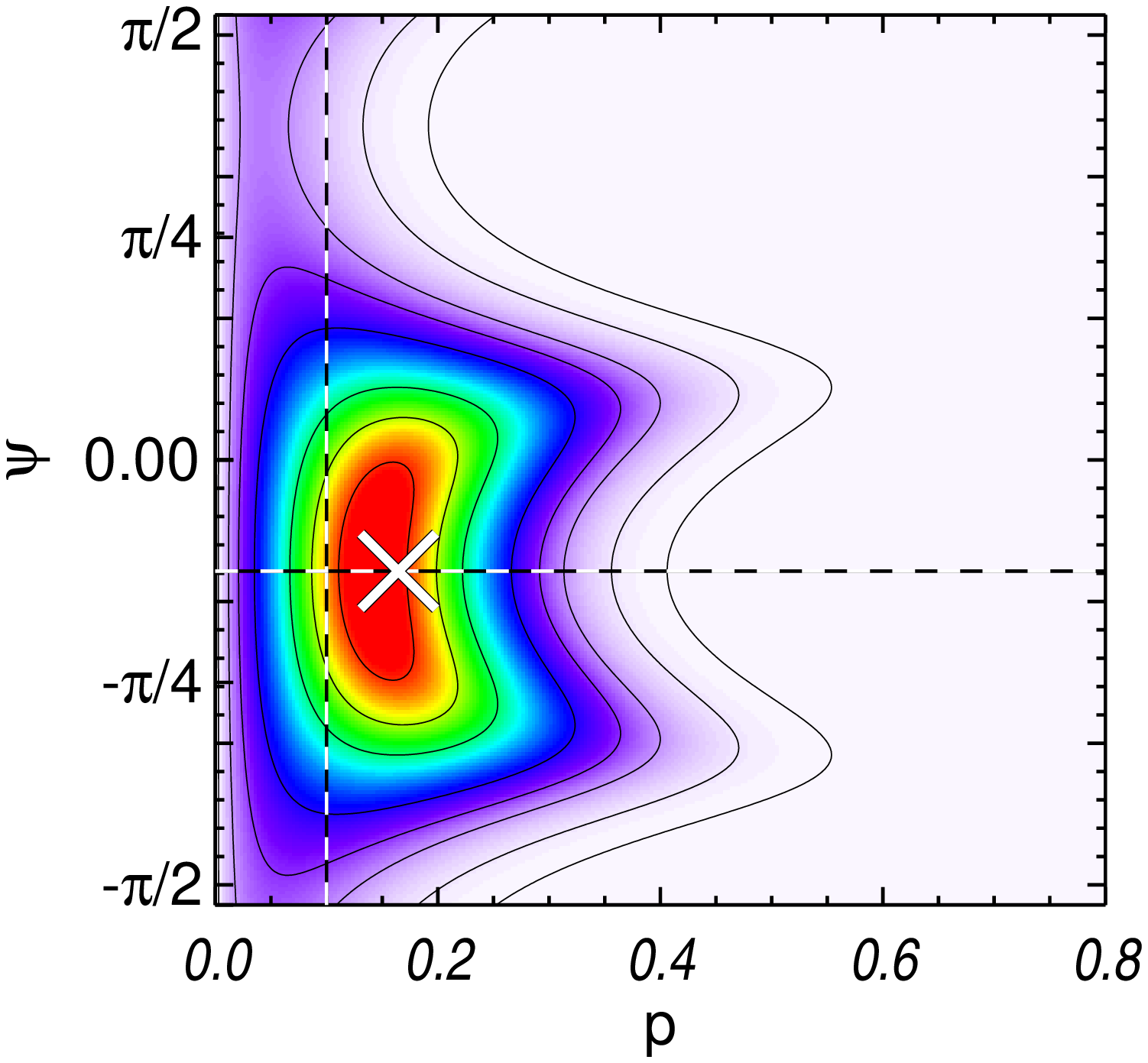}
\end{minipage} \\

 \end{tabular} 
\caption{Probability density functions,
$f_{\rm 2D}(p,\psi\,|\,p_0,\psi_0,\tens{\Sigma}_{\rm p})$, plotted for
various values of $\psi_0$ (rows), spanning from $-\pi/8$ to $3\pi/8$,
and computed for four configurations of the covariance matrix (columns),
parameterized by $\varepsilon$ and $\rho$. 
The signal-to-noise on the intensity $I$ is assumed to be infinite here.
A true value of polarization $p_0\,{=}\,0.1$ has been chosen, and with S/N
$p_0/\sigma_{\rm p,G}$=1.  White crosses indicate 
the mean likelihood estimates of the PDF ($\overline{p},\overline{\psi}$).
The contour levels are provided at 0.1, 1, 5, 10, 20, 50, 70, and 90\%
of the maximum of the distribution.}
\label{fig:pdf_impact_psi}
\end{figure*}

In the usual canonical case, $\varepsilon\,{=}\,1$ and $\rho\,{=}\,0$,
the PDF remains strictly symmetric whatever the value of the initial true
polarization angle $\psi_0$.  However, when changing the true polarization
angle $\psi_0$, as shown in Fig.~\ref{fig:pdf_impact_psi}, 
the PDF may become asymmetrical once the ellipticity $\varepsilon\,{\neq}\,1$
or the correlation $\rho\,{\neq}\,0$.
This will induce a statistical bias in the measurement of the polarization
angle $\psi$, which could be positive or negative depending on the covariance
matrix and the true value $\psi_0$, as discussed in
Sect.~\ref{sec:impact_fullcov}.

Examples of 2D probability density functions,
$f_{\rm 2D}(p,\psi\,|\,I_0,p_0,\psi_0,\tens{\Sigma})$, for 
finite values of $I_0/\sigma_{\rm I}$ (1, 2, and 5), and various
$\varepsilon$ and $\rho$ situations,
are shown in Fig.~\ref{fig:pdf_impact_finitesnri}, for the case
$\rho_{\rm Q}\,{=}\,\rho_{\rm U}\,{=}\,0$.
The true polarization parameters are $p_0\,{=}\,0.1$ and 
$\psi_0\,{=}\,0^\circ$, and the polarization signal-to-noise ratio is set to
$p_0/\sigma_{\rm p,G}\,{=}\,1$, so these plots may be directly compared to
the third row of Fig.~\ref{fig:pdf_impact_epsirho}. 
The effect of varying $I_0/\sigma_{\rm I}$ on the global shape of the PDF
seems rather small, but the position of the maximum likelihood in
$(p,\psi)$ is noticeably changed to lower values of $p$ when
$I_0/\sigma_{\rm I}\,{\lesssim}\,2$,
while the mean likelihood appears to be increased.

\section{General PDF of $p$ and $\psi$}
\label{sec:f_1d_full}

In the context of communication network science
\citet{Aalo2007} derived full expressions for the probability density
functions of envelope and phase quantities
in the general case. These expressions can be directly translated 
to express the PDF of the polarization fraction and angle, $p$ and $\psi$. 

We can apply the rotation of the covariance introduced in
Sect.~\ref{sec:notations} by an angle $\theta$, given by Eq.~\ref{eq:theta},
to remove the correlation term between the Stokes parameters. We define the
mean and the variance of the normalized Stokes parameters in this new frame by
\begin{equation}
\mu_1 = p_0 \cos (2\psi_0 - \theta), \quad
\mu_2 = p_0 \sin (2\psi_0 - \theta)
\end{equation}
and 
\begin{equation}
\sigma_1^2 = ( \sigma_{\rm Q}^2\cos^2\theta + \sigma_{\rm U}^2 \sin^2\theta
 + \rho\sigma_{\rm Q}\sigma_{\rm U}\sin2\theta ) \, / \, I_0^2, \quad
\sigma_2^2 = ( \sigma_{\rm Q}^2\sin^2\theta + \sigma_{\rm U}^2 \cos^2\theta
 - \rho\sigma_{\rm Q}\sigma_{\rm U}\sin 2\theta ) \, / \, I_0^2.
\end{equation}
The probability density function of $p$ is now written as
\begin{align}
f_{\rm p}(p\,|\,p_0,\psi_0,\tens{\Sigma}_{\rm p})
 = \frac{p}{2\sigma_1\sigma_2} \exp
 \left\{-\frac{1}{2}\left[\frac{\mu_1^2}{\sigma_1^2}
 + \frac{\mu_2^2}{\sigma_2^2} + \frac{p^2}{2}
 \left(\frac{1}{\sigma_1^2} + \frac{1}{\sigma_2^2} \right) \right] \right\}
\qquad\qquad\qquad\qquad\qquad\qquad\qquad\qquad\qquad\qquad\qquad\qquad
 \nonumber \\
\qquad\qquad\qquad\qquad
 \times \sum\limits_{n=0}^{\infty}
 \frac{ \zeta_n\mathcal{I}_n\left( \frac{p^2}{4}\left( \frac{1}{\sigma_2^2}
 - \frac{1}{\sigma_1^2} \right) \right)} 
{\left[\left(\frac{\mu_1}{\sigma_1^2}\right)^2
 + \left( \frac{\mu_2}{\sigma_2^2}\right)^2 \right]^n}
 \scalebox{1.7}{\Bigg\{} \mathcal{I}_{2n}
 \left(p\sqrt{\left(\frac{\mu_1}{\sigma_1^2}\right)^2
 +\left(\frac{\mu_2}{\sigma_2^2}\right)^2}\,\right) 
 \sum\limits_{k=0}^{n}\delta_kC_k^n \left[\left(
 \frac{\mu_1}{\sigma_1^2}\right)^2\!-\left(\frac{\mu_2}{\sigma_2^2}\right)^2
 \right]^{n-k} \left(2\frac{\mu_1\mu_2}{\sigma_1^2\sigma_2^2}\right)^k
 \scalebox{1.7}{\Bigg\}},
\end{align}
with $\mathcal{I}_n$ the $n$th order modified Bessel function of the first
kind.  Here $\zeta_0\,{=}\,1$ and $\zeta_n\,{=}\,2$ for $n\,{\neq}\,0$,
$C_k^n\,{\equiv}\,n!/k!(n-k)!$ are binomial coefficients,
and $\delta_k$ is defined by
\begin{equation}
\delta_k = \Bigg\{ \begin{array}{cl} 0 & \mathrm{for}\ k\ \mathrm{odd}, \\
 2\, (-1)^{k/2} & \mathrm{for}\ k\ \mathrm{even}. \end{array}
\end{equation}

It should be noted that the above expression converges so fast that only a
few terms of the infinite sum are required to obtain
sufficient accuracy.  On the other hand, the probability density function
of the polarization angle is given by
\begin{align}
f_{\psi}(\psi\,|\,p_0,\psi_0,\tens{\Sigma}_{\rm p})
 = \exp\left[-\frac{1}{1-\rho^2} \left(\frac{Q_0^2}{2\sigma_{\rm Q}^2}
 + \frac{U_0^2}{2\sigma_{\rm U}^2}
 - \frac{\rho Q_0 U_0}{\sigma_{\rm Q}\sigma_{\rm U}} \right) \right]
\qquad\qquad\qquad\qquad\qquad\qquad \nonumber \\
\qquad\qquad\qquad\qquad\qquad\qquad
\times\,\frac{\sqrt{1-\rho^2}}{\pi\sigma_{\rm Q}\sigma_{\rm U}\mathcal{A}(\psi)}
 \left\{1 + \frac{\sqrt{\pi}\mathcal{B}(\psi)}{\sqrt{\mathcal{A}(\psi)}}
 \exp\left[ \frac{\mathcal{B}^2(\psi)}{\mathcal{A}(\psi)} \right]
 \mathrm{erfc} \left[ -\frac{\mathcal{B}(\psi)}{\sqrt{\mathcal{A}(\psi)}}
 \right] \right\},
\end{align}
where
\begin{equation}
\mathcal{A}(\psi) = \frac{2 \cos^2 2\psi}{\sigma_{\rm Q}^2}
 + \frac{2\sin^2 2\psi}{\sigma_{\rm U}^2}
 - 4\frac{\rho \sin 2\psi \cos 2\psi}{\sigma_{\rm Q}\sigma_{\rm U}},
\end{equation}
\begin{equation}
\mathcal{B}(\psi) = \frac{1}{\sqrt{1-\rho^2}}
 \left[ \frac{\cos 2\psi}{\sigma_{\rm Q}}
 \left(\frac{Q_0}{\sigma_{\rm Q}} - \frac{\rho U_0}{\sigma_{\rm U}} \right)
 + \frac{\sin 2\psi}{\sigma_{\rm U}}
 \left( \frac{U_0}{\sigma_{\rm U}} - \frac{\rho Q_0}{\sigma_{\rm Q}} \right)
 \right],
\end{equation}
and
\begin{equation}
 \mathrm{erfc}(z) = \frac{2}{\sqrt{\pi}}
 \int\limits_z^{\infty} \exp [-x^2] dx
\end{equation}
is the complementary error function.

\section{Impact of $\rho_{\rm Q}$ and $\rho_{\rm U}$ on $\varepsilon$ and $\rho$}
\label{sec:appendix-cholesky}

The covariance matrix $\tens{\Sigma}$ is positive definite, so may be written
as a Cholesky product $\tens{\Sigma}\,{=}\,\tens{L}^{\rm T}\tens{L}$, with
\begin{equation}
\tens{L}=\left(\begin{array}{ccc}
L_{11} & 0 & 0 \\
L_{12} & L_{22} & 0 \\
L_{13} & L_{23} & L_{33} \\
\end{array}\right).
\end{equation}
The six $L_{ij}$ are independent, unlike the six parameters of the covariance
matrix, $(\sigma_{\rm I},\sigma_{\rm Q},\sigma_{\rm U},\rho,\rho_{\rm Q},
\rho_{\rm U})$, or the parameters that we use in this paper,
$(\sigma_{\rm I},\sigma_{\rm Q},\varepsilon,\rho,\rho_{\rm Q},\rho_{\rm U})$.
In the general case, these are given in terms of the $L_{ij}$ as
(assuming $I_0$=1)
\begin{align}
\rho=\frac{L_{12}L_{13}+L_{22}L_{23}} {\sqrt{\left(L_{12}^2+L_{22}^2\right)
 \left(L_{13}^2+L_{23}^2+L_{33}^2\right)}}, \quad
\varepsilon=\sqrt{\frac{L_{13}^2+L_{23}^2+L_{33}^2}{L_{12}^2+L_{22}^2}},
\nonumber \\
 \rho_{\rm Q}=\frac{L_{12}}{\sqrt{L_{12}^2+L_{22}^2}}, \quad {\rm and} \quad
 \rho_{\rm U}=\frac{L_{13}}{\sqrt{L_{13}^2+L_{23}^2+L_{33}^2}}.
\end{align}
When there is no correlation between $I$ and the $Q$ or $U$ components,
then $L_{12}\,{=}\,L_{13}$=0, which leads to the following system: 
\begin{equation}
\rho=\rho_0=\frac{L_{22}L_{23}}{\left|L_{22}\right|\sqrt{L_{23}^2+L_{33}^2}};
\quad
\varepsilon=\varepsilon_0=\frac{\sqrt{L_{23}^2+L_{33}^2}}{\left|L_{22}\right|}.
\end{equation}
The ellipticity and the correlation coefficient are therefore modified by the
presence of the correlation between $I$ and ($Q,U$).  A little algebra
leads to expressions for $\varepsilon$ and $\rho$ as functions of
$\varepsilon_0$, $\rho_0$, $\rho_{\rm Q}$, and $\rho_{\rm U}$, namely
\begin{equation}
\varepsilon=\varepsilon_0\sqrt{\frac{1-\rho_{\rm Q}^2}{1-\rho_{\rm U}^2}}
 \qquad {\rm and} \qquad
 \rho=\rho_{\rm Q}\rho_{\rm U}+\rho_0
 \sqrt{\left(1-\rho_{\rm Q}^2\right)\left(1-\rho_{\rm U}^2\right)},
\end{equation}
which are Eqs.~\ref{eq:rhouq_epsirho}.

\section{Derivation of classical uncertainties}
\label{sec:classical_uncertainties}

We describe here how the expressions for the classical uncertainties of $p$
and $\psi$, introduced in Sect.~\ref{sec:classical_estimate},
are obtained from the derivatives of $p$ and $\psi$.
We first note that we generally have
\begin{equation}
\sigma^{2}_{X}=E\Big[(X-E[X])^{2}\Big]=E\Big[(dX)^{2}\Big],
\label{edx}
\end{equation}
where $dX=X-E[X]$ is an infinitesimal element.

The classical uncertainty of $p$ can therefore be given by the expression
$\sigma^{2}_{\rm p,C}\,{=}\,E\left[(dp)^{2}\right]$.
Using the expression for $p$ we obtain
\begin{align}
(dp)^2 & = \left( \dfrac{\partial p}{\partial Q}dQ +\dfrac{\partial p}
 {\partial U}dU+\dfrac{\partial p}{\partial I}dI \right)^2 \nonumber \\
 & = \left(\dfrac{\partial p}{\partial Q}\right)^{2}(dQ)^{2}
 +\left(\dfrac{\partial p}{\partial U}\right)^{2}(dU)^{2}
 +\left(\dfrac{\partial p}{\partial I}\right)^{2}(dI)^{2}
 +\, 2\dfrac{\partial p}{\partial Q}
 \dfrac{\partial p}{\partial U}dQdU+2\dfrac{\partial p}{\partial Q}
 \dfrac{\partial p}{\partial I}dQdI+2\dfrac{\partial p}{\partial U}
 \dfrac{\partial p}{\partial I}dUdI,
\end{align}
where the partial derivatives are
\begin{equation}
\dfrac{\partial p}{\partial Q} = \dfrac{1}{2}
 \dfrac{2Q}{I\sqrt{Q^{2}+U^{2}}}=\dfrac{Q}{pI^{2}}, \qquad
\dfrac{\partial p}{\partial U} = \dfrac{1}{2}
 \dfrac{2U}{I\sqrt{Q^{2}+U^{2}}}= \dfrac{U}{pI^{2}}, \qquad {\rm and} \qquad
\dfrac{\partial p}{\partial I} = -\dfrac{\sqrt{Q^{2}+U^{2}}}{I^{2}}
 =-\dfrac{p}{I}.
\end{equation}
This leads to the following expression for the classical uncertainty:
\begin{align}
\sigma^{2}_{\rm p,C} &=\dfrac{1}{p^{2}I^{4}}\,
 E\Big[ Q^{2}(dQ)^{2}+U^{2}(dU)^{2}+p^{4}I^{2}(dI)^{2}
 +2QUdQdU-2QIp^{2}dQdI-2UIp^{2}dUdI \Big] \nonumber \\
 &= \dfrac{1}{p^{2}I^{4}}\Big( Q^{2}E\left[ (Q-E[Q])^{2}\right]
 + U^{2}E\left[ (U-E[U])^{2}\right]
 + p^{4}I^{2}E\left[ (I-E[I])^{2}\right] \nonumber \\
 & \qquad + 2QUE\Big[ (Q-E[Q]) (U-E[U]) \Big]
 - 2QIp^{2}E\Big[ (Q-E[Q]) (I-E[I]) \Big]
 - 2UIp^{2}E\Big[ (U-E[U]) (I-E[I]) \Big] \Big\}.
\end{align}
This finally leads to
\begin{equation}
 \sigma_{\rm p,C}^2 = \frac{1}{p^2 I^4}\,
 \Big( Q^2\sigma_{\rm Q}^2+U^2\sigma_{\rm U}^2+p^4I^2\sigma_{\rm I}^2
 +2 QU\sigma_{\rm QU}-2IQp^2\sigma_{\rm IQ}-2IUp^2\sigma_{\rm IU} \Big).
\label{eq:sigp_classical}
\end{equation}

Similarly we can derive an expression for the non-classical uncertainty of
the polarization angle, $\psi$, given by
$\sigma^{2}_{\psi,{\rm C}}\,{=}\,E\left[(d\psi)^{2}\right]$.
Using the expression of $\psi$, we obtain the partial derivatives
\begin{equation}
\dfrac{\partial \psi}{\partial U} = \frac{1}{2}\dfrac{Q}{Q^{2}+U^{2}}
\qquad {\rm and} \qquad
\dfrac{\partial \psi}{\partial Q} = - \frac{1}{2}\dfrac{U}{Q^{2}+U^{2}},
\end{equation}
as well as an expression for the classical $\psi$ uncertainty:
\begin{align}
 \sigma_{\psi,{\rm C}}^2 =
 E\left[\left(\dfrac{\partial\psi}{\partial U}dU
 +\dfrac{\partial\psi}{\partial Q}dQ\right)^2\right]
 = E \left[\left(\dfrac{QdU-UdQ}{2p^{2}I^{2}}\right)^2\right]
 &= E \left[ \dfrac{Q^{2}dU^{2}+U^{2}dQ^{2}-2QUdQdU}{4p^{4}I^{4}}\right]
 \nonumber \\
 &= \dfrac{Q^2\sigma_{\rm UU}+U^2\sigma_{\rm QQ}
 -2QU\sigma_{\rm QU}}{4p^{4}I^{4}}.
\label{eq:sigpsi_cla_1}
\end{align}
Using Eq.~\ref{eq:sigp_classical} and assuming
$\sigma_{\rm II}\,{=}\,\sigma_{\rm IQ}\,{=}\,\sigma_{\rm IU}\,{=}\,0$, 
we find
\begin{equation}
 p^2I^{4}=\dfrac{Q^2 \sigma_{\rm Q}^2+U^2\sigma_{\rm U}^2+2QU\sigma_{\rm QU}}
 {\sigma_{\rm p,C}^2},
\end{equation}
and replacing this expression in Eq.~\ref{eq:sigpsi_cla_1} finally leads to
\begin{equation}
 \sigma_{\psi,{\rm C}} = \sqrt{
 \dfrac{Q^{2}\sigma_{\rm U}^2+U^{2}\sigma_{\rm Q}^2-2QU\sigma_{\rm QU}}
 {Q^{2}\sigma_{\rm Q}^2+U^{2}\sigma_{\rm U}^2+2QU\sigma_{\rm QU}} }
 \times \dfrac{\sigma_{\rm p,C}}{2p}.
\end{equation}

Notice that the above two expressions for the classical estimates have been
obtained in the small-error limit, 
and therefore they are formally inapplicable to the large uncertainty
regime.  In Sect.~\ref{sec:sigma_p} we discuss the extent to which they can
provide reasonable proxies for the errors, even at low S/N.

\begin{figure*}[tp]
\hspace{-1cm}
\begin{tabular}{cccc}
&
\begin{minipage}[c]{.15\linewidth}
 \begin{tabular}{l}
 $I_0/\sigma_{\rm I}=1$ 
 \end{tabular} 
\end{minipage} &
\begin{minipage}[c]{.15\linewidth}
 \begin{tabular}{l}
 $I_0/\sigma_{\rm I}=2$ 
 \end{tabular} 
\end{minipage} &
\begin{minipage}[c]{.15\linewidth}
 \begin{tabular}{l}
 $I_0/\sigma_{\rm I}=3$ 
 \end{tabular} 
\end{minipage} \\ \\
\begin{minipage}[c]{.18\linewidth}
$\quad$
 \begin{tabular}{l}
 $\varepsilon = 1$ \\
 $\rho = 0$ \\
 \end{tabular} \\ 
$\Bigg($
 \begin{tabular}{l}
 $\varepsilon_{\rm eff} = 1$ \\
 $\theta = 0$ \\
 \end{tabular} 
$\Bigg)$
\end{minipage} &
\begin{minipage}[c]{.29\linewidth}
\end{minipage} & 
\begin{minipage}[c]{.29\linewidth}
\end{minipage} &
\begin{minipage}[c]{.29\linewidth}
\end{minipage} \\

\begin{minipage}[c]{.18\linewidth}
$\quad$
 \begin{tabular}{l}
 $\varepsilon = 1/2$ \\
 $\rho = 0$ \\
 \end{tabular} \\
$\Bigg($
 \begin{tabular}{l}
 $\varepsilon_{\rm eff} = 2$ \\
 $\theta = \pi $ \\
 \end{tabular} 
$\Bigg)$
\end{minipage} &
\begin{minipage}[c]{.29\linewidth}
\end{minipage} &
\begin{minipage}[c]{.29\linewidth}
\end{minipage} &
\begin{minipage}[c]{.29\linewidth}
\end{minipage} \\

\begin{minipage}[c]{.18\linewidth}
$\quad$
 \begin{tabular}{l}
 $\varepsilon = 2$ \\
 $\rho = 0$ \\
 \end{tabular} \\
$\Bigg($
 \begin{tabular}{l}
 $\varepsilon_{\rm eff} = 2$ \\
 $\theta = 0$ \\
 \end{tabular} 
$\Bigg)$
\end{minipage} &
\begin{minipage}[c]{.29\linewidth}
\end{minipage} &
\begin{minipage}[c]{.29\linewidth}
\end{minipage} &
\begin{minipage}[c]{.29\linewidth}
\end{minipage} \\

\begin{minipage}[c]{.18\linewidth}
$\quad$
 \begin{tabular}{l}
 $\varepsilon = 1$ \\
 $\rho = -1/2$ \\
 \end{tabular} \\
$\Bigg($
 \begin{tabular}{l}
 $\varepsilon_{\rm eff} \sim 1.73$ \\
 $\theta = -\pi/4$ \\
 \end{tabular} 
$\Bigg)$
\end{minipage} &
\begin{minipage}[c]{.29\linewidth}
\end{minipage} &
\begin{minipage}[c]{.29\linewidth}
\end{minipage} &
\begin{minipage}[c]{.29\linewidth}
\end{minipage} \\

\begin{minipage}[c]{.18\linewidth}
$\quad$
 \begin{tabular}{l}
 $\varepsilon = 1$ \\
 $\rho = 1/2$ \\
 \end{tabular} \\
$\Bigg($
 \begin{tabular}{l}
 $\varepsilon_{\rm eff} \sim 1.73$ \\
 $\theta = \pi/4$ \\
 \end{tabular} 
$\Bigg)$
\end{minipage} &
\begin{minipage}[c]{.29\linewidth}
\end{minipage} &
\begin{minipage}[c]{.29\linewidth}
\end{minipage} &
\begin{minipage}[c]{.29\linewidth}
\end{minipage} \\
\end{tabular} 
\caption{Probability density functions,
$f_{\rm 2D}(p,\psi\,|\,I_0,\,p_0,\psi_0,\tens{\Sigma})$, with finite S/N on
intensity, $I_0/\sigma_{\rm I}=1$, 2, and 5 (columns from left to right), 
computed for a given set of polarization parameters, $\psi_0\,{=}\,0^\circ$
and $p_0\,{=}\,0.1$ (dashed lines), and a signal-to-noise ratio on the
polarized intensity set to $p_0/\sigma_{\rm p,G}\,{=}\,1$.
Correlation coefficients $\rho_{\rm Q}$ and $\rho_{\rm U}$ are set to zero.
Various configurations of the covariance matrix are shown (rows).
White crosses indicate the mean likelihood estimates of the PDF
($\overline{p},\overline{\psi}$).
The contour levels are provided at 0.1, 1, 5, 10, 20, 50, 70, and 90\%
of the maximum of the distribution.  Note that the polarization fraction is
here defined over both the negative and positive ranges, due to the noise
of the intensity.}

\label{fig:pdf_impact_finitesnri}
\end{figure*}

\end{document}